\documentclass{article}%
\usepackage{amssymb}
\usepackage{amsmath}
\usepackage{graphicx}
\usepackage{amsfonts}%
\providecommand{\U}[1]{\protect\rule{.1in}{.1in}}

\newtheorem{proposition}{Proposition}

\begin{document}

\title{Quantum mechanics over sets}
\author{David Ellerman\\Department of Philosophy \\U. of California/Riverside}
\maketitle

\begin{abstract}
In the tradition of toy models of quantum mechanics in vector spaces over
finite fields (e.g., Schumacher and Westmoreland's "modal quantum theory"),
one finite field stands out, $%
\mathbb{Z}
_{2}$, since vectors over $%
\mathbb{Z}
_{2}$ have an interpretation as natural mathematical objects, i.e., sets. This
engages a sets-to-vector-spaces bridge that is part of the mathematical
folklore to translate both ways between set concepts and vector space
concepts. Using that bridge, the mathematical framework of
(finite-dimensional) quantum mechanics can be transported down to sets
resulting in \textit{quantum mechanics over sets} or \textit{QM/sets}. This
approach leads to a different treatment of Dirac's brackets than in "modal
quantum theory" (MQT), and that gives a full probability calculus (unlike MQT
that only has zero-one modalities of impossible and possible). That, in turn,
leads to a rather fulsome theory of QM over sets that includes "logical"
models of the double-slit experiment, Bell's Theorem, quantum information
theory, quantum computing, and much else. Indeed, QM/sets is proposed as the
"logic" of QM in the old-fashioned sense of "logic" as giving the simplified
essentials of a theory. QM/sets is also a key part of a broader research
program to provide an interpretation of QM based on the notion of "objective
indefiniteness," a program that grew out the recent development of the logic
of partitions mathematically dual to the usual Boolean logic of subsets.

\end{abstract}
\tableofcontents

\part{The lifting program and the probability calculus}

\section{Toy models of QM over finite fields}

In the tradition of "toy models" for quantum mechanics (QM), Schumacher and
Westmoreland \cite{schum:modal} and Hanson et al. \cite{hansonsabry:dqt} have
recently investigated models of quantum mechanics over finite fields. One
finite field stands out over the rest, $%
\mathbb{Z}
_{2}$, since vectors in a vector space over $%
\mathbb{Z}
_{2}$ have a natural interpretation, namely as \textit{sets} that are subsets
of a universe set. But in any vector space over a finite field, there is no
inner product so the first problem in constructing a toy model of QM in this
context is the definition of Dirac's brackets. Which aspects of the usual
treatment of the brackets should be retained and which aspects should be dropped?

Schumacher and Westmoreland (S\&W) chose to have their brackets continue to
have values in the base field, e.g., $%
\mathbb{Z}
_{2}=\left\{  0,1\right\}  $, so their "theory does not make use of the idea
of probability."\cite[p. 919]{schum:modal} Instead, the values of $0$ and $1$
are respectively interpreted modally as\textit{\ impossible} and
\textit{possible} and hence their name of "modal quantum theory." A number of
results from full QM carry over to their modal quantum theory, e.g.,
no-cloning, superdense coding, and teleportation, but without a probability
calculus, other results such as Bell's Theorem do not carry over: "in the
absence of probabilities and expectation values the Bell approach will not
work." \cite[p. 921]{schum:modal} Hence they develop a variation using the
modal concepts from a toy model by Hardy. \cite{hardy:nonlocality}

But all these limitations can be overcome by the different treatment of the
brackets taken here which yields a full probability calculus for a model of
\textit{quantum mechanics over sets} (QM/sets) using the $%
\mathbb{Z}
_{2}$ base field. Binary coding theory also uses vector spaces over $%
\mathbb{Z}
_{2}$, and one of the principal functions, the Hamming distance function
\cite{mceliece:coding}, takes non-negative integer values. Applied to two
subsets $S,T$ of a given universe set $U$, the Hamming distance function is
the cardinality $\left\vert S+T\right\vert $ of their symmetric difference
(i.e., the number of places where the two binary strings differ). In full QM,
the bracket $\left\langle \psi|\varphi\right\rangle $ is taken as the size of
the "overlap" between the two states. Hence it is natural in QM/sets to define
the bracket $\left\langle S|T\right\rangle $ applied to subsets $S,T\subseteq
U$ as the size of their overlap, i.e., the cardinality $\left\vert S\cap
T\right\vert $ of their intersection.

The usual QM formalism (always finite dimensional), e.g., the norm as the
square root of the brackets $\left\vert \psi\right\vert =\sqrt{\left\langle
\psi|\psi\right\rangle }$, can be developed in this context, and then Born's
Rule yields a probability calculus. And it is essentially a familiar calculus,
logical probability theory for a finite universe set of outcomes developed by
Laplace, Boole, and others. The only difference from that classical calculus
is the vector space formulation which allows different (equicardinal) bases or
universe sets of outcomes and thus it is "non-commutative." This allows the
development of the QM/sets version of many QM results such as Bell's Theorem,
the indeterminacy principle, double-slit experiments, and much else in the
context of finite sets. And that, in turn, helps to illuminate some of the
seemingly "weird" aspects of full QM.

By developing a sets-version of QM, the concepts and relationships of full QM
are represented in a pared-down ultra-simple version that can be seen as
representing the essential "logic" of QM. It represents the "logic of QM" in
that old sense of "logic" as giving the basic essentials of a theory (even
reduced to "zero-oneness"), not in the sense of giving the behavior of
propositions in a theory (which is the usual "quantum logic"). This approach
to full QM \cite{ell:objindef} arises out of the recent development of the
logic of partitions (\cite{ell:partitions} and \cite{ell:intropartlogic}) that
is (category-theoretically) dual to the ordinary Boolean logic of subsets
(which is usually mis-specified as the special case of propositional logic).

\section{The lifting sets-to-vector-spaces program}

\subsection{The basis principle}

There is a natural bridge (or ladder) between QM/sets and full QM based on the
mathematical relation between sets and vector spaces that is part of the
mathematical folklore. A subset can be viewed as a vector in a vector space
over $%
\mathbb{Z}
_{2}$, and a vector expressed in a basis can be viewed as a linearized set
where each (basis-) element in the set has a coefficient in the base field of
scalars. Using this conceptual bridge (or ladder), set-based concepts as in
QM/sets can be transported or "lifted" to vector space concepts as in QM, and
vector space concepts may be "delifted" or transported back to set concepts.
QM/sets is the delifted version of the mathematical machinery of QM, and,
conversely, the machinery of QM/sets lifts to give the mathematics of QM (but,
of course, not the specifically physical assumptions such as the Hamiltonian
or the DeBroglie relations connecting energy and frequency or momentum and wavelength).

The bridge from set concepts to vector space concepts has the guiding:

\begin{center}
\textbf{Basis Principle:}

\textit{Apply the set concept to a basis set and then linearly generate the
lifted vector space concept.\footnote{Intuitions can be guided by the
linearization map which takes a set $U$ to the (free) vector space $%
\mathbb{C}
^{U}$ where $u\in U$ lifts to the basis vector $\delta_{u}=\chi_{\left\{
u\right\}  }:U\rightarrow%
\mathbb{C}
$. But some choices are involved in the lifting program. For instance, the set
attribute $f:U\rightarrow%
\mathbb{R}
$ could be taken as defining the linear functional $%
\mathbb{C}
^{U}\rightarrow%
\mathbb{C}
$ that takes $\delta_{u}$ to $f\left(  u\right)  $ or the linear operator $%
\mathbb{C}
^{U}\rightarrow%
\mathbb{C}
^{U}$ that takes $\delta_{u}$ to $f\left(  u\right)  \delta_{u}$. We will see
that the latter is the right choice.}}
\end{center}

\noindent For instance, what is the vector space lift of the set concept of
cardinality? We apply the set concept of cardinality to a basis set of a
vector space where it yields the notion of \textit{dimension} of the vector
space (after checking that all bases have equal cardinality). Thus the lift of
set-cardinality is not the cardinality of a vector space but its
dimension.\footnote{In QM, the extension of concepts on finite dimensional
Hilbert space to infinite dimensional ones is well-known. Since our expository
purpose is conceptual rather than mathematical, we will stick to finite
dimensional spaces.} Thus the null set $\emptyset$ with cardinality $0$ lifts
to the trivial zero vector space with dimension $0$.

\subsection{Lifting partitions to vector spaces}

Given a universe set $U$, a \textit{partition} $\pi$ of $U$ is a set of
non-empty subsets or blocks (or cells) $\left\{  B\right\}  $ of $U$ that are
pairwise disjoint and whose union is $U$. In category-theoretic terms, a
partition is a direct sum decomposition of a set, and that concept will lift,
in the sets-to-vector-spaces lifting program, to the concept of a direct sum
decomposition of a vector space. We obtain this lifting by applying the basis
principle. Apply a set partition to a basis set of a vector space. Each block
$B$ of the set partition of the basis set linearly generates a subspace
$W_{B}\subseteq V$, and the subspaces together form a \textit{direct sum
decomposition}: $V=\sum_{B}\oplus W_{B}$. Thus the proper lifted notion of a
partition for a vector space is \textit{not} a set partition of a space
compatible with the vector space structure as would be defined by a subspace
$W\subseteq V$ where $v\thicksim v^{\prime}$ if $v-v^{\prime}\in W$. A
\textit{vector space partition} is a direct sum decomposition of the vector
space--which is not at all a set partition of the vector space.

\subsection{Lifting partition joins to vector spaces}

The main partition operation from partition logic that we need to lift to
vector spaces is the join operation. Two set partitions cannot be joined
unless they are \textit{compatible} in the sense of being defined on the same
universe set. This notion of compatibility lifts to vector spaces, via the
basis principle, by defining two vector space partitions (i.e., two direct sum
decompositions) $\omega=\left\{  W_{\lambda}\right\}  $ and $\xi=\{X_{\mu}\}$
on $V$ as being \textit{compatible} if there is a basis set for $V$ so that
the two vector space partitions arise from two set partitions of that common
or simultaneous basis set.

If two set partitions $\pi=\left\{  B\right\}  $ and $\sigma=\left\{
C\right\}  $ are compatible, then their \textit{join} $\pi\vee\sigma$ is
defined as the set partition whose blocks are the non-empty intersections
$B\cap C$. Similarly the lifted concept is that if two vector space partitions
$\omega=\left\{  W_{\lambda}\right\}  $ and $\xi=\{X_{\mu}\}$ are compatible,
then their \textit{join} $\omega\vee\xi$ is defined as the vector space
partition whose subspaces are the non-zero intersections $W_{\lambda}\cap
X_{\mu}$. And by the definition of compatibility, we could generate the
subspaces of the join $\omega\vee\xi$ by the blocks in the join of the two set
partitions of the common basis set.

\subsection{Lifting numerical attributes to linear operators}

A set partition might be seen as an abstract rendition of the inverse image
partition $\left\{  f^{-1}\left(  r\right)  \right\}  $ defined by some
concrete numerical attribute $f:U\rightarrow%
\mathbb{R}
$ on $U$. What is the lift of an attribute? At first glance, the basis
principle would seem to imply: define a set numerical attribute on a basis set
(with values in the base field) and then linearly generate a functional from
the vector space to the base field. But a functional does not define a vector
space partition; it only defines the set partition of the vector space
compatible with the vector space operations that is determined by the kernel
of the functional. Hence we need to try a more careful application of the
basis principle.

It is helpful to first give a suggestive reformulation of a set attribute
$f:U\rightarrow%
\mathbb{R}
$. If $f$ is constant on a subset $S\subseteq U$ with a value $r$, then we
might symbolize this as:

\begin{center}
$f\upharpoonright S=rS$
\end{center}

\noindent and suggestively call $S$ an "eigenvector" and $r$ an "eigenvalue."
\noindent The multiplication $rS$ is only formal and should be read as: the
function $f$ has the value $r$ on the subset $S$. For any "eigenvalue" $r$,
define power set $\wp(f^{-1}\left(  r\right)  )$ = "eigenspace of $r$" as the
set of all the "eigenvectors" with that "eigenvalue." Since the "eigenspaces"
span the set $U$, the attribute $f:U\rightarrow%
\mathbb{R}
$ can be represented by:

\begin{center}
$f=\sum_{r}r\chi_{f^{-1}\left(  r\right)  }:U\rightarrow%
\mathbb{R}
$

"Spectral decomposition" of set attribute $f:U\rightarrow%
\mathbb{R}
$
\end{center}

\noindent\lbrack where $\chi_{f^{-1}\left(  r\right)  }$ is the characteristic
function for the set $f^{-1}\left(  r\right)  $ and where the index $r$ runs
over the image or "spectrum" of the function $f:U\rightarrow%
\mathbb{R}
$].\footnote{There are two ways to think of the "set version" of a concept: as
a straight set concept with no mention of vector spaces over $%
\mathbb{Z}
_{2}$, or as a vector space over $%
\mathbb{Z}
_{2}$ concept (which already starts to combine set and vector space concepts).
For instance, the pure set concept of the partition given by an attribute
$f:U\rightarrow%
\mathbb{R}
$ is the set partition $\left\{  f^{-1}\left(  r\right)  \right\}  _{r}$ and
the "direct sum" is the set disjoint union $U=%
{\textstyle\biguplus_{r}}
f^{-1}\left(  r\right)  $. But this can be recast in $%
\mathbb{Z}
_{2}^{\left\vert U\right\vert }$ as the vector space direct sum: $\wp\left(
U\right)  =\sum_{r}\oplus\wp\left(  f^{-1}\left(  r\right)  \right)  $ of the
vector space partition $\left\{  \wp\left(  f^{-1}\left(  r\right)  \right)
\right\}  _{r}$.} Thus a set attribute determines a set partition and has a
constant value on the blocks of the set partition, so by the basis principle,
that lifts to a vector space concept that determines a vector space partition
and has a constant value on the blocks of the vector space partition.

The suggestive terminology gives the lift. The lift of $f\upharpoonright S=rS
$ is the eigenvector equation $Lv=\lambda v$ where $L$ is a linear operator on
$V$. The lift of $r$ is the eigenvalue $\lambda$ and the lift of an $S$ such
that $f\upharpoonright S=rS$ is an eigenvector $v$ such that $Lv=\lambda v$.
The lift of an "eigenspace" $\wp(f^{-1}\left(  r\right)  )$ is the eigenspace
$W_{\lambda}$ of an eigenvalue $\lambda$. The lift of the simplest attributes,
which are the characteristic functions $\chi_{f^{-1}\left(  r\right)  }$, are
the projection operators $P_{\lambda}$ that project to the eigenspaces
$W_{\lambda}$. The characteristic property of the characteristic functions
$\chi:U\rightarrow%
\mathbb{R}
$ is that they are idempotent in the sense that $\chi\left(  u\right)
\chi\left(  u\right)  =\chi\left(  u\right)  $ for all $u\in U$, and the
lifted characteristic property of the projection operators $P:V\rightarrow V$
is that they are idempotent in the sense that $P^{2}:V\rightarrow V\rightarrow
V=P:V\rightarrow V$. Finally, the "spectral decomposition" of a set attribute
lifts to the spectral decomposition of a \textit{vector space attribute}:

\begin{center}
$f=\sum_{r}r\chi_{f^{-1}\left(  r\right)  }:U\rightarrow%
\mathbb{R}
$ lifts to $L=\sum_{\lambda}\lambda P_{\lambda}:V\rightarrow V$

Lift of a set attribute to a vector space attribute
\end{center}

\noindent Thus a vector space attribute is just a linear operator whose
eigenspaces span the whole space which is called a \textit{diagonalizable}
\textit{linear operator} \cite{hk:la}. Then we see that the proper lift of a
set attribute using the basis principle does indeed define a vector space
partition, namely that of the eigenspaces of a diagonalizable linear operator,
and that the values of the attribute are constant on the blocks of the vector
space partition--as desired. To keep the eigenvalues of the linear operator
real, quantum mechanics restricts the vector space attributes to
\textit{Hermitian} (or \textit{self-adjoint}) linear operators, which
represent \textit{observables}, on a Hilbert space.

Hermann Weyl is one of the few quantum physicists who, in effect, outlined the
lifting program connecting QM/sets and QM. He called a partition a "grating"
or "sieve," and then considered both set partitions and vector space
partitions (direct sum decompositions) as the respective types of
gratings.\cite[pp. 255-257]{weyl:phil} He started with a numerical attribute
on a set, which defined the set partition or "grating" \cite[p. 255]%
{weyl:phil} with blocks having the same attribute-value. Then he moved to the
quantum case where the set or "aggregate of $n$ states has to be replaced by
an $n$-dimensional Euclidean vector space" \cite[p. 256]{weyl:phil} (note the
lift from cardinality $n$ sets to dimension $n$ vector spaces). The
appropriate notion of a vector space partition or "grating" is a "splitting of
the total vector space into mutually orthogonal subspaces" so that "each
vector $\overrightarrow{x}$ splits into $r$ component vectors lying in the
several subspaces" \cite[p. 256]{weyl:phil}, i.e., a vector space partition
(direct sum decomposition of the space).%

\begin{center}
\includegraphics[
height=3.0726in,
width=5.2619in
]%
{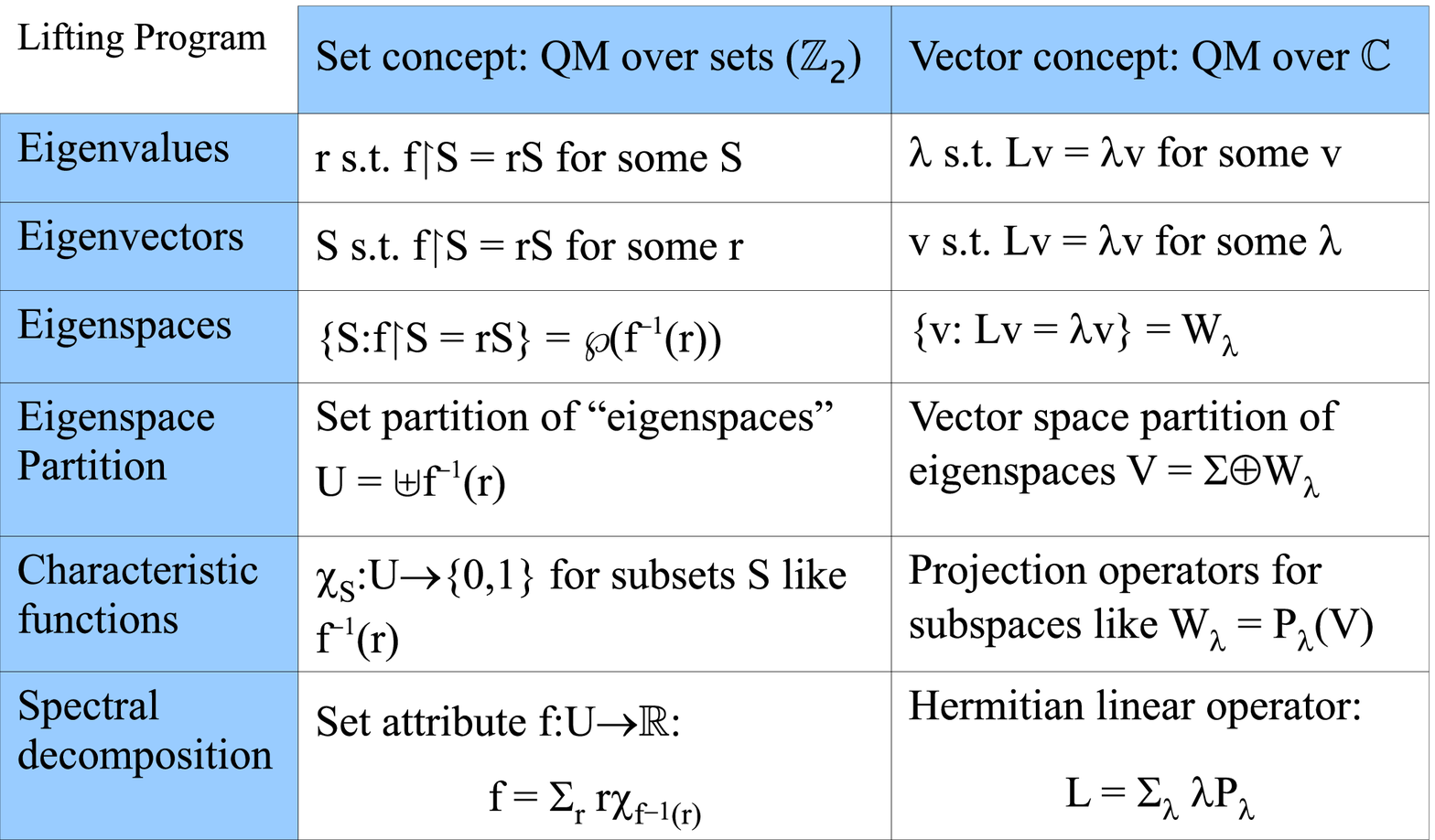}%
\\
Figure 1: Set numerical attributes lift to linear operators
\end{center}

\subsection{Lifting compatible attributes to commuting operators}

Since two set attributes $f:U\rightarrow%
\mathbb{R}
$ and $g:U^{\prime}\rightarrow%
\mathbb{R}
$ define two inverse image partitions $\left\{  f^{-1}\left(  r\right)
\right\}  $ and $\left\{  g^{-1}\left(  s\right)  \right\}  $ on their
domains, we need to extend the concept of compatible partitions to the
attributes that define the partitions. That is, two attributes $f:U\rightarrow%
\mathbb{R}
$ and $g:U^{\prime}\rightarrow%
\mathbb{R}
$ are \textit{compatible} if they have the same domain $U=U^{\prime}$. We have
previously lifted the notion of compatible set partitions to compatible vector
space partitions. Since real-valued set attributes lift to Hermitian linear
operators, the notion of compatible set attributes just defined would lift to
two linear operators being \textit{compatible} if their eigenspace partitions
are compatible. It is a standard fact of QM math (e.g., \cite[pp.
102-3]{hughes:interp} or \cite[p. 177]{hk:la}) that two (Hermitian) linear
operators $L,M:V\rightarrow V$ are compatible if and only if they commute,
$LM=ML$. Hence the \textit{commutativity} of linear operators is the lift of
the compatibility (i.e., defined on the same set) of set attributes. Thus the
join of two eigenspace partitions is defined iff the operators commute. Weyl
also pointed this out: "Thus combination [join] of two gratings [vector space
partitions] presupposes commutability...". \cite[p. 257]{weyl:phil}

Given two compatible set attributes $f:U\rightarrow%
\mathbb{R}
$ and $g:U\rightarrow%
\mathbb{R}
$, the join of their "eigenspace" partitions has as blocks the non-empty
intersections $f^{-1}\left(  r\right)  \cap g^{-1}\left(  s\right)  $. Each
block in the join of the "eigenspace" partitions could be characterized by the
ordered pair of "eigenvalues" $\left(  r,s\right)  $. An "eigenvector" of $f$,
$S\subseteq f^{-1}\left(  r\right)  $, and of $g$, $S\subseteq g^{-1}\left(
s\right)  $, would be a "simultaneous eigenvector": $S\subseteq f^{-1}\left(
r\right)  \cap g^{-1}\left(  s\right)  $.

In the lifted case, two commuting Hermitian linear operator $L$ and $M$ have
compatible eigenspace partitions $W_{L}=\left\{  W_{\lambda}\right\}  $ (for
the eigenvalues $\lambda$ of $L$) and $W_{M}=\left\{  W_{\mu}\right\}  $ (for
the eigenvalues $\mu$ of $M$). The blocks in the join $W_{L}\vee W_{M}$ of the
two compatible eigenspace partitions are the non-zero subspaces $\left\{
W_{\lambda}\cap W_{\mu}\right\}  $ which can be characterized by the ordered
pairs of eigenvalues $\left(  \lambda,\mu\right)  $. The nonzero vectors $v\in
W_{\lambda}\cap W_{\mu}$ are \textit{simultaneous eigenvectors} for the two
commuting operators, and there is a basis for the space consisting of
simultaneous eigenvectors.\footnote{One must be careful not to assume that the
simultaneous eigenvectors are the eigenvectors for the operator $LM=ML$ due to
the problem of degeneracy.}

A set of compatible set attributes is said to be \textit{complete} if the join
of their partitions is the discrete partition (the blocks have cardinality
$1$). Each element of $U$ is then characterized by the ordered $n $-tuple
$\left(  r,...,s\right)  $ of attribute values.

In the lifted case, a set of commuting linear operators is said to be
\textit{complete} if the join of their eigenspace partitions is nondegenerate,
i.e., the blocks have dimension $1$. The eigenvectors that generate those
one-dimensional blocks of the join are characterized by the ordered $n$-tuples
$\left(  \lambda,...,\mu\right)  $ of eigenvalues so the eigenvectors are
usually denoted as the eigenkets $\left\vert \lambda,...,\mu\right\rangle $ in
the Dirac notation. These \textit{Complete Sets of Commuting Operators} are
Dirac's CSCOs \cite{dirac:principles}.

\subsection{Summary of the QM/sets-to-QM bridge}

The lifting program or bridge developed so far is summarized in the following table.%

\begin{center}
\includegraphics[
height=3.6679in,
width=4.9356in
]%
{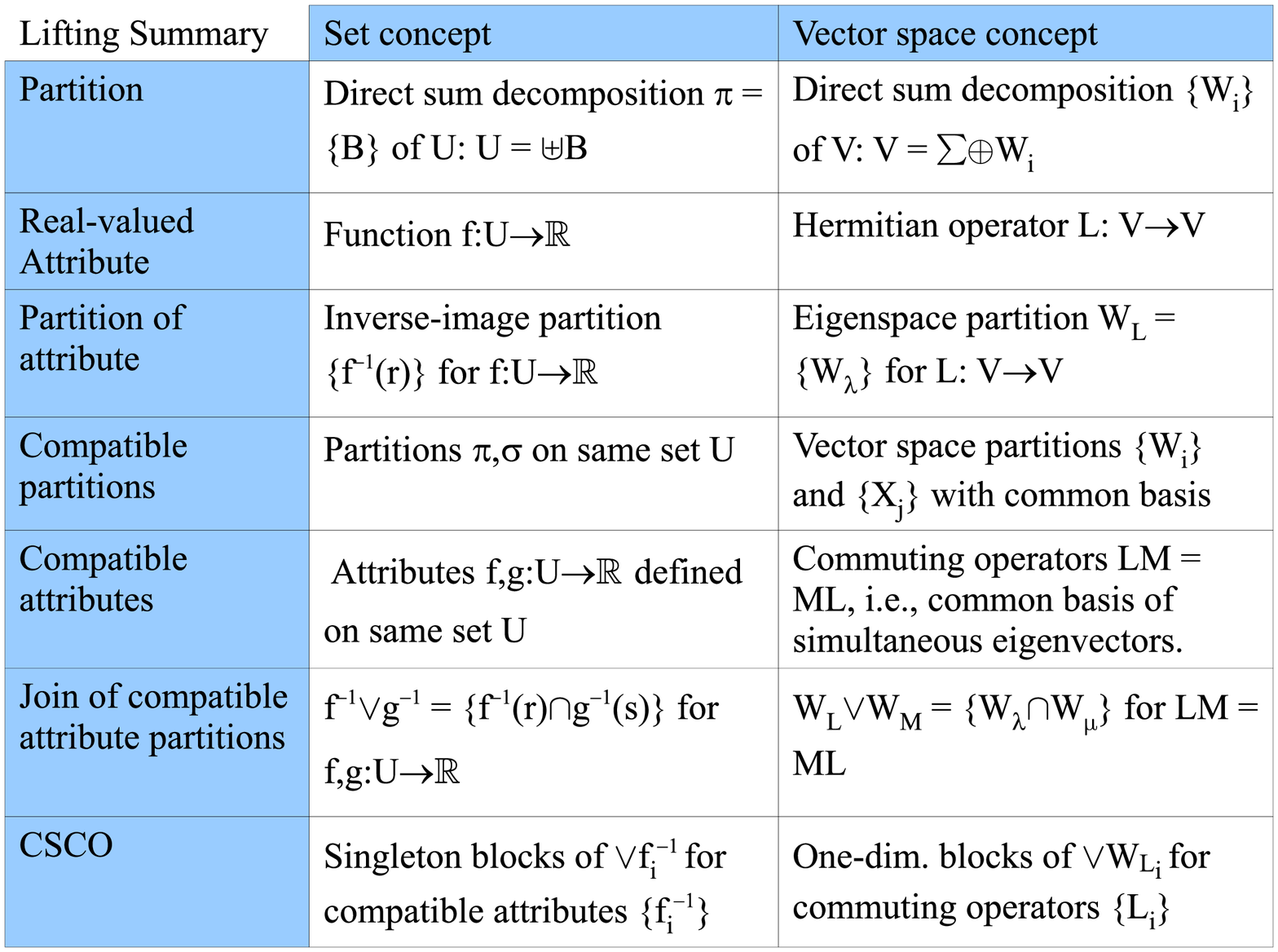}%
\\
Figure 2: Summary of Lifting Program
\end{center}

\section{The probability calculus in QM/sets}

\subsection{Vector spaces over $%
\mathbb{Z}
_{2}$}

The set version of QM is said to be "over $%
\mathbb{Z}
_{2}$" since the power set $\wp\left(  U\right)  $ (for a finite non-empty
universe set $U$) is a vector space over $%
\mathbb{Z}
_{2}=\left\{  0,1\right\}  $ where the subset addition $S+T$ is the
\textit{symmetric difference} (or inequivalence) of subsets, i.e.,
$S+T=S\not \equiv T=S\cup T-S\cap T$ for $S,T\subseteq U$. Given a finite
universe set $U=\left\{  u_{1},...,u_{n}\right\}  $ of cardinality $n$, the
$U$-basis in $%
\mathbb{Z}
_{2}^{n}$ is the set of singletons $\left\{  u_{1}\right\}  ,\left\{
u_{2}\right\}  ,...,\left\{  u_{n}\right\}  $ and a vector in $%
\mathbb{Z}
_{2}^{n}$ is specified in the $U$-basis by its $%
\mathbb{Z}
_{2}$-valued characteristic function $\chi_{S}:U\rightarrow%
\mathbb{Z}
_{2}$ for an subset $S\subseteq U$ (e.g., a string of $n$ binary numbers).
Similarly, a vector $v$ in $%
\mathbb{C}
^{n}$ is specified in terms of an orthonormal basis $\left\{  \left\vert
v_{i}\right\rangle \right\}  $ by a $%
\mathbb{C}
$-valued function $\left\langle \_|v\right\rangle :\left\{  v_{i}\right\}
\rightarrow%
\mathbb{C}
$ assigning a complex amplitude $\left\langle v_{i}|v\right\rangle $ to each
basis vector. One of the key pieces of mathematical machinery in QM, namely
the inner product, does not exist in vector spaces over finite fields but
basis-dependent "brackets" can still be defined and a norm or absolute value
can be defined to play a similar role in the probability algorithm of
QM/sets.\footnote{Often scare quotes, as in "brackets," are used to indicate
the named concept in QM/sets as opposed to full QM--although this may also be
clear from the context.}

Seeing $\wp\left(  U\right)  $ as the vector space $%
\mathbb{Z}
_{2}^{|U|}$ allows different bases in which the vectors can be expressed (as
well as the basis-free notion of a vector as a ket, since only the bra is
basis-dependent). Consider the simple case of $U=\left\{  a,b,c\right\}  $
where the $U$-basis is $\left\{  a\right\}  $, $\left\{  b\right\}  $, and
$\left\{  c\right\}  $. But the three subsets $\left\{  a,b\right\}  $,
$\left\{  b,c\right\}  $, and $\left\{  a,b,c\right\}  $ also form a basis
since: $\left\{  a,b\right\}  +\left\{  a,b,c\right\}  =\left\{  c\right\}  $;
$\left\{  b,c\right\}  +\left\{  c\right\}  =\left\{  b\right\}  $; and
$\left\{  a,b\right\}  +\left\{  b\right\}  =\left\{  a\right\}  $. These new
basis vectors could be considered as the basis-singletons in another
equicardinal universe $U^{\prime}=\left\{  a^{\prime},b^{\prime},c^{\prime
}\right\}  $ where $a^{\prime}=\left\{  a,b\right\}  $, $b^{\prime}=\left\{
b,c\right\}  $, and $c^{\prime}=\left\{  a,b,c\right\}  $. In the following
\textit{ket table}, each row is a ket of $V=%
\mathbb{Z}
_{2}^{3}$ expressed in the $U$-basis, the $U^{\prime}$-basis, and a
$U^{\prime\prime}$-basis.

\begin{center}%
\begin{tabular}
[c]{|c|c|c|}\hline
$U=\left\{  a,b,c\right\}  $ & $U^{\prime}=\left\{  a^{\prime},b^{\prime
},c^{\prime}\right\}  $ & $U^{\prime\prime}=\left\{  a^{\prime\prime
},b^{\prime\prime},c^{\prime\prime}\right\}  $\\\hline\hline
$\left\{  a,b,c\right\}  $ & $\left\{  c^{\prime}\right\}  $ & $\left\{
a^{\prime\prime},b^{\prime\prime},c^{\prime\prime}\right\}  $\\\hline
$\left\{  a,b\right\}  $ & $\left\{  a^{\prime}\right\}  $ & $\left\{
b^{\prime\prime}\right\}  $\\\hline
$\left\{  b,c\right\}  $ & $\left\{  b^{\prime}\right\}  $ & $\left\{
b^{\prime\prime},c^{\prime\prime}\right\}  $\\\hline
$\left\{  a,c\right\}  $ & $\left\{  a^{\prime},b^{\prime}\right\}  $ &
$\left\{  c^{\prime\prime}\right\}  $\\\hline
$\left\{  a\right\}  $ & $\left\{  b^{\prime},c^{\prime}\right\}  $ &
$\left\{  a^{\prime\prime}\right\}  $\\\hline
$\left\{  b\right\}  $ & $\left\{  a^{\prime},b^{\prime},c^{\prime}\right\}  $
& $\left\{  a^{\prime\prime},b^{\prime\prime}\right\}  $\\\hline
$\left\{  c\right\}  $ & $\left\{  a^{\prime},c^{\prime}\right\}  $ &
$\left\{  a^{\prime\prime},c^{\prime\prime}\right\}  $\\\hline
$\emptyset$ & $\emptyset$ & $\emptyset$\\\hline
\end{tabular}

Vector space isomorphism: $%
\mathbb{Z}
_{2}^{3}\cong\wp\left(  U\right)  \cong\wp\left(  U^{\prime}\right)  \cong%
\wp\left(  U^{\prime\prime}\right)  $ where row = ket.
\end{center}

\subsection{The brackets}

In a Hilbert space, the inner product is used to define the amplitudes
$\left\langle v_{i}|v\right\rangle $ and the norm $\left\vert v\right\vert
=\sqrt{\left\langle v|v\right\rangle }$, and the probability algorithm can be
formulated using this norm. In a vector space over $%
\mathbb{Z}
_{2}$, the Dirac notation can still be used but in a basis-dependent form
(like matrices as opposed to operators) that defines a real-valued norm even
though there is no inner product. The kets $\left\vert S\right\rangle $ for
$S\subseteq U$ are basis-free but the corresponding bras are basis-dependent.
For $u\in U$, the "\textit{bra"} $\left\langle \left\{  u\right\}  \right\vert
_{U}:\wp\left(  U\right)  \rightarrow%
\mathbb{R}
$ is defined by the "\textit{bracket"}:

\begin{center}
$\left\langle \left\{  u\right\}  |_{U}S\right\rangle =\left\{
\begin{array}
[c]{c}%
1\text{ if }u\in S\\
0\text{ if }u\notin S
\end{array}
\right.  =\chi_{S}\left(  u\right)  $
\end{center}

\noindent Then $\left\langle \left\{  u_{i}\right\}  |_{U}\left\{
u_{j}\right\}  \right\rangle =\chi_{\left\{  u_{j}\right\}  }\left(
u_{i}\right)  =\chi_{\left\{  u_{i}\right\}  }\left(  u_{j}\right)
=\delta_{ij} $ is the set-version of $\left\langle v_{i}|v_{j}\right\rangle
=\delta_{ij}$ (for an orthonormal basis $\left\{  \left\vert v_{i}%
\right\rangle \right\}  $). Assuming a finite $U$, the "bracket" linearly
extends to the more general basis-dependent form (where $\left\vert
S\right\vert $ is the cardinality of $S$):

\begin{center}
$\left\langle T|_{U}S\right\rangle =\left\vert T\cap S\right\vert $ for
$T,S\subseteq U$.\footnote{Thus $\left\langle T|_{U}S\right\rangle =\left\vert
T\cap S\right\vert $ takes values outside the base field of $%
\mathbb{Z}
_{2}$ just like the Hamming distance function $\left\vert T+S\right\vert $ on
vector spaces over $%
\mathbb{Z}
_{2}$ in coding theory \cite[p. 66]{mceliece:coding} as applied to pairs of
sets represented as binary strings.}
\end{center}

This basis principle can be run in reverse to "delift" a vector space concept
to sets. Consider an orthonormal basis set $\left\{  \left\vert v_{i}%
\right\rangle \right\}  $ in a finite dimensional Hilbert space. Given two
subsets $T,S\subseteq\left\{  \left\vert v_{i}\right\rangle \right\}  $ of the
basis set, consider the unnormalized superpositions $\psi_{T}=\sum_{\left\vert
v_{i}\right\rangle \in T}\left\vert v_{i}\right\rangle $ and $\psi_{S}%
=\sum_{\left\vert v_{i}\right\rangle \in S}\left\vert v_{i}\right\rangle $.
Then their inner product in the Hilbert space is $\left\langle \psi_{T}%
|\psi_{S}\right\rangle =\left\vert T\cap S\right\vert $, which "delifts"
(crossing the bridge in the other direction) to $\left\langle T|_{U}%
S\right\rangle =\left\vert T\cap S\right\vert $ for subsets $T,S\subseteq U$
of the $U$-basis of $%
\mathbb{Z}
_{2}^{\left\vert U\right\vert }$. In both cases, the bracket gives the size of
the overlap.

\subsection{Ket-bra resolution}

The basis-dependent "\textit{ket-bra}" $\left\vert \left\{  u\right\}
\right\rangle \left\langle \left\{  u\right\}  \right\vert _{U}$ is the
"one-dimensional" projection operator:

\begin{center}
$\left\vert \left\{  u\right\}  \right\rangle \left\langle \left\{  u\right\}
\right\vert _{U}=\left\{  u\right\}  \cap():\wp\left(  U\right)
\rightarrow\wp\left(  U\right)  $
\end{center}

\noindent and the "ket-bra identity" holds as usual:

\begin{center}
$\sum_{u\in U}\left\vert \left\{  u\right\}  \right\rangle \left\langle
\left\{  u\right\}  \right\vert _{U}=\sum_{u\in U}\left(  \left\{  u\right\}
\cap()\right)  =I:\wp\left(  U\right)  \rightarrow\wp\left(  U\right)  $
\end{center}

\noindent where the summation is the symmetric difference of sets in $%
\mathbb{Z}
_{2}^{n}$. The overlap $\left\langle T|_{U}S\right\rangle $ can be resolved
using the "ket-bra identity" in the same basis: $\left\langle T|_{U}%
S\right\rangle =\sum_{u}\left\langle T|_{U}\left\{  u\right\}  \right\rangle
\left\langle \left\{  u\right\}  |_{U}S\right\rangle $. Similarly a ket
$\left\vert S\right\rangle $ can be resolved in the $U$-basis;

\begin{center}
$\left\vert S\right\rangle =\sum_{u\in U}\left\vert \left\{  u\right\}
\right\rangle \left\langle \left\{  u\right\}  |_{U}S\right\rangle =\sum_{u\in
U}\left\langle \left\{  u\right\}  |_{U}S\right\rangle \left\vert \left\{
u\right\}  \right\rangle =\sum_{u\in U}\left\vert \left\{  u\right\}  \cap
S\right\vert \left\{  u\right\}  $
\end{center}

\noindent where a subset $S\subseteq U$ is just expressed as the sum of the
singletons $\left\{  u\right\}  \subseteq S$. That is ket-bra resolution in
sets. The ket $\left\vert S\right\rangle $ is the same as the ket $\left\vert
S^{\prime}\right\rangle $ for some subset $S^{\prime}\subseteq U^{\prime}$ in
another $U^{\prime}$-basis, but when the basis-dependent bra $\left\langle
\left\{  u\right\}  \right\vert _{U}$ is applied to the ket $\left\vert
S\right\rangle =\left\vert S^{\prime}\right\rangle $, then it is the subset
$S\subseteq U$, not $S^{\prime}\subseteq U^{\prime}$, that comes outside the
ket symbol $\left\vert \ \right\rangle $ in $\left\langle \left\{  u\right\}
|_{U}S\right\rangle =\left\vert \left\{  u\right\}  \cap S\right\vert
$.\footnote{The term "$\left\{  u\right\}  \cap S^{\prime}$" is not even
defined since it is the intersection of subsets of two different universes.
One of the luxuries of having a basis independent inner product in QM over $%
\mathbb{C}
$ is being able to ignore bases in the bra-ket notation.}

\subsection{The norm}

Then the (basis-dependent) $U$\textit{-norm} $\left\Vert S\right\Vert _{U}%
:\wp\left(  U\right)  \rightarrow%
\mathbb{R}
$ is defined, as usual, as the square root of the bracket:\footnote{We use the
double-line notation $\left\Vert S\right\Vert _{U}$ for the norm of a set to
distinguish it from the single-line notation $\left\vert S\right\vert $ for
the cardinality of a set, whereas the customary absolute value notation for
the norm of a vector in full QM is $\left\vert v\right\vert $.}

\begin{center}
$\left\Vert S\right\Vert _{U}=\sqrt{\left\langle S|_{U}S\right\rangle }%
=\sqrt{|S|}$
\end{center}

\noindent for $S\in\wp\left(  U\right)  $ which is the set-version of the
basis-free norm $\left\vert \psi\right\vert =\sqrt{\left\langle \psi
|\psi\right\rangle }$ (since the inner product does not depend on the basis).
Note that a ket has to be expressed in the $U$-basis to apply the
basis-dependent definition so in the above example, $\left\Vert \left\{
a^{\prime}\right\}  \right\Vert _{U}=\sqrt{2}$ since $\left\{  a^{\prime
}\right\}  =\left\{  a,b\right\}  $ in the $U$-basis.

\subsection{The Born rule}

For a specific basis $\left\{  \left\vert v_{i}\right\rangle \right\}  $ and
for any nonzero vector $v$ in a finite dimensional complex vector space,
$\left\vert v\right\vert ^{2}=\sum_{i}\left\langle v_{i}|v\right\rangle
\left\langle v_{i}|v\right\rangle ^{\ast}$ ($^{\ast}$ is complex conjugation)
whose set version would be: $\left\Vert S\right\Vert _{U}^{2}=\sum_{u\in
U}\left\langle \left\{  u\right\}  |_{U}S\right\rangle ^{2} $. Since

\begin{center}
$\left\vert v\right\rangle =\sum_{i}\left\langle v_{i}|v\right\rangle
\left\vert v_{i}\right\rangle $ and $\left\vert S\right\rangle =\sum_{u\in
U}\left\langle \left\{  u\right\}  |_{U}S\right\rangle \left\vert \left\{
u\right\}  \right\rangle $,
\end{center}

\noindent applying the Born rule by squaring the coefficients $\left\langle
v_{i}|v\right\rangle $ and $\left\langle \left\{  u\right\}  |_{U}%
S\right\rangle $ (and normalizing) gives the probabilities of the
eigen-elements $v_{i}$ or $\left\{  u\right\}  $ given a state $v$ or $S$ in
QM and QM/sets:

\begin{center}
$\sum_{i}\frac{\left\langle v_{i}|v\right\rangle \left\langle v_{i}%
|v\right\rangle ^{\ast}}{\left\vert v\right\vert ^{2}}=1$ and $\sum_{u}%
\frac{\left\langle \left\{  u\right\}  |_{U}S\right\rangle ^{2}}{\left\Vert
S\right\Vert _{U}^{2}}=\sum_{u}\frac{\left\vert \left\{  u\right\}  \cap
S\right\vert }{\left\vert S\right\vert }=1$
\end{center}

\noindent where $\frac{\left\langle v_{i}|v\right\rangle \left\langle
v_{i}|v\right\rangle ^{\ast}}{\left\vert v\right\vert ^{2}}$ is a `mysterious'
quantum probability while $\frac{\left\langle \left\{  u\right\}
|_{U}S\right\rangle ^{2}}{\left\Vert S\right\Vert _{U}^{2}}=\frac{\left\vert
\left\{  u\right\}  \cap S\right\vert }{\left\vert S\right\vert }$ is the
unmysterious Laplacian equal probability $\Pr\left(  \left\{  u\right\}
|S\right)  $ rule for getting $u$ when sampling $S$.\footnote{Note that there
is no notion of a normalized vector in a vector space over $%
\mathbb{Z}
_{2}$ (another consequence of the lack of an inner product). The normalization
is, as it were, postponed to the probability algorithm which is computed in
the rationals.}

\subsection{Spectral decomposition on sets}

An observable, i.e., a Hermitian operator, on a Hilbert space determines its
home basis set of orthonormal eigenvectors. In a similar manner, a real-valued
attribute $f:U\rightarrow%
\mathbb{R}
$ defined on $U$ has the $U$-basis as its "home basis set." As previously
noted, the connection between the numerical attributes $f:U\rightarrow%
\mathbb{R}
$ of QM/sets and the Hermitian operators of QM is established by "seeing" the
function $f$ as a formal operator: $f\upharpoonright():\wp\left(  U\right)
\rightarrow\wp\left(  U\right)  $. Applied to the basis elements $\left\{
u\right\}  \subseteq U$, we may write $f\upharpoonright\left\{  u\right\}
=f\left(  u\right)  \left\{  u\right\}  =r\left\{  u\right\}  $ as the
set-version of an eigenvalue equation applied to an eigenvector where the
multiplication $r\left\{  u\right\}  $ is only formal (read $r\left\{
u\right\}  $ as: the function $f$ takes the value $r$ on $\left\{  u\right\}
$). Then for any subset $S\subseteq f^{-1}\left(  r\right)  $ where $f$ is
constant, we may also formally write: $f\upharpoonright S=rS$ as an
"eigenvalue equation" satisfied by all the "eigenvectors" $S$ in the
"eigenspace" $\wp\left(  f^{-1}\left(  r\right)  \right)  $, a subspace of
$\wp\left(  U\right)  $, for the "eigenvalue" $r$. Since $f^{-1}\left(
r\right)  \cap():\wp\left(  U\right)  \rightarrow\wp\left(  U\right)  $ is the
projection operator\footnote{Since $\wp\left(  U\right)  $ is now interpreted
as a vector space, it should be noted that the projection operator
$T\cap():\wp\left(  U\right)  \rightarrow\wp\left(  U\right)  $ is not only
idempotent but linear, i.e., $\left(  T\cap S_{1}\right)  +(T\cap S_{2}%
)=T\cap\left(  S_{1}+S_{2}\right)  $. Indeed, this is the distributive law
when $\wp\left(  U\right)  $ is interpreted as a Boolean ring.} to the
"eigenspace" $\wp\left(  f^{-1}\left(  r\right)  \right)  $ for the
"eigenvalue" $r$, we have the spectral decomposition for a Hermitian operator
$L=\sum_{\lambda}\lambda P_{\lambda} $ in QM and for a $U$-attribute
$f:U\rightarrow%
\mathbb{R}
$ in QM/sets:

\begin{center}
$L=\sum_{\lambda}\lambda P_{\lambda}:V\rightarrow V$ and $f\upharpoonright
()=\sum_{r}r\left(  f^{-1}\left(  r\right)  \cap()\right)  :\wp\left(
U\right)  \rightarrow\wp\left(  U\right)  $

Spectral decomposition of operators in QM and QM/sets.
\end{center}

When the base field increases from $%
\mathbb{Z}
_{2}$ to $%
\mathbb{R}
$ or $%
\mathbb{C}
$, then the formal multiplication $r\left(  f^{-1}\left(  r\right)
\cap()\right)  $ is internalized as an actual multiplication, and the
projection operator $f^{-1}\left(  r\right)  \cap()$ on sets becomes a
projection operator on a vector space over $%
\mathbb{R}
$ or $%
\mathbb{C}
$. Thus the operator representation $L=\sum_{\lambda}\lambda P_{\lambda}$ of
an observable numerical attribute is just the internalization of a numerical
attribute made possible by the enriched base field $%
\mathbb{R}
$ or $%
\mathbb{C}
$. Similarly, the set brackets $\left\langle T|_{U}S\right\rangle $ taking
values outside the base field $%
\mathbb{Z}
_{2}$ become internalized as an inner product with the same enrichment of the
base field. It is the comparative "poverty" of the base field $%
\mathbb{Z}
_{2}$ that requires the QM/sets "brackets" to take "de-internalized" or
"externalized" values outside the base field and for a formal multiplication
to used in the operator presentation $f\upharpoonright()=\sum_{r}r\left(
f^{-1}\left(  r\right)  \cap()\right)  $ of a numerical attribute
$f:U\rightarrow%
\mathbb{R}
$.\footnote{In the engineering literature, eigenvalues are seen as "stretching
or shrinking factors" but that is \textit{not} their role in QM. The whole
machinery of eigenvectors [e.g., $f\upharpoonright\left\{  u\right\}
=r\left\{  u\right\}  $], eigenspaces [e.g., $\wp\left(  f^{-1}\left(
r\right)  \right)  $], and eigenvalues [e.g., $f(u)=r$] in QM is a way of
representing a numerical attribute [e.g., $f:U\rightarrow%
\mathbb{R}
$] \textit{inside} a vector space that has a rich enough base field.} Or put
the other way around, the only numerical attributes that can be internally
represented in $\wp\left(  U\right)  \cong%
\mathbb{Z}
_{2}^{n}$ are the characteristic functions $\chi_{S}:U\rightarrow%
\mathbb{Z}
_{2}$ that are internally represented in the $U$-basis as the projection
operators $S\cap():\wp\left(  U\right)  \rightarrow\wp\left(  U\right)  $.

\subsection{Completeness and orthogonality of projection operators}

The usual completeness and orthogonality conditions on eigenspaces also have
set-versions in QM over $%
\mathbb{Z}
_{2}$:

\begin{enumerate}
\item completeness: $\sum_{\lambda}P_{\lambda}=I:V\rightarrow V$ has the
set-version: $\sum_{r}f^{-1}\left(  r\right)  \cap()=I:\wp\left(  U\right)
\rightarrow\wp\left(  U\right)  $, and

\item orthogonality: for $\lambda\neq\lambda^{\prime}$, $P_{\lambda}%
P_{\lambda^{\prime}}=0:V\rightarrow V$ (where $0$ is the zero operator) has
the set-version: for $r\neq r^{\prime}$, $\left[  f^{-1}\left(  r\right)
\cap()\right]  \left[  f^{-1}\left(  r^{\prime}\right)  \cap()\right]
=\emptyset\cap():\wp\left(  U\right)  \rightarrow\wp\left(  U\right)
$.\footnote{Note that in spite of the lack of an inner product, the
orthogonality of projection operators $S\cap()$ is perfectly well defined in
QM/sets where it boils down to the disjointness of subsets, i.e., the
cardinality of their overlap (instead of their inner product) being $0$.}
\end{enumerate}

\subsection{Measuring attributes on sets}

\noindent The Pythagorean results (for the complete and orthogonal projection operators):

\begin{center}
$\left\vert v\right\vert ^{2}=\sum_{\lambda}\left\vert P_{\lambda}\left(
v\right)  \right\vert ^{2}$ and $\left\Vert S\right\Vert _{U}^{2}=\sum
_{r}\left\Vert f^{-1}\left(  r\right)  \cap S\right\Vert _{U}^{2}$,
\end{center}

\noindent give the probabilities for measuring attributes. Since

\begin{center}
$\left\vert S\right\vert =\left\Vert S\right\Vert _{U}^{2}=\sum_{r}\left\Vert
f^{-1}\left(  r\right)  \cap S\right\Vert _{U}^{2}=\sum_{r}\left\vert
f^{-1}\left(  r\right)  \cap S\right\vert $
\end{center}

\noindent we have in QM and in QM/sets:

\begin{center}
$\sum_{\lambda}\frac{\left\vert P_{\lambda}\left(  v\right)  \right\vert ^{2}%
}{\left\vert v\right\vert ^{2}}=1$ and $\sum_{r}\frac{\left\Vert f^{-1}\left(
r\right)  \cap S\right\Vert _{U}^{2}}{\left\Vert S\right\Vert _{U}^{2}}%
=\sum_{r}\frac{\left\vert f^{-1}\left(  r\right)  \cap S\right\vert
}{\left\vert S\right\vert }=1$
\end{center}

\noindent where $\frac{\left\vert P_{\lambda}\left(  v\right)  \right\vert
^{2}}{\left\vert v\right\vert ^{2}}$ is the quantum probability of getting
$\lambda$ in an $L$-measurement of $v$ while $\frac{\left\vert f^{-1}\left(
r\right)  \cap S\right\vert }{\left\vert S\right\vert }$ has the rather
unmysterious interpretation of the probability $\Pr\left(  r|S\right)  $ of
the random variable $f:U\rightarrow%
\mathbb{R}
$ having the "eigen-value" $r$ when sampling $S\subseteq U$. Thus the
set-version of the Born rule is not some weird "quantum" notion of probability
on sets but the perfectly ordinary Laplace-Boole rule for the conditional
probability $\frac{\left\vert f^{-1}\left(  r\right)  \cap S\right\vert
}{\left\vert S\right\vert }$, given $S\subseteq U$, of a random variable
$f:U\rightarrow%
\mathbb{R}
$ having the value $r$.

\subsection{Contextuality}

Given a ket $\left\vert S\right\rangle $, the probability of getting another
ket $\left\vert \left\{  a\right\}  \right\rangle $ as an outcome of a
measurement in QM/sets will depend on the context in terms of the measurement
basis. In the previous ket table, comparing sets in the $U$-basis and
$U^{\prime\prime}$-basis, we see that $\left\{  a,b\right\}  =\left\{
b^{\prime\prime}\right\}  $ (or in the ket notation: $\left\vert \left\{
a,b\right\}  \right\rangle =\left\vert \left\{  b^{\prime\prime}\right\}
\right\rangle $) and $\left\{  a\right\}  =\left\{  a^{\prime\prime}\right\}
$. Taking $S=\left\{  a,b\right\}  $, the probability of getting $\left\{
a\right\}  $ in a $U$-basis measurement is: $\Pr\left(  \left\{  a\right\}
|S\right)  =|\left\{  a\right\}  \cap\left\{  a,b\right\}  |/\left\vert
\left\{  a,b\right\}  \right\vert =1/2$. But taking the same ket $\left\vert
\left\{  a,b\right\}  \right\rangle =\left\vert \left\{  b^{\prime\prime
}\right\}  \right\rangle $ as the given state and measuring in the
$U^{\prime\prime}$-basis, the probability of getting the ket $\left\vert
\left\{  a\right\}  \right\rangle =\left\vert \left\{  a^{\prime\prime
}\right\}  \right\rangle $ is: $\Pr\left(  \left\{  a^{\prime\prime}\right\}
|\left\{  b^{\prime\prime}\right\}  \right)  =\left\vert \left\{
a^{\prime\prime}\right\}  \cap\left\{  b^{\prime\prime}\right\}  \right\vert
/\left\vert \left\{  b^{\prime\prime}\right\}  \right\vert =0$.

\subsection{The objective indefiniteness interpretation}

On top of the mathematics of QM/sets, there is an objective indefiniteness
interpretation which is just the set-version of the objective indefiniteness
interpretation of QM developed elsewhere \cite{ell:objindef}. The
collecting-together of some elements $u\in U$ into a subset $S\subseteq U$ is
interpreted as the superposition of the "eigen-elements" $u\in S$ to form an
"indefinite element" $S$ (with the vector sum $S=\sum_{u\in U}\left\langle
\left\{  u\right\}  |_{U}S\right\rangle \left\{  u\right\}  $ in the vector
space $\wp\left(  U\right)  $ over $%
\mathbb{Z}
_{2}$ giving the superposition).\footnote{In logic, a \textit{choice function
}is a function $\varepsilon()$ that applied to a non-empty subset $S\subseteq
U$ picks out an element $\varepsilon\left(  S\right)  =u\in S$ (or
equivalently a singleton $\varepsilon\left(  S\right)  =\left\{  u\right\}
\subseteq S$). The indeterminancy of a choice function is, as it were, where
stochasticity enters QM. For finite sets, we might consider a probabilistic
choice function that would pick out any element (or singleton) of $S$ with the
equal probability $1/\left\vert S\right\vert $. A (non-degenerate)
"measurement" in QM/sets is a "physical" version of a probabilistic choice
function; it goes from an indefinite element $S$ to some definite element
$\left\{  u\right\}  \subseteq S$ with the probability $1/\left\vert
S\right\vert $.}

The indefinite element $S$ is being "measured" using the "observable" $f$
where the probability $\Pr\left(  r|S\right)  $ of getting the "eigenvalue" $r
$ is $\frac{\left\vert f^{-1}\left(  r\right)  \cap S\right\vert }{\left\vert
S\right\vert }$ and where the "damned quantum jump" goes from $S$ to the
"projected resultant state" $f^{-1}\left(  r\right)  \cap S$ which is in the
"eigenspace" $\wp\left(  f^{-1}\left(  r\right)  \right)  $ for that
"eigenvalue" $r$. That state represents a more-definite element $f^{-1}\left(
r\right)  \cap S$ that now has the definite $f$-value of $r$--so a second
measurement would yield the same "eigenvalue" $r$ and the same vector
$f^{-1}\left(  r\right)  \cap\left[  f^{-1}\left(  r\right)  \cap S\right]
=f^{-1}\left(  r\right)  \cap S$ using the idempotency of the set-version of
projection operators (all as in the standard Dirac-von-Neumann treatment of
measurement). These questions of interpretation will not be emphasized here
where the focus is on the mathematical relationship between QM/sets and full QM.

\subsection{Summary of the probability calculus}

These set-versions and more (the average value of an attribute is treated
later) are summarized in the following table for a finite $U$ and a finite
dimensional Hilbert space $V$ with $\left\{  \left\vert v_{i}\right\rangle
\right\}  $ as any orthonormal basis.

\begin{center}%
\begin{tabular}
[c]{|c|c|}\hline
Vector space over $%
\mathbb{Z}
_{2}$: QM/sets & Hilbert space case: QM over $%
\mathbb{C}
$\\\hline\hline
Projections: $S\cap():\wp\left(  U\right)  \rightarrow\wp\left(  U\right)  $ &
$P:V\rightarrow V$\\\hline
Spectral Decomp.: $f\upharpoonright()=\sum_{r}r\left(  f^{-1}\left(  r\right)
\cap()\right)  $ & $L=\sum_{\lambda}\lambda P_{\lambda}$\\\hline
Compl.: $\sum_{r}f^{-1}\left(  r\right)  \cap()=I:\wp\left(  U\right)
\rightarrow\wp\left(  U\right)  $ & $\sum_{\lambda}P_{\lambda}=I$\\\hline
Orthog.: $r\neq r^{\prime}$, $\left[  f^{-1}\left(  r\right)  \cap()\right]
\left[  f^{-1}\left(  r^{\prime}\right)  \cap()\right]  =\emptyset\cap()$ &
$\lambda\neq\lambda^{\prime}$, $P_{\lambda}P_{\lambda^{\prime}}=0$\\\hline
Brackets: $\left\langle S|_{U}T\right\rangle =\left\vert S\cap T\right\vert $
= overlap for $S,T\subseteq U$ & $\left\langle \psi|\varphi\right\rangle =$
"overlap" of $\psi$ and $\varphi$\\\hline
Ket-bra: $\sum_{u\in U}\left\vert \left\{  u\right\}  \right\rangle
\left\langle \left\{  u\right\}  \right\vert _{U}=\sum_{u\in U}\left(
\left\{  u\right\}  \cap()\right)  =I$ & $\sum_{i}\left\vert v_{i}%
\right\rangle \left\langle v_{i}\right\vert =I$\\\hline
Resolution: $\left\langle S|_{U}T\right\rangle =\sum_{u}\left\langle
S|_{U}\left\{  u\right\}  \right\rangle \left\langle \left\{  u\right\}
|_{U}T\right\rangle $ & $\left\langle \psi|\varphi\right\rangle =\sum
_{i}\left\langle \psi|v_{i}\right\rangle \left\langle v_{i}|\varphi
\right\rangle $\\\hline
Norm: $\left\Vert S\right\Vert _{U}=\sqrt{\left\langle S|_{U}S\right\rangle
}=\sqrt{\left\vert S\right\vert }$ where $S\subseteq U$ & $\left\vert
\psi\right\vert =\sqrt{\left\langle \psi|\psi\right\rangle }$\\\hline
Pythagoras: $\left\Vert S\right\Vert _{U}^{2}=\sum_{u\in U}\left\langle
\left\{  u\right\}  |_{U}S\right\rangle ^{2}=\left\vert S\right\vert $ &
$\left\vert \psi\right\vert ^{2}=\sum_{i}\left\langle v_{i}|\psi\right\rangle
^{\ast}\left\langle v_{i}|\psi\right\rangle $\\\hline
Laplace: $S\neq\emptyset$, $\sum_{u\in U}\frac{\left\langle \left\{
u\right\}  |_{U}S\right\rangle ^{2}}{\left\Vert S\right\Vert _{U}^{2}}%
=\sum_{u\in S}\frac{1}{\left\vert S\right\vert }=1$ & $\left\vert
\psi\right\rangle \neq0$, $\sum_{i}\frac{\left\langle v_{i}|\psi\right\rangle
^{\ast}\left\langle v_{i}|\psi\right\rangle }{\left\vert \psi\right\vert ^{2}%
}=\frac{\left\vert \left\langle v_{i}|\psi\right\rangle \right\vert ^{2}%
}{\left\vert \psi\right\vert ^{2}}=1$\\\hline
Born: $\left\vert S\right\rangle =\sum_{u\in U}\left\langle \left\{
u\right\}  |_{U}S\right\rangle \left\vert \left\{  u\right\}  \right\rangle $,
$\Pr\left(  u|S\right)  =\frac{\left\langle \left\{  u\right\}  |_{U}%
S\right\rangle ^{2}}{\left\Vert S\right\Vert _{U}^{2}}$ & $\left\vert
\psi\right\rangle =\sum_{i}\left\langle v_{i}|\psi\right\rangle \left\vert
v_{i}\right\rangle $, $\Pr\left(  v_{i}|\psi\right)  =\frac{\left\vert
\left\langle v_{i}|\psi\right\rangle \right\vert ^{2}}{\left\vert
\psi\right\vert ^{2}}$\\\hline
$\left\Vert S\right\Vert _{U}^{2}=\sum_{r}\left\Vert f^{-1}\left(  r\right)
\cap S\right\Vert _{U}^{2}=\sum_{r}\left\vert f^{-1}\left(  r\right)  \cap
S\right\vert =\left\vert S\right\vert $ & $\left\vert \psi\right\vert
^{2}=\sum_{\lambda}\left\vert P_{\lambda}\left(  \psi\right)  \right\vert
^{2}$\\\hline
$S\neq\emptyset$, $\sum_{r}\frac{\left\Vert f^{-1}\left(  r\right)  \cap
S\right\Vert _{U}^{2}}{\left\Vert S\right\Vert _{U}^{2}}=\sum_{r}%
\frac{\left\vert f^{-1}\left(  r\right)  \cap S\right\vert }{\left\vert
S\right\vert }=1$ & $\left\vert \psi\right\rangle \neq0$, $\sum_{\lambda}%
\frac{\left\vert P_{\lambda}\left(  \psi\right)  \right\vert ^{2}}{\left\vert
\psi\right\vert ^{2}}=1$\\\hline
Measurement: $\Pr(r|S)=\frac{\left\Vert f^{-1}\left(  r\right)  \cap
S\right\Vert _{U}^{2}}{\left\Vert S\right\Vert _{U}^{2}}=\frac{\left\vert
f^{-1}\left(  r\right)  \cap S\right\vert }{\left\vert S\right\vert }$ &
$\Pr\left(  \lambda|\psi\right)  =\frac{\left\vert P_{\lambda}\left(
\psi\right)  \right\vert ^{2}}{\left\vert \psi\right\vert ^{2}}$\\\hline
Average of attribute: $\left\langle f\right\rangle _{S}=\frac{\left\langle
S|_{U}f\upharpoonright()|S\right\rangle }{\left\langle S|_{U}S\right\rangle }$
& $\left\langle L\right\rangle _{\psi}=\frac{\left\langle \psi|L|\psi
\right\rangle }{\left\langle \psi|\psi\right\rangle }$.\\\hline
\end{tabular}

Probability mathematics for QM over $%
\mathbb{Z}
_{2}$ and for QM over $%
\mathbb{C}
$
\end{center}

\section{Measurement in QM/sets}

\subsection{Measurement as partition join operation}

In QM/sets, numerical attributes $f:U\rightarrow%
\mathbb{R}
$ can be considered as equiprobable random variables on a set of outcomes $U$.
The inverse images of attributes (or random variables) define set partitions
$\left\{  f^{-1}\left(  r\right)  \right\}  $ on the set of outcomes $U$.
Considered abstractly, the partitions on a set $U$ are partially ordered by
refinement where a partition $\pi=\left\{  B\right\}  $ \textit{refines} a
partition $\sigma=\left\{  C\right\}  $, written $\sigma\preceq\pi$, if for
any block $B\in\pi$, there is a block $C\in\sigma$ such that $B\subseteq C$.
The principal logical operation needed here is the \textit{partition join}:
$\pi\vee\sigma$ is the partition whose blocks are the non-empty intersections
$B\cap C$ for $B\in\pi$ and $C\in\sigma$.

Each partition $\pi$ can be represented as a binary relation
$\operatorname*{dit}\left(  \pi\right)  \subseteq U\times U$ on $U$ where the
ordered pairs $\left(  u,u^{\prime}\right)  $ in $\operatorname*{dit}\left(
\pi\right)  $ are the \textit{distinctions} or \textit{dits} of $\pi$ in the
sense that $u$ and $u^{\prime}$ are in distinct blocks of $\pi$. These dit
sets $\operatorname*{dit}\left(  \pi\right)  $ as binary relations might be
called "partition relations" but they are also the "apartness relations" in
computer science. An ordered pair $\left(  u,u^{\prime}\right)  $ is an
\textit{indistinction} or \textit{indit} of $\pi$ if $u$ and $u^{\prime}$ are
in the same block of $\pi$. The set of indits, $\operatorname*{indit}\left(
\pi\right)  $, as a binary relation is just the equivalence relation
associated with the partition $\pi$.

In the duality between the ordinary Boolean logic of subsets (usually
mis-specified as "propositional" logic) and the logic of partitions
(\cite{ell:partitions} or \cite{ell:intropartlogic}), the elements of a subset
and the distinctions of a partition are dual concepts. The partial ordering of
subsets in the powerset Boolean algebra $\wp\left(  U\right)  $ is the
inclusion of elements and the refinement ordering of partitions on $U$ is just
the inclusion of dit sets, i.e., $\sigma\preceq\pi$ iff $\operatorname*{dit}%
\left(  \sigma\right)  \subseteq\operatorname*{dit}\left(  \pi\right)  $. The
partial ordering in each case is a lattice where the top of the Boolean
lattice is the subset $U$ of all possible elements and the top of the lattice
of partitions is the \textit{discrete partition} $\mathbf{1}=\left\{  \left\{
u\right\}  \right\}  _{u\in U}$ of singletons which makes all possible
distinctions: $\operatorname*{dit}\left(  \mathbf{1}\right)  =U\times
U-\Delta$ (where $\Delta=\left\{  \left(  u,u\right)  :u\in U\right\}  $ is
the diagonal). The bottom of the Boolean lattice is the empty set $\emptyset$
of no elements and the bottom of the lattice of partitions is the
\textit{indiscrete partition }(or \textit{blob}) $\mathbf{0}=\left\{
U\right\}  $ which makes no distinctions.

The two lattices can be illustrated in the case of $U=\left\{  a,b,c\right\}
$.%

\begin{center}
\includegraphics[
height=1.8182in,
width=3.6903in
]%
{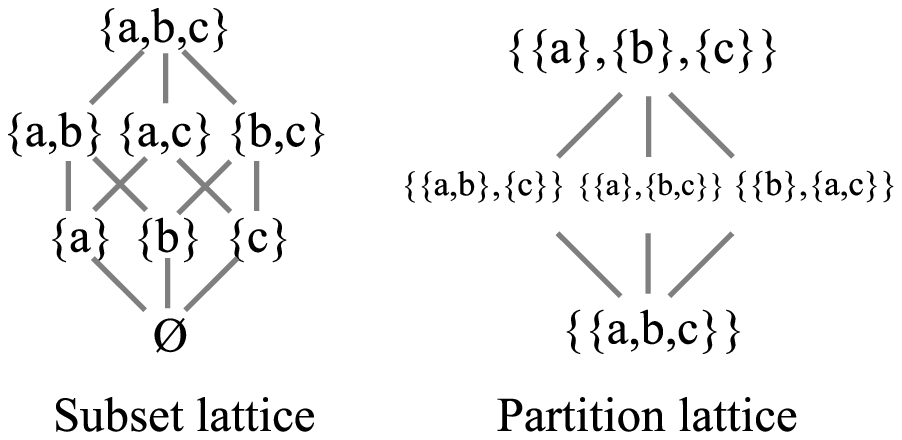}%
\\
Figure 3: Subset and partition lattices
\end{center}

In the correspondences between QM/sets and QM, a block in a partition on $U$
[i.e., a vector in $\wp\left(  U\right)  $] corresponds to pure state in QM (a
state vector in a quantum state space), and a partition on $U$ can be thought
of as a mixture of orthogonal pure states with the probabilities given by the
probability calculus on QM/sets. Given a "pure state" $S\subseteq U$, the
possible results of a non-degenerate $U$-measurement are the blocks of the
discrete partition $\left\{  \left\{  u\right\}  \right\}  _{u\in S}$ on $S$
with each singleton being equiprobable. Each such measurement would have one
of the potential "eigenstates" $\left\{  u\right\}  \subseteq S$ as the actual result.

In QM, measurements make distinctions that turn a pure state into a mixture.
The abstract essentials of measurement are represented in QM/sets as a
distinction-creating processes of turning a "pure state" $S$ into a "mixed
state" partition on $S$ (with "distinctions" as defined above in partition
logic). The distinction-creating process of "measurement" in QM/sets is the
partition join of the indiscrete partition $\left\{  S\right\}  $ (taking $S$
as the universe) and the inverse-image partition $\left\{  f^{-1}\left(
r\right)  \right\}  $ of the numerical attribute $f:U\rightarrow%
\mathbb{R}
$ restricted to $S$. Again Weyl gets it right. Weyl refers to a partition as a
"grating" or "sieve" and then notes that "Measurement means application of a
sieve or grating" \cite[p. 259]{weyl:phil}, e.g., the application (i.e., join)
of the set-grating $\left\{  f^{-1}\left(  r\right)  \right\}  _{r}$ to the
"pure state" $\left\{  S\right\}  $ to give the "mixed state" $\left\{  S\cap
f^{-1}\left(  r\right)  \right\}  _{r}$.

\subsection{Nondegenerate measurements}

In the simple example illustrated below, we start at the one block or "state"
of the indiscrete partition or blob which is the completely indistinct element
$\left\{  a,b,c\right\}  $. A measurement always uses some attribute that
defines an inverse-image partition on $U=\left\{  a,b,c\right\}  $. In the
case at hand, there are "essentially" four possible attributes that could be
used to "measure" the indefinite element $\left\{  a,b,c\right\}  $ (since
there are four partitions that refine the blob).

For an example of a "nondegenerate measurement," consider any attribute
$f:U\rightarrow%
\mathbb{R}
$ which has the discrete partition as its inverse image, such as the ordinal
number of a letter in the alphabet: $f\left(  a\right)  =1$, $f\left(
b\right)  =2$, and $f\left(  c\right)  =3$. This attribute or "observable" has
three "eigenvectors": $f\upharpoonright\left\{  a\right\}  =1\left\{
a\right\}  $, $f\upharpoonright\left\{  b\right\}  =2\left\{  b\right\}  $,
and $f\upharpoonright\left\{  c\right\}  =3\left\{  c\right\}  $ with the
corresponding "eigenvalues." The "eigenvectors" are $\left\{  a\right\}  $,
$\left\{  b\right\}  $, and $\left\{  c\right\}  $, the blocks in the discrete
partition of $U$. Starting in the "pure state" $S=\left\{  a,b,c\right\}  $, a
$U$-measurement using the observable $f$ gives the "mixed state":

\begin{center}
$\left\{  U\right\}  \vee\left\{  f^{-1}\left(  r\right)  \right\}
_{r=1,2,3}=\mathbf{0}\vee\mathbf{1}=\mathbf{1}$.
\end{center}

\noindent Each such measurement would return an "eigenvalue" $r$ with the
probability of $\Pr\left(  r|S\right)  =\frac{\left\vert f^{-1}\left(
r\right)  \cap S\right\vert }{\left\vert S\right\vert }=\frac{1}{3}$.

A "projective measurement" makes distinctions in the measured "state" that are
sufficient to induce the "quantum jump" or "projection" to the "eigenvector"
associated with the observed "eigenvalue." If the observed "eigenvalue" was
$3$, then the "state" $\left\{  a,b,c\right\}  $ "projects" to $f^{-1}\left(
3\right)  \cap\left\{  a,b,c\right\}  =\left\{  c\right\}  \cap\left\{
a,b,c\right\}  =\left\{  c\right\}  $ as pictured below.%

\begin{center}
\includegraphics[
height=1.457in,
width=2.2349in
]%
{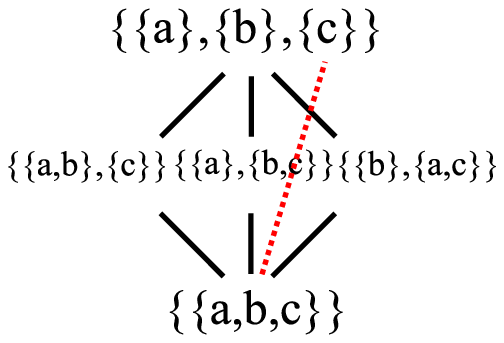}%
\end{center}

\begin{center}
Figure 4: "Nondegenerate measurement"
\end{center}

It might be emphasized that this is an objective state reduction (or "collapse
of the wave packet") from the single indefinite element $\left\{
a,b,c\right\}  $ to the single definite element $\left\{  c\right\}  $, not a
subjective removal of ignorance as if the "state" had all along been $\left\{
c\right\}  $. For instance, Pascual Jordan in 1934 argued that:

\begin{quotation}
\noindent the electron is forced to a decision. We compel it to assume a
definite position; previously, in general, it was neither here nor there; it
had not yet made its decision for a definite position... . ... [W]e ourselves
produce the results of the measurement. (quoted in \cite[p. 161]{jammer:phil})
\end{quotation}

This might be illustrated using Weyl's notion of a partition as a "sieve or
grating" \cite[p. 259]{weyl:phil} that is applied in a measurement. We might
think of a grating as a series of regular polygonal shapes that might be
imposed on an indefinite blob of dough. In a measurement, the blob of dough
falls through one of the polygonal holes with equal probability and then takes
on that shape.%

\begin{center}
\includegraphics[
height=2.1096in,
width=2.4251in
]%
{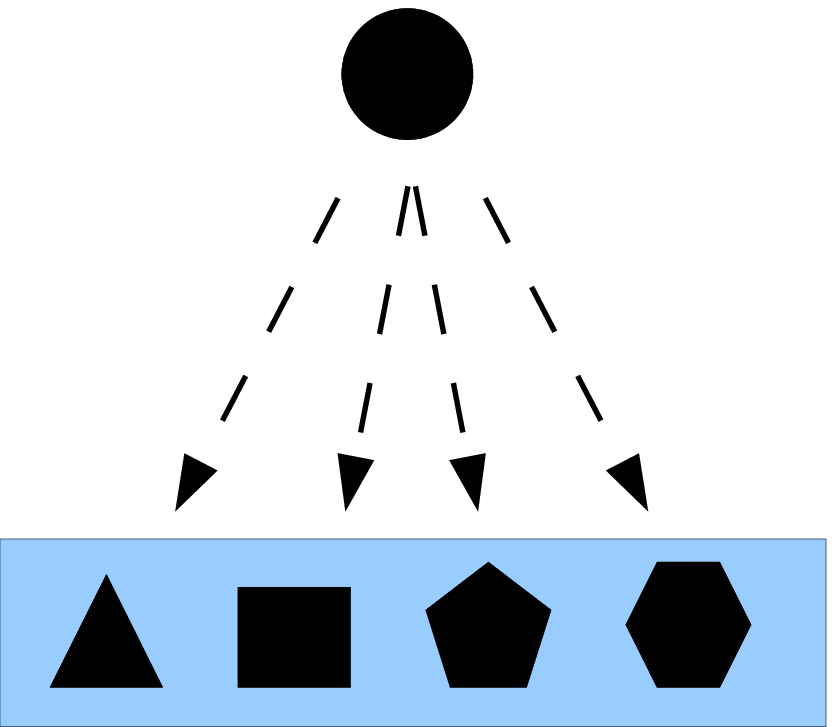}%
\end{center}

\begin{center}
Figure 5: Measurement as randomly giving an indefinite blob of dough a regular
polygonal shape.
\end{center}

\subsection{Degenerate measurements}

For an example of a "degenerate measurement," we choose an attribute with a
non-discrete inverse-image partition such as $\pi=\left\{  \left\{  a\right\}
,\left\{  b,c\right\}  \right\}  $. Hence the attribute could just be the
characteristic function $\chi_{\left\{  b,c\right\}  }$ with the two
"eigenspaces" $\wp(\left\{  a\right\}  )$ and $\wp(\left\{  b,c\right\}  )$
and the two "eigenvalues" $0$ and $1$ respectively. Since one of the two
"eigenspaces" is not a singleton of an eigen-element, the "eigenvalue" of $1$
is a set version of a "degenerate eigenvalue." This attribute $\chi_{\left\{
b,c\right\}  }$ has four (non-zero) "eigenvectors": $\chi_{\left\{
b,c\right\}  }\upharpoonright\left\{  b,c\right\}  =1\left\{  b,c\right\}  $,
$\chi_{\left\{  b,c\right\}  }\upharpoonright\left\{  b\right\}  =1\left\{
b\right\}  $, $\chi_{\left\{  b,c\right\}  }\upharpoonright\left\{  c\right\}
=1\left\{  c\right\}  $, and $\chi_{\left\{  b,c\right\}  }\upharpoonright
\left\{  a\right\}  =0\left\{  a\right\}  $.

The "measuring apparatus" makes distinctions by "joining" the "observable" partition

\begin{center}
$\mathbf{\chi}_{\left\{  b,c\right\}  }^{-1}=\left\{  \chi_{\left\{
b,c\right\}  }^{-1}\left(  1\right)  ,\chi_{\left\{  b,c\right\}  }%
^{-1}\left(  0\right)  \right\}  =\left\{  \left\{  b,c\right\}  ,\{a\right\}
\}$
\end{center}

\noindent with the "pure state" which is the single block representing the
indefinite element $S=U=\left\{  a,b,c\right\}  $. A measurement apparatus of
that "observable" returns one of "eigenvalues" with certain probabilities:

\begin{center}
$\Pr(0|S)=\frac{\left\vert \left\{  a\right\}  \cap\left\{  a,b,c\right\}
\right\vert }{\left\vert \left\{  a,b,c\right\}  \right\vert }=\frac{1}{3}$
and $\Pr\left(  1|S\right)  =\frac{\left\vert \left\{  b,c\right\}
\cap\left\{  a,b,c\right\}  \right\vert }{\left\vert \left\{  a,b,c\right\}
\right\vert }=\frac{2}{3}$.
\end{center}

Suppose it returns the "eigenvalue" $1$. Then the indefinite element $\left\{
a,b,c\right\}  $ "jumps" to the "projection" $\chi_{\left\{  b,c\right\}
}^{-1}\left(  1\right)  \cap\left\{  a,b,c\right\}  =\left\{  b,c\right\}  $
of the "state" $\left\{  a,b,c\right\}  $ to that "eigenvector" \cite[p.
221]{cohen-t:QM1}.

Since this is a "degenerate" result (i.e., the "eigenspaces" don't all have
"dimension" one), another measurement is needed to make more distinctions.
Measurements by attributes that give either of the other two partitions,
$\left\{  \left\{  a,b\right\}  ,\{c\right\}  \}$ or $\left\{  \left\{
b\right\}  ,\left\{  a,c\right\}  \right\}  $, suffice to distinguish
$\left\{  b,c\right\}  $ into $\left\{  b\right\}  $ or $\left\{  c\right\}
$, so either attribute together with the attribute $\chi_{\left\{
b,c\right\}  }$ would form a \textit{complete set of compatible attributes}
(i.e., the set version of a CSCO). The join of the two attributes' partitions
gives the discrete partition. Taking the other attribute as $\chi_{\left\{
a,b\right\}  }$, the join of the two attributes' partitions is discrete:

\begin{center}
$\mathbf{\chi}_{\left\{  b,c\right\}  }^{-1}\vee\mathbf{\chi}_{\left\{
a,b\right\}  }^{-1}=\left\{  \left\{  a\right\}  ,\left\{  b,c\right\}
\right\}  \vee\left\{  \left\{  a,b\right\}  ,\{c\right\}  \}=\left\{
\left\{  a\right\}  ,\left\{  b\right\}  ,\left\{  c\right\}  \right\}
=\mathbf{1}$.
\end{center}

\noindent Hence all the "eigenstate" singletons can be characterized by the
ordered pairs of the "eigenvalues" of these two "observables": $\left\{
a\right\}  =\left\vert 0,1\right\rangle $, $\left\{  b\right\}  =\left\vert
1,1\right\rangle $, and $\left\{  c\right\}  =\left\vert 1,0\right\rangle $
(using Dirac's ket-notation to give the ordered pairs).

The second "projective measurement" of the indefinite "superposition" element
$\left\{  b,c\right\}  $ using the attribute $\chi_{\left\{  a,b\right\}  }$
with the "eigenspace" partition $\chi_{\left\{  a,b\right\}  }^{-1}=\left\{
\left\{  a,b\right\}  ,\{c\right\}  \}$ would induce a jump to either
$\left\{  b\right\}  $ or $\left\{  c\right\}  $ with the probabilities:

\begin{center}
$\Pr\left(  1|\left\{  b,c\right\}  \right)  =\frac{\left\vert \left\{
a,b\right\}  \cap\left\{  b,c\right\}  \right\vert }{\left\vert \left\{
b,c\right\}  \right\vert }=\frac{1}{2}$ and $\Pr\left(  0|\left\{
b,c\right\}  \right)  =\frac{\left\vert \left\{  c\right\}  \cap\left\{
b,c\right\}  \right\vert }{\left\vert \left\{  b,c\right\}  \right\vert
}=\frac{1}{2}$.
\end{center}

\noindent If the measured "eigenvalue" is $0$, then the "state" $\left\{
b,c\right\}  $ "projects" to $\chi_{\left\{  a,b\right\}  }^{-1}\left(
0\right)  \cap\left\{  b,c\right\}  =\left\{  c\right\}  $ as pictured below.%

\begin{center}
\includegraphics[
height=1.457in,
width=2.2349in
]%
{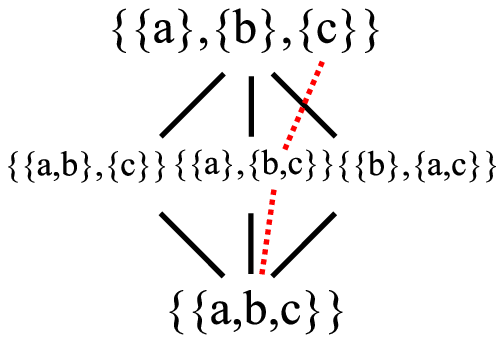}%
\end{center}

\begin{center}
Figure 6: "Degenerate measurement"
\end{center}

\noindent The two "projective measurements" of $\left\{  a,b,c\right\}  $
using the complete set of compatible (both defined on $U$) attributes
$\chi_{\left\{  b,c\right\}  }$ and $\chi_{\left\{  a,b\right\}  }$ produced
the respective "eigenvalues" $1$ and $0$, and the resulting "eigenstate" was
characterized by the "eigenket" $\left\vert 1,0\right\rangle =\{c\}$.

\section{"Time" evolution in QM/sets}

The different "de-internalized" treatment of the "brackets" in QM/sets gives a
probability calculus, unlike Schumacher and Westmoreland's "modal quantum
theory." \cite{schum:modal} But both theories agree that evolution of the
quantum states over $%
\mathbb{Z}
_{2}$ is given by non-singular linear transformations. These transformations
are, of course, reversible like the unitary transformations of full QM but
"unitary" is not defined in the absence of an inner product. QM/sets
nevertheless has basis-dependent "brackets" and those "brackets" are preserved
if we change the basis along with the non-singular transformation. Let $A:%
\mathbb{Z}
_{2}^{n}\rightarrow%
\mathbb{Z}
_{2}^{n}$ be a non-singular transformation where the images of the $U$-basis
$A\left\vert \left\{  u\right\}  \right\rangle $ are taken as the basis
vectors $\left\{  u^{\prime}\right\}  $ of a $U^{\prime}$-basis. Then for
$S,T\subseteq U$, we have the following preservation of the "brackets":

\begin{center}
$\left\langle T|_{U}S\right\rangle =\left\langle AT|_{AU}AS\right\rangle
=\left\langle T^{\prime}|_{U^{\prime}}S^{\prime}\right\rangle $
\end{center}

\noindent where $AT=T^{\prime}\subseteq U^{\prime}=AU$ and $AS=S^{\prime
}\subseteq U^{\prime}=AU$ .

In the objective indefiniteness interpretation of QM based on partition logic
\cite{ell:objindef}, von Neumann's type 1 processes (measurements) and type 2
processes (unitary evolution) \cite{vonn:mfqm} are modeled respectively as the
processes that make distinctions or that don't make any distinctions in the
strong sense of preserving the degree of indistinctness $\left\langle
\varphi|\psi\right\rangle $ between quantum states. That characterization of
evolution carries over to QM/sets since it is precisely the non-singular
transformations that preserve distinctness of QM/sets quantum states, i.e.,
distinctness of non-zero vectors in $%
\mathbb{Z}
_{2}^{n}$.

By rendering QM concepts in the simple context of sets, QM/sets gives an
understanding of the basic logic of the QM concept. Much effort has been
expended in the philosophy of QM to understand measurement. We have seen that
by rendering QM measurement in the context of sets that it is the
distinction-making process of applying the partition $\left\{  f^{-1}\left(
r\right)  \right\}  _{r}$ of an observable attribute to a pure state partition
$\left\{  S\right\}  $ (i.e., taking the partition join) to get the mixed
state partition $\left\{  f^{-1}\left(  r\right)  \cap S\right\}  _{r}$. Now
we see that time evolution in QM (i.e., a degree-of-indistinctness
$\left\langle \psi|\varphi\right\rangle $ preserving transformation) is
modeled in QM/sets by distinction-preserving non-singular transformations.
This explains von Neumann's classification of the two types of quantum
processes: the distinction-making or type 1 processes (measurement) and the
distinction-preserving or type 2 processes (time evolution). In this manner,
QM/sets brings out the essence or "logic" of the full QM concepts of
measurement and time evolution.

\section{Interference without "waves" in QM/sets}

The role of the so-called "waves" in ordinary quantum mechanics can be further
clarified by viewing quantum dynamics in QM/sets. In QM over $%
\mathbb{C}
$, suppose the Hamiltonian $H$ has an orthonormal basis of energy eigenstate
$\left\{  \left\vert E_{j}\right\rangle \right\}  $. Then the application of
the unitary propagation operator $U\left(  t\right)  $ from $t=0$ to time $t$
applied to $\left\vert \psi_{0}\right\rangle =\sum_{j}c_{j}\left\vert
E_{j}\right\rangle $ has the action:

\begin{center}
$U\left(  t\right)  \left\vert \psi_{0}\right\rangle =\left\vert \psi
_{t}\right\rangle =e^{iHt}\left\vert \psi_{0}\right\rangle =\sum_{j}%
c_{j}e^{iHt}\left\vert E_{j}\right\rangle =\sum_{j}c_{j}e^{iE_{j}t}\left\vert
E_{j}\right\rangle $.
\end{center}

\noindent Thus $U\left(  t\right)  $ transforms the orthonormal basis
$\left\{  \left\vert E_{j}\right\rangle \right\}  $ into the orthonormal basis
$\left\{  \left\vert E_{j}^{\prime}\right\rangle \right\}  =\left\{
e^{iE_{j}t}\left\vert E_{j}\right\rangle \right\}  $.\footnote{Indeed, a
\textit{unitary} operator on an inner product space can be defined as a linear
operator that transforms an orthonormal basis into an orthonormal basis.} Even
though this unitary transformation introduces different relative phases for
the different energy eigenstates in $U\left(  t\right)  \left\vert \psi
_{0}\right\rangle $, the probabilities for an energy measurement do not change
since $\left\vert c_{j}\right\vert ^{2}=\left\vert c_{j}e^{iE_{j}t}\right\vert
^{2}$. The effects of time evolution show when the evolved state $U\left(
t\right)  \left\vert \psi_{0}\right\rangle $ is measured in \textit{another}
basis $\left\{  \left\vert a_{k}\right\rangle \right\}  $. Suppose for each
$j$, $\left\vert E_{j}\right\rangle =\sum_{k}\alpha_{k}^{j}\left\vert
a_{k}\right\rangle $ so that:

\begin{center}
$U\left(  t\right)  \left\vert \psi_{0}\right\rangle =\left\vert \psi
_{t}\right\rangle =\sum_{j}c_{j}e^{iE_{j}t}\left\vert E_{j}\right\rangle
=\sum_{j}c_{j}e^{iE_{j}t}\sum_{k}\alpha_{k}^{j}\left\vert a_{k}\right\rangle
=\sum_{k}\left(  \sum_{j}c_{j}e^{iE_{j}t}\alpha_{k}^{j}\right)  \left\vert
a_{k}\right\rangle $.
\end{center}

\noindent Then under time evolution, there is interference in the coefficient
$\sum_{j}c_{j}e^{iE_{j}t}\alpha_{k}^{j}$ of each eigenstate $\left\vert
a_{k}\right\rangle $. Since the complex exponentials $e^{iE_{j}t}$ can be
mathematically interpreted as "waves," this is the interference characteristic
of wave-like behavior in the evolution of the quantum state $\left\vert
\psi_{0}\right\rangle $.

But there is interference without waves in QM/sets where many of the
characteristic phenomena of QM can nevertheless be reproduced (see later
sections on the two-slit experiment and Bell's Theorem). Suppose we start with
a state $S\subseteq U=\left\{  u_{1},...,u_{n}\right\}  $ which is represented
in the $U$-basis as $\left\vert S\right\rangle =\sum_{j}\left\langle
u_{j}|_{U}S\right\rangle \left\vert u_{j}\right\rangle =\sum_{j}%
b_{j}\left\vert u_{j}\right\rangle $ where $\left\langle u_{j}|_{U}%
S\right\rangle =b_{j}\in%
\mathbb{Z}
_{2}$. Then the "dynamics" of a nonsingular transformation $A:%
\mathbb{Z}
_{2}^{n}\rightarrow%
\mathbb{Z}
_{2}^{n}$ takes the basis $\left\{  \left\vert u_{j}\right\rangle \right\}  $
to another basis $\left\{  \left\vert u_{j}^{\prime}\right\rangle \right\}  $
(where $A\left\vert u_{j}\right\rangle =\left\vert u_{j}^{\prime}\right\rangle
$) which is the set or binary vector space version of $U\left(  t\right)  $
taking the orthonormal basis $\left\{  \left\vert E_{j}\right\rangle \right\}
$ to the orthonormal basis $\left\{  \left\vert E_{j}^{\prime}\right\rangle
\right\}  $ where $\left\vert E_{j}^{\prime}\right\rangle =e^{iE_{j}%
t}\left\vert E_{j}\right\rangle $. Thus $\left\vert S\right\rangle $ is
transformed, by linearity, into $\left\vert S^{\prime}\right\rangle =\sum
_{j}b_{j}\left\vert u_{j}^{\prime}\right\rangle $ with the same $b_{j}$'s so
that $\Pr\left(  u_{j}|S\right)  =\frac{b_{j}^{2}}{\left\vert S\right\vert
}=\frac{b_{j}^{2}}{\left\vert S^{\prime}\right\vert }=\Pr\left(  u_{j}%
^{\prime}|S^{\prime}\right)  $ and $\left\langle S|_{U}T\right\rangle
=\left\langle S^{\prime}|_{U^{\prime}}T^{\prime}\right\rangle $ (where for
$T\subseteq U$, $A\left\vert T\right\rangle =\left\vert T^{\prime
}\right\rangle $ for some $T^{\prime}\subseteq U^{\prime}$). But the state
$\left\vert S^{\prime}\right\rangle =\sum_{j}b_{j}\left\vert u_{j}^{\prime
}\right\rangle $ could be measured in another $U^{\prime\prime}$-basis
$\left\{  \left\vert u_{j}^{\prime\prime}\right\rangle \right\}  $ where
$\left\vert u_{j}^{\prime}\right\rangle =\sum_{k}\alpha_{k}^{j}\left\vert
u_{k}^{\prime\prime}\right\rangle $ so that:

\begin{center}
$A\left\vert S\right\rangle =\left\vert S^{\prime}\right\rangle =\sum_{j}%
b_{j}\left\vert u_{j}^{\prime}\right\rangle =\sum_{j}b_{j}\sum_{k}\alpha
_{k}^{j}\left\vert u_{k}^{\prime\prime}\right\rangle =\sum_{k}\left(  \sum
_{j}b_{j}\alpha_{k}^{j}\right)  \left\vert u_{j}^{\prime\prime}\right\rangle $.
\end{center}

\noindent Then under time evolution, there is interference in the coefficient
$\sum_{j}b_{j}\alpha_{k}^{j}$ of each eigenstate $\left\vert u_{j}%
^{\prime\prime}\right\rangle $. This suffices to give the interference
phenomena that are ordinarily seen as characteristic of wave-like behavior but
there is not even the mathematics of waves in QM/sets. The mathematics of
waves (complex exponentials $e^{i\varphi}$) comes into the mathematics of
quantum mechanics \textit{only over }$%
\mathbb{C}
$; real exponentials either grow or decay but don't behave as waves.

The following table summarizes the results using the minimal superpositions:
$\left\vert S\right\rangle =b_{1}\left\vert u_{1}\right\rangle +b_{2}%
\left\vert u_{2}\right\rangle $ and $\left\vert \psi_{0}\right\rangle
=c_{1}\left\vert E_{1}\right\rangle +c_{2}\left\vert E_{2}\right\rangle $.

\begin{center}%
\begin{tabular}
[c]{|c|c|}\hline
QM/sets & QM\\\hline\hline
$\left\vert u_{j}\right\rangle \overset{A}{\rightarrow}\left\vert
u_{j}^{\prime}\right\rangle $ & $\left\vert E_{j}\right\rangle
\overset{U}{\rightarrow}\left\vert E_{j}^{\prime}\right\rangle =e^{ig_{j}%
t}\left\vert E_{j}\right\rangle $\\\hline
$\left\vert S\right\rangle =b_{1}\left\vert u_{1}\right\rangle +b_{2}%
\left\vert u_{2}\right\rangle \rightarrow b_{1}\left\vert u_{1}^{\prime
}\right\rangle +b_{2}\left\vert u_{2}^{\prime}\right\rangle $ & $\left\vert
\psi_{0}\right\rangle =c_{1}\left\vert E_{1}\right\rangle +c_{2}\left\vert
E_{2}\right\rangle \rightarrow c_{1}\left\vert E_{1}^{\prime}\right\rangle
+c_{2}\left\vert E_{2}^{\prime}\right\rangle $\\\hline
$\left\vert u_{j}^{\prime}\right\rangle =\sum_{k}\left\langle u_{k}%
^{\prime\prime}|_{U^{\prime\prime}}u_{j}^{\prime}\right\rangle \left\vert
u_{k}^{\prime\prime}\right\rangle =\sum_{k}\alpha_{k}^{j}\left\vert
u_{k}^{\prime\prime}\right\rangle $ & $\left\vert E_{j}\right\rangle =\sum
_{k}\alpha_{k}^{j}\left\vert a_{k}\right\rangle $; $\left\vert E_{j}^{\prime
}\right\rangle =e^{ig_{j}t}\sum_{k}\alpha_{k}^{j}\left\vert a_{k}\right\rangle
$\\\hline
$b_{1}\left\vert u_{1}\right\rangle +b_{2}\left\vert u_{2}\right\rangle
\rightarrow\sum_{k}\left(  b_{1}\alpha_{k}^{1}+b_{2}\alpha_{k}^{2}\right)
\left\vert u_{k}^{\prime\prime}\right\rangle $ & $c_{1}\left\vert
E_{1}\right\rangle +c_{2}\left\vert E_{2}\right\rangle \rightarrow\sum
_{k}\left(  c_{1}e^{ig_{1}t}\alpha_{k}^{1}+c_{2}e^{ig_{2}t}\alpha_{k}%
^{2}\right)  \left\vert a_{k}\right\rangle $\\\hline
\end{tabular}

Table showing the role in interference in QM/sets and in QM
\end{center}

Thus QM/sets allows us to tease the QM behavior due to interference apart from
the specifically wave-version of that interference in QM over $%
\mathbb{C}
$. The root of the interference is superposition, i.e., the different $j$'s in
the coefficients $\sum_{j}c_{j}e^{iE_{j}t}\alpha_{k}^{j}$ in QM or $\sum
_{j}b_{j}\alpha_{k}^{j}$ in QM/sets, and superposition is the mathematical
representation of indefiniteness. It is indefiniteness that is the basic
feature, and a particle in a superposition state for a certain observable will
have the evolution of that indefiniteness expressed by coefficients $\sum
_{j}c_{j}e^{iE_{j}t}\alpha_{k}^{j}$ using complex exponentials (i.e., the
mathematics of waves) so the indefiniteness will then appear as "wave-like"
behavior--even though there are no physical waves in QM.

\section{Double-slit experiment in QM/sets}

QM/sets represents the logical essence of full QM without any of the physical
assumptions. Hence to delift the double-slit experiment to QM/sets, we need to
imagine the elements of some $U$-basis as "positions" and an non-singular
matrix $A$ as giving the dynamic evolution for one "time" period.

Consider the dynamics given in terms of the $U$-basis where: $\left\{
a\right\}  \rightarrow\left\{  a,b\right\}  $; $\left\{  b\right\}
\rightarrow\left\{  a,b,c\right\}  $; and $\left\{  c\right\}  \rightarrow
\left\{  b,c\right\}  $ in one time period. This is represented by the
non-singular one-period change of state matrix:

\begin{center}
$A=%
\begin{bmatrix}
\left\langle \left\{  a\right\}  |_{U}\left\{  a,b\right\}  \right\rangle  &
\left\langle \left\{  a\right\}  |_{U}\left\{  a,b,c\right\}  \right\rangle  &
\left\langle \left\{  a\right\}  |_{U}\left\{  b,c\right\}  \right\rangle \\
\left\langle \left\{  b\right\}  |_{U}\left\{  a,b\right\}  \right\rangle  &
\left\langle \left\{  b\right\}  |_{U}\left\{  a,b,c\right\}  \right\rangle  &
\left\langle \left\{  b\right\}  |_{U}\left\{  b,c\right\}  \right\rangle \\
\left\langle \left\{  c\right\}  |_{U}\left\{  a,b\right\}  \right\rangle  &
\left\langle \left\{  c\right\}  |_{U}\left\{  a,b,c\right\}  \right\rangle  &
\left\langle \left\{  c\right\}  |_{U}\left\{  b,c\right\}  \right\rangle
\end{bmatrix}
=%
\begin{bmatrix}
1 & 1 & 0\\
1 & 1 & 1\\
0 & 1 & 1
\end{bmatrix}
$.
\end{center}

If we take the $U$-basis vectors as "vertical position" eigenstates, we can
device a QM/sets version of the double-slit experiment which models "all of
the mystery of quantum mechanics" \cite[p. 130]{fey-phylaw}. Taking $\left\{
a\right\}  $, $\left\{  b\right\}  $, and $\left\{  c\right\}  $ as three
vertical positions, we have a vertical diaphragm with slits at $\left\{
a\right\}  $ and $\left\{  c\right\}  $. Then there is a screen or wall to the
right of the slits so that a "particle" will travel from the diaphragm to the
wall in one time period according to the $A$-dynamics.%

\begin{center}
\includegraphics[
height=1.8182in,
width=2.5139in
]%
{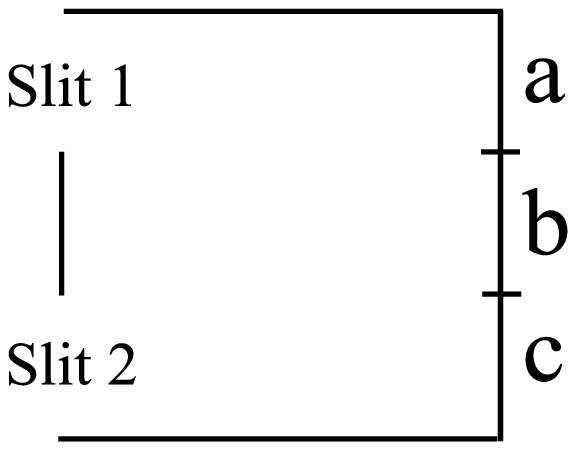}%
\\
Figure 7: Two-slit setup
\end{center}

We start with or prepare the state of a "particle" being at the slits in the
indefinite position state $\left\{  a,c\right\}  $. Then there are two cases.

\textbf{First case of distinctions at slits}: The first case is where we
measure the $U$-state at the slits and then let the resultant position
eigenstate evolve by the $A$-dynamics to hit the wall at the right where the
position is measured again. The probability that the particle is at slit 1 or
at slit 2 is:

\begin{center}
$\Pr\left(  \left\{  a\right\}  \text{ at slits }|\left\{  a,c\right\}  \text{
at slits}\right)  =\frac{\left\langle \left\{  a\right\}  |_{U}\left\{
a,c\right\}  \right\rangle ^{2}}{\left\Vert \left\{  a,c\right\}  \right\Vert
_{U}^{2}}=\frac{\left\vert \left\{  a\right\}  \cap\left\{  a,c\right\}
\right\vert }{\left\vert \left\{  a,c\right\}  \right\vert }=\frac{1}{2}$;

$\Pr\left(  \left\{  c\right\}  \text{ at slits }|\left\{  a,c\right\}  \text{
at slits}\right)  =\frac{\left\langle \left\{  c\right\}  |_{U}\left\{
a,c\right\}  \right\rangle ^{2}}{\left\Vert \left\{  a,c\right\}  \right\Vert
_{U}^{2}}=\frac{\left\vert \left\{  c\right\}  \cap\left\{  a,c\right\}
\right\vert }{\left\vert \left\{  a,c\right\}  \right\vert }=\frac{1}{2}$.
\end{center}

If the particle was measured at slit 1, i.e., was in the post-measurement
eigenstate $\left\{  a\right\}  $, then it evolves in one time period by the
$A $-dynamics to $\left\{  a,b\right\}  $ where the position measurements
yield the probabilities of being at $\left\{  a\right\}  $ or at $\left\{
b\right\}  $ as:

\begin{center}
$\Pr\left(  \left\{  a\right\}  \text{ at wall }|\left\{  a\right\}  \text{ at
slits}\right)  =\Pr\left(  \left\{  a\right\}  \text{ at wall }|\left\{
a,b\right\}  \text{ at wall}\right)  =\frac{\left\langle \left\{  a\right\}
|_{U}\left\{  a,b\right\}  \right\rangle ^{2}}{\left\Vert \left\{
a,b\right\}  \right\Vert _{U}^{2}}=\frac{\left\vert \left\{  a\right\}
\cap\left\{  a,b\right\}  \right\vert }{\left\vert \left\{  a,b\right\}
\right\vert }=\frac{1}{2}$,

$\Pr\left(  \left\{  b\right\}  \text{ at wall }|\left\{  a\right\}  \text{ at
slits}\right)  =\Pr\left(  \left\{  b\right\}  \text{ at wall }|\left\{
a,b\right\}  \text{ at wall}\right)  =\frac{\left\langle \left\{  b\right\}
|_{U}\left\{  a,b\right\}  \right\rangle ^{2}}{\left\Vert \left\{
a,b\right\}  \right\Vert _{U}^{2}}=\frac{\left\vert \left\{  b\right\}
\cap\left\{  a,b\right\}  \right\vert }{\left\vert \left\{  a,b\right\}
\right\vert }=\frac{1}{2}$.
\end{center}

If on the other hand the particle was found in the first measurement to be at
slit 2, i.e., was in eigenstate $\left\{  c\right\}  $, then it evolved in one
time period by the $A$-dynamics to $\left\{  b,c\right\}  $ where the position
measurements yield the probabilities of being at $\left\{  b\right\}  $ or at
$\left\{  c\right\}  $ as:

\begin{center}
$\Pr\left(  \left\{  b\right\}  \text{ at wall }|\left\{  c\right\}  \text{ at
slits}\right)  =\Pr\left(  \left\{  b\right\}  \text{ at wall }|\left\{
b,c\right\}  \text{ at wall}\right)  =\frac{\left\vert \left\{  b\right\}
\cap\left\{  b,c\right\}  \right\vert }{\left\vert \left\{  b,c\right\}
\right\vert }=\frac{1}{2}$,

$\Pr\left(  \left\{  c\right\}  \text{ at wall }|\left\{  c\right\}  \text{ at
slits}\right)  =\Pr\left(  \left\{  c\right\}  \text{ at wall }|\left\{
b,c\right\}  \text{ at wall}\right)  =\frac{\left\vert \left\{  c\right\}
\cap\left\{  b,c\right\}  \right\vert }{\left\vert \left\{  b,c\right\}
\right\vert }=\frac{1}{2}$.
\end{center}

Hence we can use the laws of probability theory to compute the probabilities
of the particle being measured at the three positions on the wall at the right
if it starts at the slits in the superposition state $\left\{  a,c\right\}  $
\textit{and} the measurements were made at the slits:

\begin{center}%
\begin{tabular}
[c]{l}%
$\Pr(\left\{  a\right\}  $ at wall $|\left\{  a,c\right\}  $ at slits$)=\frac
{1}{2}\frac{1}{2}=\frac{1}{4}$;\\
$\Pr(\left\{  b\right\}  $ at wall $|\left\{  a,c\right\}  $ at slits$)=\frac
{1}{2}\frac{1}{2}+\frac{1}{2}\frac{1}{2}=\frac{1}{2}$;\\
$\Pr(\left\{  c\right\}  $ at wall $|\left\{  a,c\right\}  $ at slits$)=\frac
{1}{2}\frac{1}{2}=\frac{1}{4}$.
\end{tabular}
%

\begin{center}
\includegraphics[
height=1.8032in,
width=2.5139in
]%
{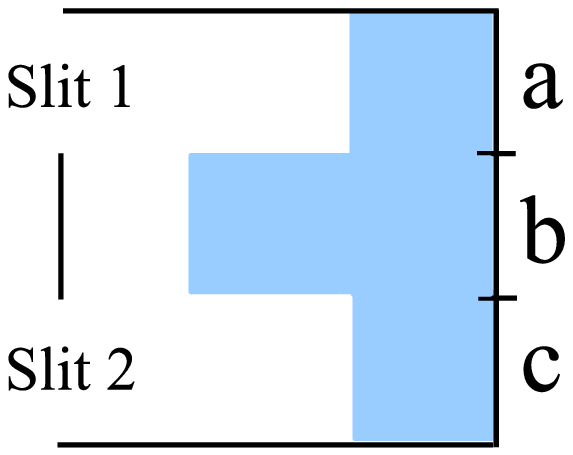}%
\end{center}

Figure 8: Final probability distribution with measurement at the slits.
\end{center}

\textbf{Second case of no distinctions at slits}: The second case is when no
measurements are made at the slits and then the superposition state $\left\{
a,c\right\}  $ evolves by the $A$-dynamics to $\left\{  a,b\right\}
+\left\langle b,c\right\rangle =\left\{  a,c\right\}  $ where the
superposition at $\left\{  b\right\}  $ cancels out. Then the final
probabilities will just be probabilities of finding $\left\{  a\right\}  $,
$\left\{  b\right\}  $, or $\left\{  c\right\}  $ when the measurement is made
only at the wall on the right is:

\begin{center}
$\Pr\left(  \left\{  a\right\}  \text{ at wall }|\left\{  a,c\right\}  \text{
at slits}\right)  =\Pr(\left\{  a\right\}  $ at wall $|\left\{  a,c\right\}  $
at wall$)=\Pr\left(  \left\{  a\right\}  |\left\{  a,c\right\}  \right)
=\frac{\left\vert \left\{  a\right\}  \cap\left\{  a,c\right\}  \right\vert
}{\left\vert \left\{  a,c\right\}  \right\vert }=\frac{1}{2}$;

$\Pr\left(  \left\{  b\right\}  \text{ at wall }|\left\{  a,c\right\}  \text{
at slits}\right)  =\Pr(\left\{  b\right\}  $ at wall $|\left\{  a,c\right\}  $
at wall$)=\Pr\left(  \left\{  b\right\}  |\left\{  a,c\right\}  \right)
=\frac{\left\vert \left\{  b\right\}  \cap\left\{  a,c\right\}  \right\vert
}{\left\vert \left\{  a,c\right\}  \right\vert }=0$;

$\Pr\left(  \left\{  c\right\}  \text{ at wall }|\left\{  a,c\right\}  \text{
at slits}\right)  =\Pr(\left\{  c\right\}  $ at wall $|\left\{  a,c\right\}  $
at wall$)=\Pr\left(  \left\{  c\right\}  |\left\{  a,c\right\}  \right)
=\frac{\left\vert \left\{  c\right\}  \cap\left\{  a,c\right\}  \right\vert
}{\left\vert \left\{  a,c\right\}  \right\vert }=\frac{1}{2}$.%

\begin{center}
\includegraphics[
height=1.8032in,
width=2.5139in
]%
{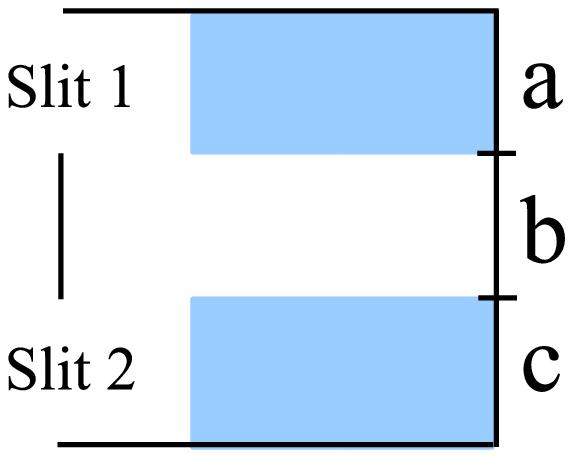}%
\end{center}

Figure 9: Final probability distribution with no measurement at slits
\end{center}

Since no "collapse" took place at the slits due to no distinctions being made
there, the indistinct element $\left\{  a,c\right\}  $ evolved (rather than
one or the other of the distinct elements $\left\{  a\right\}  $ or $\left\{
c\right\}  $). The action of $A$ is the same on $\left\{  a\right\}  $ and
$\left\{  c\right\}  $ as when they evolve separately since $A$ is a linear
operator but the two results are now added together \textit{as part of the
evolution}. This allows the "interference" of the two results and thus the
cancellation of the $\left\{  b\right\}  $ term in $\left\{  a,b\right\}
+\left\langle b,c\right\rangle =\left\{  a,c\right\}  $. The addition is, of
course, mod $2$ (where $-1=+1$) so, in "wave language," the two "wave crests"
that add at the location $\left\{  b\right\}  $ cancel out. When this
indistinct element $\left\{  a,c\right\}  $ "hits the wall" on the right,
there is an equal probability of that distinction-measurement yielding either
of those eigenstates. Figure 9 shows the simplest example of the "light and
dark bands" characteristic of superposition and interference illustrating "all
of the mystery of quantum mechanics".

This model gives the simplest logical essence of the two-slit experiment
without the complex-valued wave functions that distract from the essential
point; the difference between the separate evolutions resulting from
measurement at the slits, and the combined evolution of the superposition
$\left\{  a,c\right\}  $ that allows interference without "waves".

\section{Entanglement in QM/sets}

A QM concept that generates much interest is entanglement. Hence it might be
useful to consider "entanglement" in QM/sets.

First we need to establish the connections across the set-vector-space bridge
by lifting the set notion of the direct (or Cartesian) product $X\times Y$ of
two sets $X$ and $Y$. Using the basis principle, we apply the set concept to
the two basis sets $\left\{  v_{1},...,v_{m}\right\}  $ and $\left\{
w_{1},...,w_{n}\right\}  $ of two vector spaces $V$ and $W$ (over the same
base field) and then we see what it generates. The set direct product of the
two basis sets is the set of all ordered pairs $\left(  v_{i},w_{j}\right)  $,
which we will write as $v_{i}\otimes w_{j}$, and then we generate the vector
space, denoted $V\otimes W$, over the same base field from those basis
elements $v_{i}\otimes w_{j}$. That vector space is the \textit{tensor
product}, and it is not in general the direct product $V\times W$ of the
vector spaces. The cardinality of $X\times Y$ is the product of the
cardinalities of the two sets, and the dimension of the tensor product
$V\otimes W$ is the product of the dimensions of the two spaces (while the
dimension of the direct product $V\times W$ is the sum of the two dimensions).

A vector $z\in V\otimes W$ is said to be \textit{separated} if there are
vectors $v\in V$ and $w\in W$ such that $z=v\otimes w$; otherwise, $z$ is said
to be \textit{entangled}. Since vectors delift to subsets, a subset
$S\subseteq X\times Y$ is said to be \textit{separated} or a \textit{product}
if there exists subsets $S_{X}\subseteq X$ and $S_{Y}\subseteq Y$ such that
$S=S_{X}\times S_{Y}$; otherwise $S\subseteq X\times Y$ is said to be
\textit{entangled.} In general, let $S_{X}$ be the support or projection of
$S$ on $X $, i.e., $S_{X}=\left\{  x:\exists y\in Y,\left(  x,y\right)  \in
S\right\}  $ and similarly for $S_{Y}$. Then $S$ is separated iff
$S=S_{X}\times S_{Y}$.

For any subset $S\subseteq X\times Y$, where $X$ and $Y$ are finite sets, a
natural measure of its "entanglement" can be constructed by first viewing $S$
as the support of the equiprobable or Laplacian joint probability distribution
on $S$. If $\left\vert S\right\vert =N$, then define $\Pr\left(  x,y\right)
=\frac{1}{N}$ if $\left(  x,y\right)  \in S$ and $\Pr\left(  x,y\right)  =0$ otherwise.

The marginal distributions\footnote{The marginal distributions are the set
versions of the reduced density matrices of QM.} are defined in the usual way:

\begin{center}
$\Pr\left(  x\right)  =\sum_{y}\Pr\left(  x,y\right)  $

$\Pr\left(  y\right)  =\sum_{x}\Pr\left(  x,y\right)  $.
\end{center}

\noindent A joint probability distribution $\Pr\left(  x,y\right)  $ on
$X\times Y$ is \textit{independent} if for all $\left(  x,y\right)  \in
X\times Y$,

\begin{center}
$\Pr\left(  x,y\right)  =\Pr\left(  x\right)  \Pr\left(  y\right)  $.

Independent distribution
\end{center}

\noindent Otherwise $\Pr\left(  x,y\right)  $ is said to be
\textit{correlated}.

\begin{proposition}
A subset $S\subseteq X\times Y$ is "entangled" iff the equiprobable
distribution on $S$ is correlated (non-independent).
\end{proposition}

Proof: If $S$ is "separated", i.e., $S=S_{X}\times S_{Y}$, then $\Pr\left(
x\right)  =|S_{Y}|/N$ for $x\in S_{X}$ and $\Pr\left(  y\right)  =\left\vert
S_{X}\right\vert /N$ for $y\in S_{Y}$ where $\left\vert S_{X}\right\vert
\left\vert S_{Y}\right\vert =N$. Then for $\left(  x,y\right)  \in S$,

\begin{center}
$\Pr\left(  x,y\right)  =\frac{1}{N}=\frac{N}{N^{2}}=\frac{\left\vert
S_{X}\right\vert \left\vert S_{Y}\right\vert }{N^{2}}=\Pr\left(  x\right)
\Pr\left(  y\right)  $
\end{center}

\noindent and $\Pr(x,y)=0=\Pr\left(  x\right)  \Pr\left(  y\right)  $ for
$\left(  x,y\right)  \notin S$ so the equiprobable distribution is
independent. If $S$ is "entangled," i.e., $S\neq S_{X}\times S_{Y}$, then
$S\subsetneqq S_{X}\times S_{Y}$ so let $\left(  x,y\right)  \in S_{X}\times
S_{Y}-S$. Then $\Pr\left(  x\right)  ,\Pr\left(  y\right)  >0$ but $\Pr\left(
x,y\right)  =0$ so it is not independent, i.e., is correlated. $\square$

Consider the set version of one qubit space where $U=\left\{  a,b\right\}  $.
The product set $U\times U$ has $15$ nonempty subsets. Each factor $U$ of
$U\times U$ has $3$ nonempty subsets so $3\times3=9$ of the $15$ subsets are
separated subsets leaving $6$ entangled subsets.

\begin{center}%
\begin{tabular}
[c]{|c|}\hline
$S\subseteq U\times U$\\\hline\hline
$\left\{  \left(  a,a\right)  ,\left(  b,b\right)  \right\}  $\\\hline
$\left\{  \left(  a,b\right)  ,\left(  b,a\right)  \right\}  $\\\hline
$\left\{  \left(  a,a\right)  ,(a,b),\left(  b,a\right)  \right\}  $\\\hline
$\left\{  \left(  a,a\right)  ,(a,b),\left(  b,b\right)  \right\}  $\\\hline
$\left\{  (a,b),\left(  b,a\right)  ,\left(  b,b\right)  \right\}  $\\\hline
$\left\{  (a,a),\left(  b,a\right)  ,\left(  b,b\right)  \right\}  $\\\hline
\end{tabular}

The six entangled subsets
\end{center}

The first two are the "Bell states" which are the two graphs of bijections
$U\longleftrightarrow U$ and have the maximum entanglement if entanglement is
measured by the logical divergence $d\left(  \Pr(x,y)||\Pr\left(  x\right)
\Pr\left(  y\right)  \right)  $\cite{ell:distinctions}. All the $9$ separated
states have zero entanglement by the same measure.

For an entangled subset $S$, a sampling $x$ of left-hand system will change
the probability distribution for a sampling of the right-hand system $y$,
$\Pr\left(  y|x\right)  \neq\Pr\left(  y\right)  $. In the case of maximal
"entanglement" (e.g., the "Bell states"), when $S$ is the graph of a bijection
between $U$ and $U$, the value of $y$ is \textit{determined} by the value of
$x$ (and vice-versa).

\section{Bell's Theorem in QM/sets}

A simple version of a Bell inequality can be derived in the case of $%
\mathbb{Z}
_{2}^{2}$ where the only three bases are: $U=\left\{  a,b\right\}  $,
$U^{\prime}=\left\{  a^{\prime},b^{\prime}\right\}  $, and $U^{\prime\prime
}=\left\{  a^{\prime\prime},b^{\prime\prime}\right\}  $, with the relations
given in the ket table:

\begin{center}%
\begin{tabular}
[c]{|c|c|c|c|}\hline
kets & $U$-basis & $U^{\prime}$-basis & $U^{\prime\prime}$-basis\\\hline\hline
$\left\vert 1\right\rangle $ & $\left\{  a,b\right\}  $ & $\left\{  a^{\prime
}\right\}  $ & $\left\{  a^{\prime\prime}\right\}  $\\\hline
$\left\vert 2\right\rangle $ & $\left\{  b\right\}  $ & $\left\{  b^{\prime
}\right\}  $ & $\left\{  a^{\prime\prime},b^{\prime\prime}\right\}  $\\\hline
$\left\vert 3\right\rangle $ & $\left\{  a\right\}  $ & $\left\{  a^{\prime
},b^{\prime}\right\}  $ & $\left\{  b^{\prime\prime}\right\}  $\\\hline
$\left\vert 4\right\rangle $ & $\emptyset$ & $\emptyset$ & $\emptyset$\\\hline
\end{tabular}

Ket table for $\wp\left(  U\right)  \cong\wp\left(  U^{\prime}\right)
\cong\wp\left(  U^{\prime\prime}\right)  \cong%
\mathbb{Z}
_{2}^{2}$.
\end{center}

Attributes defined on the three universe sets $U$, $U^{\prime}$, and
$U^{\prime\prime}$, such as say $\chi_{\left\{  a\right\}  }$, $\chi_{\left\{
b^{\prime}\right\}  }$, and $\chi_{\left\{  a^{\prime\prime}\right\}  }$, are
incompatible as can be seen in several ways. For instance the set partitions
defined on $U$ and $U^{\prime}$, namely $\left\{  \left\{  a\right\}
,\left\{  b\right\}  \right\}  $ and $\left\{  \left\{  a^{\prime}\right\}
,\left\{  b^{\prime}\right\}  \right\}  $, cannot be obtained as two different
ways to partition the same set since $\left\{  a\right\}  =\left\{  a^{\prime
},b^{\prime}\right\}  $ and $\left\{  a^{\prime}\right\}  =\left\{
a,b\right\}  $, i.e., an "eigenstate" in one basis is a superposition in the
other. The same holds in the other pairwise comparison of $U$ and
$U^{\prime\prime}$ and of $U^{\prime}$ and $U^{\prime\prime}$.

Given a ket in $%
\mathbb{Z}
_{2}^{2}\cong\wp\left(  U\right)  \cong\wp\left(  U^{\prime}\right)  \cong%
\wp\left(  U^{\prime\prime}\right)  $, and using the usual equiprobability
assumption on sets, the probabilities of getting the different outcomes for
the various "observables" in the different given states are given in the
following table.

\begin{center}%
\begin{tabular}
[c]{|l||c|c||c|c||c|c|}\hline
Given state
$\backslash$
Outcome of test & $a$ & $b$ & $a^{\prime} $ & $b^{\prime}$ & $a^{\prime\prime
}$ & $b^{\prime\prime}$\\\hline\hline
$\left\{  a,b\right\}  =\left\{  a^{\prime}\right\}  =\left\{  a^{\prime
\prime}\right\}  $ & $\frac{1}{2}$ & $\frac{1}{2}$ & $1$ & $0$ & $1$ &
$0$\\\hline
$\left\{  b\right\}  =\left\{  b^{\prime}\right\}  =\left\{  a^{\prime\prime
},b^{\prime\prime}\right\}  $ & $0$ & $1$ & $0$ & $1$ & $\frac{1}{2}$ &
$\frac{1}{2}$\\\hline
$\left\{  a\right\}  =\left\{  a^{\prime},b^{\prime}\right\}  =\left\{
b^{\prime\prime}\right\}  $ & $1$ & $0$ & $\frac{1}{2}$ & $\frac{1}{2}$ & $0$
& $1$\\\hline
\end{tabular}

State-outcome probability table.
\end{center}

The delift of the tensor product of vector spaces is the Cartesian or direct
product of sets, and the delift of the vectors in the tensor product are the
subsets of direct product of sets (as seen in the above treatment of
entanglement in QM/sets). Thus in the $U$-basis, the basis elements are the
elements of $U\times U$ and the "vectors" are all the subsets in $\wp\left(
U\times U\right)  $. But we could obtain the same "space" as $\wp\left(
U^{\prime}\times U^{\prime}\right)  $ and $\wp\left(  U^{\prime\prime}\times
U^{\prime\prime}\right)  $, and we can construct a ket table where each row is
a ket expressed in the different bases. And these calculations in terms of
sets could also be carried out in terms of vector spaces over $%
\mathbb{Z}
_{2}$ where the rows of the ket table are the kets in the tensor product:

\begin{center}
$%
\mathbb{Z}
_{2}^{2}\otimes%
\mathbb{Z}
_{2}^{2}\cong\wp\left(  U\times U\right)  \cong\wp\left(  U^{\prime}\times
U^{\prime}\right)  \cong\wp\left(  U^{\prime\prime}\times U^{\prime\prime
}\right)  $.
\end{center}

Since $\left\{  a\right\}  =\left\{  a^{\prime},b^{\prime}\right\}  =\left\{
b^{\prime\prime}\right\}  $ and $\left\{  b\right\}  =\left\{  b^{\prime
}\right\}  =\left\{  a^{\prime\prime},b^{\prime\prime}\right\}  $, the subset
$\left\{  a\right\}  \times\left\{  b\right\}  =\left\{  \left(  a,b\right)
\right\}  \subseteq U\times U$ is expressed in the $U^{\prime}\times
U^{\prime}$-basis as $\left\{  a^{\prime},b^{\prime}\right\}  \times\left\{
b^{\prime}\right\}  =\left\{  \left(  a^{\prime},b^{\prime}\right)  ,\left(
b^{\prime},b^{\prime}\right)  \right\}  $, and in the $U^{\prime\prime}\times
U^{\prime\prime}$-basis it is $\left\{  b^{\prime\prime}\right\}
\times\left\{  a^{\prime\prime},b^{\prime\prime}\right\}  =\left\{  \left(
b^{\prime\prime},a^{\prime\prime}\right)  ,\left(  b^{\prime\prime}%
,b^{\prime\prime}\right)  \right\}  $. Hence one row in the ket table has:

\begin{center}
$\left\{  \left(  a,b\right)  \right\}  =\left\{  \left(  a^{\prime}%
,b^{\prime}\right)  ,\left(  b^{\prime},b^{\prime}\right)  \right\}  =\left\{
\left(  b^{\prime\prime},a^{\prime\prime}\right)  ,\left(  b^{\prime\prime
},b^{\prime\prime}\right)  \right\}  $.
\end{center}

\noindent Since the full ket table has $16$ rows, we will just give a partial
table that suffices for our calculations.

\begin{center}%
\begin{tabular}
[c]{|c|c|c|}\hline
$U\times U$ & \multicolumn{1}{|c|}{$U^{\prime}\times U^{\prime}$} &
$U^{\prime\prime}\times U^{\prime\prime}$\\\hline\hline
$\left\{  \left(  a,a\right)  \right\}  $ & $\left\{  \left(  a^{\prime
},a^{\prime}\right)  ,\left(  a^{\prime},b^{\prime}\right)  ,\left(
b^{\prime},a^{\prime}\right)  ,\left(  b^{\prime},b^{\prime}\right)  \right\}
$ & $\left\{  \left(  b^{\prime\prime},b^{\prime\prime}\right)  \right\}
$\\\hline
$\left\{  \left(  a,b\right)  \right\}  $ & $\left\{  \left(  a^{\prime
},b^{\prime}\right)  ,\left(  b^{\prime},b^{\prime}\right)  \right\}  $ &
$\left\{  \left(  b^{\prime\prime},a^{\prime\prime}\right)  ,\left(
b^{\prime\prime},b^{\prime\prime}\right)  \right\}  $\\\hline
$\left\{  \left(  b,a\right)  \right\}  $ & $\left\{  \left(  b^{\prime
},a^{\prime}\right)  ,\left(  b^{\prime},b^{\prime}\right)  \right\}  $ &
$\left\{  \left(  a^{\prime\prime},b^{\prime\prime}\right)  ,\left(
b^{\prime\prime},b^{\prime\prime}\right)  \right\}  $\\\hline
$\left\{  \left(  b,b\right)  \right\}  $ & $\left\{  \left(  b^{\prime
},b^{\prime}\right)  \right\}  $ & $\left\{  \left(  a^{\prime\prime
},a^{\prime\prime}\right)  ,\left(  a^{\prime\prime},b^{\prime\prime}\right)
,\left(  b^{\prime\prime},a^{\prime\prime}\right)  ,\left(  b^{\prime\prime
},b^{\prime\prime}\right)  \right\}  $\\\hline
$\left\{  \left(  a,a\right)  ,\left(  a,b\right)  \right\}  $ & $\left\{
\left(  a^{\prime},a^{\prime}\right)  ,\left(  b^{\prime},a^{\prime}\right)
\right\}  $ & $\left\{  \left(  b^{\prime\prime},a^{\prime\prime}\right)
\right\}  $\\\hline
$\left\{  \left(  a,a\right)  ,\left(  b,a\right)  \right\}  $ & $\left\{
\left(  a^{\prime},a^{\prime}\right)  ,\left(  a^{\prime},b^{\prime}\right)
\right\}  $ & $\left\{  \left(  a^{\prime\prime},b^{\prime\prime}\right)
\right\}  $\\\hline
$\left\{  \left(  a,a\right)  ,\left(  b,b\right)  \right\}  $ & $\left\{
\left(  a^{\prime},a^{\prime}\right)  ,\left(  a^{\prime},b^{\prime}\right)
,\left(  b^{\prime},a^{\prime}\right)  \right\}  $ & $\left\{  \left(
a^{\prime\prime},a^{\prime\prime}\right)  ,\left(  a^{\prime\prime}%
,b^{\prime\prime}\right)  ,\left(  b^{\prime\prime},a^{\prime\prime}\right)
\right\}  $\\\hline
$\left\{  \left(  a,b\right)  ,\left(  b,a\right)  \right\}  $ & $\left\{
\left(  a^{\prime},b^{\prime}\right)  ,\left(  b^{\prime},a^{\prime}\right)
\right\}  $ & $\left\{  \left(  a^{\prime\prime},b^{\prime\prime}\right)
,\left(  b^{\prime\prime},a^{\prime\prime}\right)  \right\}  $\\\hline
\end{tabular}

Partial ket table for $\wp\left(  U\times U\right)  \cong\wp\left(  U^{\prime
}\times U^{\prime}\right)  \cong\wp\left(  U^{\prime\prime}\times
U^{\prime\prime}\right)  $
\end{center}

As before, we can classify each subset as separated or entangled and we can
furthermore see how that is independent of the basis. For instance $\left\{
\left(  a,a\right)  ,\left(  a,b\right)  \right\}  $ is separated since:

\begin{center}
$\left\{  \left(  a,a\right)  ,\left(  a,b\right)  \right\}  =\left\{
a\right\}  \times\left\{  a,b\right\}  =\left\{  \left(  a^{\prime},a^{\prime
}\right)  ,\left(  b^{\prime},a^{\prime}\right)  \right\}  =\left\{
a^{\prime},b^{\prime}\right\}  \times\left\{  a^{\prime}\right\}  =\left\{
\left(  b^{\prime\prime},a^{\prime\prime}\right)  \right\}  =\left\{
b^{\prime\prime}\right\}  \times\left\{  a^{\prime\prime}\right\}  $.
\end{center}

An example of an entangled state is:

\begin{center}
$\left\{  \left(  a,a\right)  ,\left(  b,b\right)  \right\}  =\left\{  \left(
a^{\prime},a^{\prime}\right)  ,\left(  a^{\prime},b^{\prime}\right)  ,\left(
b^{\prime},a^{\prime}\right)  \right\}  =\left\{  \left(  a^{\prime\prime
},a^{\prime\prime}\right)  ,\left(  a^{\prime\prime},b^{\prime\prime}\right)
,\left(  b^{\prime\prime},a^{\prime\prime}\right)  \right\}  $.
\end{center}

\noindent Taking this entangled state as the initial state, the probability of
getting the state $\left\{  a\right\}  $ by performing a $U$-basis measurement
on the left-hand system is:

\begin{center}
$\Pr\left(  \left\{  \left(  a,-\right)  \right\}  |\left\{  \left(
a,a\right)  ,\left(  b,b\right)  \right\}  \right)  =\frac{\left\vert \left\{
\left(  a,a\right)  \right\}  \right\vert }{\left\vert \left\{  \left(
a,a\right)  ,\left(  b,b\right)  \right\}  \right\vert }=\frac{1}{2}$.
\end{center}

The probability of getting the state $\left\{  a^{\prime}\right\}  $ by
performing a $U^{\prime}$-basis measurement on the left-hand system is:

\begin{center}
$\Pr\left(  \left\{  \left(  a^{\prime},-\right)  \right\}  |\left\{  \left(
a^{\prime},a^{\prime}\right)  ,\left(  a^{\prime},b^{\prime}\right)  ,\left(
b^{\prime},a^{\prime}\right)  \right\}  \right)  =\frac{\left\vert \left\{
\left(  a^{\prime},a^{\prime}\right)  ,\left(  a^{\prime},b^{\prime}\right)
\right\}  \right\vert }{\left\vert \left\{  \left(  a^{\prime},a^{\prime
}\right)  ,\left(  a^{\prime},b^{\prime}\right)  ,\left(  b^{\prime}%
,a^{\prime}\right)  \right\}  \right\vert }=\frac{2}{3}$.
\end{center}

The probability of getting the state $\left\{  a^{\prime\prime}\right\}  $ by
performing a $U^{\prime\prime}$-basis measurement on the left-hand system is:

\begin{center}
$\Pr\left(  \left\{  \left(  a^{\prime\prime},-\right)  \right\}  |\left\{
\left(  a^{\prime\prime},a^{\prime\prime}\right)  ,\left(  a^{\prime\prime
},b^{\prime\prime}\right)  ,\left(  b^{\prime\prime},a^{\prime\prime}\right)
\right\}  \right)  =\frac{\left\vert \left\{  \left(  a^{\prime\prime
},a^{\prime\prime}\right)  ,\left(  a^{\prime\prime},b^{\prime\prime}\right)
\right\}  \right\vert }{\left\vert \left\{  \left(  a^{\prime\prime}%
,a^{\prime\prime}\right)  ,\left(  a^{\prime\prime},b^{\prime\prime}\right)
,\left(  b^{\prime\prime},a^{\prime\prime}\right)  \right\}  \right\vert
}=\frac{2}{3}$.
\end{center}

The probability of each of these outcomes occurring (if each is done instead
of either of the others) is the product of the conditional probabilities. Then
there is a probability distribution on $U\times U^{\prime}\times
U^{\prime\prime}$, all conditionalized by the same entangled state, where:

\begin{center}
$\Pr\left(  a,a^{\prime},a^{\prime\prime}\right)  $

$=\Pr\left(  \left\{  \left(  a,-\right)  \right\}  |\left\{  \left(
a,a\right)  ,\left(  b,b\right)  \right\}  \right)  $

$\times\Pr\left(  \left\{  \left(  a^{\prime},-\right)  \right\}  |\left\{
\left(  a^{\prime},a^{\prime}\right)  ,\left(  a^{\prime},b^{\prime}\right)
,\left(  b^{\prime},a^{\prime}\right)  \right\}  \right)  $

$\times\Pr\left(  \left\{  \left(  a^{\prime\prime},-\right)  \right\}
|\left\{  \left(  a^{\prime\prime},a^{\prime\prime}\right)  ,\left(
a^{\prime\prime},b^{\prime\prime}\right)  ,\left(  b^{\prime\prime}%
,a^{\prime\prime}\right)  \right\}  \right)  $

$=\frac{1}{2}\frac{2}{3}\frac{2}{3}=\frac{2}{9}$.
\end{center}

\noindent In this way, a probability distribution $\Pr\left(  x,y,z\right)  $
is defined on $U\times U^{\prime}\times U^{\prime\prime}$.

A Bell inequality can be obtained from this joint probability distribution
over the outcomes $U\times U^{\prime}\times U^{\prime\prime}$ of measuring
these three incompatible attributes \cite{d'esp:sciam}. Consider the following marginals:%

\begin{align*}
\Pr\left(  a,a^{\prime}\right)   &  =\Pr\left(  a,a^{\prime},a^{\prime\prime
}\right)  +\Pr\left(  a,a^{\prime},b^{\prime\prime}\right)  \checkmark\\
\Pr\left(  b^{\prime},b^{\prime\prime}\right)   &  =\Pr\left(  a,b^{\prime
},b^{\prime\prime}\right)  \checkmark+\Pr\left(  b,b^{\prime},b^{\prime\prime
}\right) \\
\Pr\left(  a,b^{\prime\prime}\right)   &  =\Pr\left(  a,a^{\prime}%
,b^{\prime\prime}\right)  \checkmark+\Pr\left(  a,b^{\prime},b^{\prime\prime
}\right)  \checkmark\text{.}%
\end{align*}

\noindent The two terms in the last marginal are each contained in one of the
two previous marginals (as indicated by the check marks) and all the
probabilities are non-negative, so we have the following inequality:

\begin{center}
$\Pr\left(  a,a^{\prime}\right)  +\Pr\left(  b^{\prime},b^{\prime\prime
}\right)  \geq\Pr\left(  a,b^{\prime\prime}\right)  $

Bell inequality.
\end{center}

All this has to do with measurements on the left-hand system. But the "Bell
state" is left-right symmetrical so the same probabilities would be obtained
if we used a right-hand system measurement:

$\Pr\left(  \left\{  \left(  a,-\right)  \right\}  |\left\{  \left(
a,a\right)  ,\left(  b,b\right)  \right\}  \right)  =\Pr\left(  \left\{
\left(  -,a\right)  \right\}  |\left\{  \left(  a,a\right)  ,\left(
b,b\right)  \right\}  \right)  =\frac{1}{2}$;

$\Pr\left(  \left\{  \left(  b,-\right)  \right\}  |\left\{  \left(
a,a\right)  ,\left(  b,b\right)  \right\}  \right)  =\Pr\left(  \left\{
\left(  -,b\right)  \right\}  |\left\{  \left(  a,a\right)  ,\left(
b,b\right)  \right\}  \right)  =\frac{1}{2}$;

$\Pr\left(  \left\{  \left(  a^{\prime},-\right)  \right\}  |\left\{  \left(
a^{\prime},a^{\prime}\right)  ,\left(  a^{\prime},b^{\prime}\right)  ,\left(
b^{\prime},a^{\prime}\right)  \right\}  \right)  =\Pr\left(  \left\{  \left(
-,a^{\prime}\right)  \right\}  |\left\{  \left(  a^{\prime},a^{\prime}\right)
,\left(  a^{\prime},b^{\prime}\right)  ,\left(  b^{\prime},a^{\prime}\right)
\right\}  \right)  =\frac{2}{3}$;

$\Pr\left(  \left\{  \left(  b^{\prime},-\right)  \right\}  |\left\{  \left(
a^{\prime},a^{\prime}\right)  ,\left(  a^{\prime},b^{\prime}\right)  ,\left(
b^{\prime},a^{\prime}\right)  \right\}  \right)  =\Pr\left(  \left\{  \left(
-,b^{\prime}\right)  \right\}  |\left\{  \left(  a^{\prime},a^{\prime}\right)
,\left(  a^{\prime},b^{\prime}\right)  ,\left(  b^{\prime},a^{\prime}\right)
\right\}  \right)  =\frac{1}{3}$;

$\Pr\left(  \left\{  \left(  a^{\prime\prime},-\right)  \right\}  |\left\{
\left(  a^{\prime\prime},a^{\prime\prime}\right)  ,\left(  a^{\prime\prime
},b^{\prime\prime}\right)  ,\left(  b^{\prime\prime},a^{\prime\prime}\right)
\right\}  \right)  =\Pr\left(  \left\{  \left(  -,a^{\prime\prime}\right)
\right\}  |\left\{  \left(  a^{\prime\prime},a^{\prime\prime}\right)  ,\left(
a^{\prime\prime},b^{\prime\prime}\right)  ,\left(  b^{\prime\prime}%
,a^{\prime\prime}\right)  \right\}  \right)  =\frac{2}{3}$; and

$\Pr\left(  \left\{  \left(  b^{\prime\prime},-\right)  \right\}  |\left\{
\left(  a^{\prime\prime},a^{\prime\prime}\right)  ,\left(  a^{\prime\prime
},b^{\prime\prime}\right)  ,\left(  b^{\prime\prime},a^{\prime\prime}\right)
\right\}  \right)  =\Pr\left(  \left\{  \left(  -,b^{\prime\prime}\right)
\right\}  |\left\{  \left(  a^{\prime\prime},a^{\prime\prime}\right)  ,\left(
a^{\prime\prime},b^{\prime\prime}\right)  ,\left(  b^{\prime\prime}%
,a^{\prime\prime}\right)  \right\}  \right)  =\frac{1}{3}$.\footnote{The same
holds for the other "Bell state": $\left\{  \left(  a,b\right)  ,\left(
b,a\right)  \right\}  $.}

\noindent This is analogous to the assumption that each sock in a pair of
socks will have the same properties.\cite[Chap. 16]{bell:unspeak} Hence the
right-hand measurements give the same probability distribution and the same inequality.

But there is an alternative interpretation to the probabilities $\Pr\left(
x,y\right)  $, $\Pr\left(  y,z\right)  $, and $\Pr\left(  x,z\right)  $
\textit{if }we assume that the outcome of a measurement on the right-hand
system is \textit{independent} of the outcome of the same measurement on the
left-hand system. Then $\Pr\left(  a,a^{\prime}\right)  $ is the probability
of a $U$-measurement on the left-hand system giving $\left\{  a\right\}  $
\textit{and then} \textit{in addition} (not instead of) a $U^{\prime}%
$-measurement on the right-hand system giving $\left\{  a^{\prime}\right\}  $,
and so forth.

This is a crucial step in the argument so it worth being very clear using subscripts.

\begin{itemize}
\item Step 1: $\Pr\left(  a,a^{\prime}\right)  _{1}$ is the probability of
getting $\left\{  a\right\}  $ in a left $U$-measurement and getting $\left\{
a^{\prime}\right\}  $ \textit{if instead} a left $U^{\prime}$-measurement was
made so:
\end{itemize}

\begin{center}
$\Pr\left(  a,a^{\prime}\right)  _{1}=\Pr\left(  \left\{  \left(  a,-\right)
\right\}  |\left\{  \left(  a,a\right)  ,\left(  b,b\right)  \right\}
\right)  \times\Pr\left(  \left\{  \left(  a^{\prime},-\right)  \right\}
|\left\{  \left(  a^{\prime},a^{\prime}\right)  ,\left(  a^{\prime},b^{\prime
}\right)  ,\left(  b^{\prime},a^{\prime}\right)  \right\}  \right)  =\frac
{1}{2}\frac{2}{3}=\frac{1}{3}$.
\end{center}

\begin{itemize}
\item Step 2: $\Pr\left(  a,a^{\prime}\right)  _{2}$ is the probability of
getting $\left\{  a\right\}  $ in a left $U$-measurement and getting $\left\{
a^{\prime}\right\}  $ if instead a \textit{right} $U^{\prime}$-measurement was
made so:
\end{itemize}

\begin{center}
$\Pr\left(  a,a^{\prime}\right)  _{2}=\Pr\left(  \left\{  \left(  a,-\right)
\right\}  |\left\{  \left(  a,a\right)  ,\left(  b,b\right)  \right\}
\right)  \times\Pr\left(  \left\{  \left(  -,a^{\prime}\right)  \right\}
|\left\{  \left(  a^{\prime},a^{\prime}\right)  ,\left(  a^{\prime},b^{\prime
}\right)  ,\left(  b^{\prime},a^{\prime}\right)  \right\}  \right)  =\frac
{1}{2}\frac{2}{3}=\frac{1}{3}$.
\end{center}

\begin{itemize}
\item Step 3: $\Pr\left(  a,a^{\prime}\right)  _{3}$ is the probability of
getting $\left\{  a\right\}  $ in a left $U$-measurement and, \textit{under
the assumption of independence of the left-right measurements}, \textit{also}
(not instead of) getting $\left\{  a^{\prime}\right\}  $ in a right
$U^{\prime}$-measurement:
\end{itemize}

\begin{center}
$\Pr\left(  a,a^{\prime}\right)  _{3}=\Pr\left(  \left\{  \left(  a,-\right)
\right\}  |\left\{  \left(  a,a\right)  ,\left(  b,b\right)  \right\}
\right)  \times\Pr\left(  \left\{  \left(  -,a^{\prime}\right)  \right\}
|\left\{  \left(  a^{\prime},a^{\prime}\right)  ,\left(  a^{\prime},b^{\prime
}\right)  ,\left(  b^{\prime},a^{\prime}\right)  \right\}  \right)  =\frac
{1}{2}\frac{2}{3}=\frac{1}{3}$.
\end{center}

Hence the joint probability distribution would be the same and the above Bell inequality:

\begin{center}
$\Pr\left(  a,a^{\prime}\right)  _{3}+\Pr\left(  b^{\prime},b^{\prime\prime
}\right)  _{3}\geq\Pr\left(  a,b^{\prime\prime}\right)  _{3}$
\end{center}

\noindent would still hold \textit{under the independence assumption} using
the step 3 probabilities in all cases. But we can use QM/sets to compute the
probabilities for those different measurements on the two systems to see if
the independence assumption is compatible with QM/sets.

To compute $\Pr\left(  a,a^{\prime}\right)  _{3}$, we first measure the
left-hand component in the $U$-basis. Since $\left\{  \left(  a,a\right)
,\left(  b,b\right)  \right\}  $ is the given state, and $\left(  a,a\right)
$ and $\left(  b,b\right)  $ are equiprobable, the probability of getting
$\left\{  a\right\}  $ (i.e., the "eigenvalue" $1$ for the "observable
$\chi_{\left\{  a\right\}  }$) is $\frac{1}{2}$. But the right-hand system is
then in the state $\left\{  a\right\}  $ and the probability of getting
$\left\{  a^{\prime}\right\}  $ (i.e., "eigenvalue" $0$ for the "observable"
$\chi_{\left\{  b^{\prime}\right\}  }$) is $\frac{1}{2}$ (as seen in the
state-outcome table). Thus the probability is $\Pr\left(  a,a^{\prime}\right)
_{3}=\frac{1}{2}\frac{1}{2}=\frac{1}{4}$.

To compute $\Pr\left(  b^{\prime},b^{\prime\prime}\right)  _{3}$, we first
perform a $U^{\prime}$-basis "measurement" on the left-hand component of the
given state $\left\{  \left(  a,a\right)  ,\left(  b,b\right)  \right\}
=\left\{  \left(  a^{\prime},a^{\prime}\right)  ,\left(  a^{\prime},b^{\prime
}\right)  ,\left(  b^{\prime},a^{\prime}\right)  \right\}  $, and we see that
the probability of getting $\left\{  b^{\prime}\right\}  $ is $\frac{1}{3}$.
Then the right-hand system is in the state $\left\{  a^{\prime}\right\}  $ and
the probability of getting $\left\{  b^{\prime\prime}\right\}  $ in a
$U^{\prime\prime}$-basis "measurement" of the right-hand system in the state
$\left\{  a^{\prime}\right\}  $ is $0$ (as seen from the state-outcome table).
Hence the probability is $\Pr\left(  b^{\prime},b^{\prime\prime}\right)
_{3}=0$.

Finally we compute $\Pr\left(  a,b^{\prime\prime}\right)  _{3}$ by first
making a $U$-measurement on the left-hand component of the given state
$\left\{  \left(  a,a\right)  ,\left(  b,b\right)  \right\}  $ and get the
result $\left\{  a\right\}  $ with probability $\frac{1}{2}$. Then the state
of the second system is $\left\{  a\right\}  $ so a $U^{\prime\prime}%
$-measurement will give the $\left\{  b^{\prime\prime}\right\}  $ result with
probability $1$ so the probability is $\Pr\left(  a,b^{\prime\prime}\right)
_{3}=\frac{1}{2}$.

Then we plug the probabilities into the Bell inequality:

\begin{center}
$\Pr\left(  a,a^{\prime}\right)  _{3}+\Pr\left(  b^{\prime},b^{\prime\prime
}\right)  _{3}\geq\Pr\left(  a,b^{\prime\prime}\right)  _{3}$

$\frac{1}{4}+0\ngeq\frac{1}{2}$

Violation of Bell inequality.
\end{center}

\noindent The violation of the Bell inequality shows that the independence
assumption about the measurement outcomes on the left-hand and right-hand
systems is incompatible with QM/sets. This result is somewhat less striking in
QM/sets than in full QM since QM/sets just shows the bare logic of the Bell
argument in the simplest space $%
\mathbb{Z}
_{2}^{2}$ without any dramatic physical assumption like a space-like
separation between the left-hand and right-hand physical systems.

\part{Quantum information and computation theory in QM/sets}

\section{Quantum information theory in QM/sets}

\subsection{Logical entropy}

Obtaining quantum information theory for QM/sets is not a simple matter of
delifting the ordinary quantum information theory (QIT). This is because much
of QIT is obtained by transporting over or lifting the notion of Shannon
entropy from classical information theory (which is then renamed "von Neumann
entropy"). Shannon entropy is a higher-level concept adapted for questions of
coding and communication; it is not a basic logical concept. Classical
information theory itself needs to be refounded on a logical basis using the
logical notion of entropy that arises naturally out of partition logic (that
is dual to the usual Boolean subset logic). That logical information theory
can then be simply reformulated using delifted machinery from QM, namely
density matrices, and thus logical information theory is reformulated as
"quantum" information theory for QM/sets.

The process is quite analogous to the way that classical logical finite
probability was reformulated as the probability calculus for QM/sets.
Conceptually, the next step beyond subset logic was the quantitative treatment
that gave logical finite probability theory. Historically, Boole presented
logical finite probability theory as this quantitative step beyond subset
logic in his book entitled: \textit{An Investigation of the Laws of Thought on
which are founded the Mathematical Theories of Logic and Probabilities}. The
universe $U$ was the finite number of possible outcomes and the subsets were
events. Quoting Poisson, Boole defined "the measure of the probability of an
event [as] the ratio of the number of cases favourable to that event, to the
total number of cases favourable and unfavourable, and all equally possible."
\cite[p. 253]{boole:lot}

Hence one obvious next quantitative step beyond partition logic is to make the
analogous conceptual moves and to see what theory emerges. The theory that
emerges is a logical version of information theory.

For a finite $U$, the finite (Laplacian) \textit{probability} $\Pr(S)$ of a
subset ("event") is the normalized counting measure on the subset:
$\Pr(S)=\left\vert S\right\vert /\left\vert U\right\vert $. Analogously, the
finite \textit{logical entropy} $h\left(  \pi\right)  $ of a partition $\pi$
is the normalized counting measure of its dit set: $h\left(  \pi\right)
=\left\vert \operatorname*{dit}\left(  \pi\right)  \right\vert /\left\vert
U\times U\right\vert $. If $U$ is an urn with each "ball" in the urn being
equiprobable, then $\Pr(S)$ is the probability of an element randomly drawn
from the urn is an element in $S$, and, similarly, $h\left(  \pi\right)  $ is
the probability that a pair of elements randomly drawn from the urn (with
replacement) is a distinction of $\pi$.

Let $\pi=\left\{  B_{1},...,B_{m}\right\}  $ with $p_{i}=\left\vert
B_{i}\right\vert /\left\vert U\right\vert $ being the probability of drawing
an element of the block $B_{i}$. The number of indistinctions
(non-distinctions) of $\pi$ is $\left\vert \operatorname*{indit}\left(
\pi\right)  \right\vert =$ $\Sigma_{i}\left\vert B_{i}\right\vert ^{2}$ so the
number of distinctions is $\left\vert \operatorname*{dit}\left(  \pi\right)
\right\vert =\left\vert U\right\vert ^{2}-\Sigma_{i}\left\vert B_{i}%
\right\vert ^{2}$ and thus since $\Sigma_{i}p_{i}=1$, the logical entropy of
$\pi$ is: $h\left(  \pi\right)  =\left[  \left\vert U\right\vert ^{2}%
-\Sigma_{i}\left\vert B_{i}\right\vert ^{2}\right]  /\left\vert U\right\vert
^{2}=1-\Sigma_{i}p_{i}^{2}=\left(  \Sigma_{i}p_{i}\right)  -\Sigma_{i}%
p_{i}^{2}=\Sigma_{i}p_{i}\left(  1-p_{i}\right)  $, so that:

\begin{center}
Logical entropy: $h\left(  \pi\right)  =\Sigma_{i}p_{i}\left(  1-p_{i}\right)
$.
\end{center}

Shannon's notion of entropy is a high-level notion adapted to communications
theory \cite{shannon:comm}. The Shannon entropy $H\left(  \pi\right)  $ of the
partition $\pi$ (with the same probabilities assigned to the blocks) is:

\begin{center}
Shannon entropy: $H\left(  \pi\right)  =\Sigma_{i}p_{i}\log\left(
1/p_{i}\right)  $
\end{center}

\noindent where the $\log$ is base $2$.

Each entropy can be seen as the probabilistic average of the "block entropies"
$h\left(  B_{i}\right)  =1-p_{i}$ and $H\left(  B_{i}\right)  =\log\left(
1/p_{i}\right)  $. To interpret the block entropies, consider a special case
where $p_{i}=1/2^{n}$ and every block is the same so there are $2^{n}$ equal
blocks like $B_{i}$ in the partition. The logical entropy of that special
equal-block partition, $\Sigma_{i}p_{i}\left(  1-p_{i}\right)  =\left(
2^{n}\right)  p_{i}\left(  1-p_{i}\right)  =\left(  2^{n}\right)  \left(
1/2^{n}\right)  \left(  1-p_{i}\right)  =1-p_{i}$, is the:

\begin{center}
Logical block entropy: $h(B_{i})=1-p_{i}$.
\end{center}

Instead of directly counting the distinctions, we could take the number of
binary equal-blocked partitions it takes to distinguish all the $2^{n}$ blocks
in that same partition. As in the game of "twenty questions," if there is a
search for an unknown designated block, then each such binary question can
reduce the number of blocks by a power of $2$ so the minimum number of binary
partitions it takes to distinguish all the $2^{n}$ blocks (and find the hidden
block no matter where it was) is $n=\log\left(  2^{n}\right)  =\log\left(
1/p_{i}\right)  $, which is the:

\begin{center}
Shannon block entropy: $H\left(  B_{i}\right)  =\log\left(  1/p_{i}\right)  $.
\end{center}

To precisely relate the block entropies, we solve each for $p_{i}$ which is
then eliminated to obtain:

\begin{center}
$h\left(  B\right)  =1-\left(  1/2^{H\left(  B\right)  }\right)  $.

Exact relation between Shannon and logical block entropies
\end{center}

\noindent The interpretation of the Shannon block entropy is then extended by
analogy to the general case where $1/p_{i}$ is not a power of $2$ so that the
Shannon entropy $H\left(  \pi\right)  =\Sigma_{i}p_{i}H\left(  B_{i}\right)  $
is then interpreted as the \textit{average} number of binary partitions needed
to make all the distinctions between the blocks of $\pi$---whereas the logical
entropy is still the \textit{exact} normalized count $h\left(  \pi\right)
=\Sigma_{i}p_{i}h\left(  B_{i}\right)  =\left\vert \operatorname*{dit}\left(
\pi\right)  \right\vert /\left\vert U\times U\right\vert $ of the distinctions
of the partition $\pi$.

The two notions of entropy boil down to two different ways to count the
distinctions of a partition. Thus the concept of a distinction from partition
logic provides a logical basis for the notion of entropy in information
theory.\footnote{For further development of logical information theory, see
Ellerman \cite{ell:distinctions}.}

\subsection{Density matrices in QM/sets}

The notion of logical entropy generalizes naturally to quantum information
theory where it also provides a new foundational notion of entropy based on
the idea of \textit{information as distinctions} that are preserved in unitary
transformations and made objectively in measurements.\cite{ell:objindef} Our
purpose here is to formulate logical entropy using the delifted notion of
density matrices which gives QIT/sets, and which then foreshadows how the
"classical" logical information theory can be lifted to give a new foundation
for the full QIT. The previous treatment of measurement in QM/sets can also be
reformulated using density matrices and logical entropy.

Given a partition $\pi=\left\{  B\right\}  $ on $U=\left\{  u_{1}%
,...,u_{n}\right\}  $, the blocks $B\in\pi$ can be thought of as
(nonoverlapping or "orthogonal") "pure states" where the "state" $B$ occurs
with the probability $p_{B}=\frac{\left\vert B\right\vert }{\left\vert
U\right\vert }$. Then we can mimic the usual procedure for forming the density
matrix $\rho\left(  \pi\right)  $ for the "orthogonal pure states" $B $ with
the probabilities $p_{B}$. The (normalized) "pure state" $B$ is represented by
the column vector $\left\vert B\right\rangle =\left[  \sqrt{q_{1}},\sqrt
{q_{2}},...,\sqrt{q_{n}}\right]  ^{t}$ where $q_{j}=1/\left\vert B\right\vert
$ if $u_{j}\in B$, and $q_{j}=0$ otherwise. Then the \textit{density matrix
}$\rho\left(  B\right)  $\textit{\ for the pure state }$B\subseteq U$ is then
(calculating in the reals):

\begin{center}
$\rho\left(  B\right)  =\left\vert B\right\rangle \left(  \left\vert
B\right\rangle \right)  ^{t}=%
\begin{bmatrix}
\sqrt{q_{1}}\\
\sqrt{q_{2}}\\
\vdots\\
\sqrt{q_{n}}%
\end{bmatrix}
\left[  \sqrt{q_{1}},\sqrt{q_{2}},...,\sqrt{q_{n}}\right]  =%
\begin{bmatrix}
q_{1} & \sqrt{q_{1}q_{2}} & \cdots & \sqrt{q_{1}q_{n}}\\
\sqrt{q_{2}q_{1}} & q_{2} & \cdots & \sqrt{q_{2}q_{n}}\\
\vdots & \vdots & \ddots & \vdots\\
\sqrt{q_{n}q_{1}} & \sqrt{q_{n}q_{2}} & \cdots & q_{n}%
\end{bmatrix}
$.
\end{center}

For instance if $U=\left\{  u_{1},u_{2},u_{3}\right\}  =\left\{
a,b,c\right\}  $ then for the blocks in the partition $\pi=\left\{  \left\{
a,b\right\}  ,\left\{  c\right\}  \right\}  $:

\begin{center}
$\rho\left(  \left\{  a,b\right\}  \right)  =%
\begin{bmatrix}
\frac{1}{2} & \frac{1}{2} & 0\\
\frac{1}{2} & \frac{1}{2} & 0\\
0 & 0 & 0
\end{bmatrix}
$ and $\rho\left(  \left\{  c\right\}  \right)  =%
\begin{bmatrix}
0 & 0 & 0\\
0 & 0 & 0\\
0 & 0 & 1
\end{bmatrix}
$.
\end{center}

\noindent Then the "mixed state" \textit{density matrix }$\rho\left(
\pi\right)  $\textit{\ of the partition} $\pi$ is the weighted sum:

\begin{center}
$\rho\left(  \pi\right)  =\sum_{B\in\pi}p_{B}\rho\left(  B\right)  $.
\end{center}

In the example, this is:

\begin{center}
$\rho\left(  \pi\right)  =\frac{2}{3}%
\begin{bmatrix}
\frac{1}{2} & \frac{1}{2} & 0\\
\frac{1}{2} & \frac{1}{2} & 0\\
0 & 0 & 0
\end{bmatrix}
+\frac{1}{3}%
\begin{bmatrix}
0 & 0 & 0\\
0 & 0 & 0\\
0 & 0 & 1
\end{bmatrix}
=%
\begin{bmatrix}
\frac{1}{3} & \frac{1}{3} & 0\\
\frac{1}{3} & \frac{1}{3} & 0\\
0 & 0 & \frac{1}{3}%
\end{bmatrix}
$.
\end{center}

\noindent While this construction mimics the usual construction of the density
matrix for orthogonal pure states, the remarkable thing is that the entries
have a direct interpretation in terms of the dits and indits of the partition
$\pi$:

\begin{center}
$\rho_{jk}\left(  \pi\right)  =\left\{
\begin{array}
[c]{c}%
\frac{1}{\left\vert U\right\vert }\text{ if }\left(  j,k\right)
\in\operatorname*{indit}\left(  \pi\right) \\
0\text{ if }\left(  j,k\right)  \in\operatorname*{dit}\left(  \pi\right)
\text{.}%
\end{array}
\right.  $
\end{center}

\noindent All the entries are real "amplitudes" whose squares are the two-draw
probabilities of drawing a pair of elements from $U$ (with replacement) that
is an indistinction of $\pi$. To foreshadow the quantum case, the non-zero
entries $\rho_{jk}\left(  \pi\right)  =\sqrt{\frac{1}{\left\vert U\right\vert
}\frac{1}{\left\vert U\right\vert }}=\frac{1}{\left\vert U\right\vert }$
indicate that $u_{j}$ and $u_{k}$ "cohere" together in a block or "pure state"
of the partition, i.e., are an indit of the partition. Since the ordered pairs
$\left(  u_{j},u_{j}\right)  $ in the diagonal $\Delta\subseteq U\times U$ are
always indits of any partition, the diagonal entries in $\rho\left(
\pi\right)  $ are always $\frac{1}{\left\vert U\right\vert }$. After
interchanging some rows and the corresponding columns, the density matrix
$\rho\left(  \pi\right)  $ would be a block-diagonal matrix with the blocks
corresponding to the blocks $B$ of the partition $\pi$.

The \textit{quantum} \textit{logical entropy} of a density matrix $\rho$ in
full QM is: $h\left(  \rho\right)  =1-\operatorname*{tr}\left[  \rho
^{2}\right]  $, and the logical entropy of a set partition $\pi$ with
equiprobable points is $h\left(  \pi\right)  =1-\sum_{B\in\pi}p_{B}^{2}$. The
following proposition shows that the above defined density matrix $\rho\left(
\pi\right)  $ in QM/sets was the right definition.

\begin{proposition}
$h\left(  \pi\right)  =1-\operatorname*{tr}\left[  \rho\left(  \pi\right)
^{2}\right]  $.
\end{proposition}

Proof: The proof is simplified if we assume that rows and columns have been
interchanged so that $\rho\left(  \pi\right)  $ is a block-diagonal matrix
with the submatrix-blocks corresponding to the blocks of partition $\pi$. If
$u_{i}\in B\in\pi,$then the $i^{th}$ diagonal element of the squared matrix
$\rho\left(  \pi\right)  ^{2}$ is $\frac{1}{\left\vert U\right\vert }\frac
{1}{\left\vert U\right\vert }+...+\frac{1}{\left\vert U\right\vert }\frac
{1}{\left\vert U\right\vert }$ ($\left\vert B\right\vert $ times)
or$\left\vert B\right\vert \left(  \frac{1}{\left\vert U\right\vert }\right)
^{2} $ and that diagonal element will occur $\left\vert B\right\vert $ times.
Hence the trace (sum of diagonal elements) is:

\begin{center}
$\operatorname*{tr}\left[  \rho\left(  \pi\right)  ^{2}\right]  =\sum_{B\in
\pi}\left\vert B\right\vert \times\left\vert B\right\vert \frac{1}{\left\vert
U\right\vert ^{2}}=\sum_{B\in\pi}\left(  \frac{|B|}{\left\vert U\right\vert
}\right)  ^{2}=\sum_{B\in\pi}p_{B}^{2}$
\end{center}

\noindent so $h\left(  \pi\right)  =1-\sum_{B}p_{B}^{2}$ equals the delifted
quantum version: $h\left(  \rho\left(  \pi\right)  \right)
=1-\operatorname*{tr}\left[  \rho\left(  \pi\right)  ^{2}\right]  $. $\square$

The logical entropy $h\left(  \pi\right)  $ of a partition is interpreted as
the total two-draw probability of drawing a distinction of the partition $\pi
$. Hence by the above proposition, $\operatorname*{tr}\left[  \rho\left(
\pi\right)  ^{2}\right]  $ is the total probability of drawing an
indistinction of $\pi$. For a pure state, we have the logical entropy
$h\left(  \rho\left(  B\right)  \right)  =1-\operatorname*{tr}\left[
\rho\left(  B\right)  ^{2}\right]  =0$ since the sum of the indistinction
probabilities $\operatorname*{tr}\left[  \rho\left(  B\right)  ^{2}\right]  $
is $1$ (all pairs are indistinctions in a pure state) while in the general
"mixed state" of a partition $\pi$ (with "orthogonal pure state" blocks
$B\in\pi$), $\operatorname*{tr}\left[  \rho\left(  \pi\right)  ^{2}\right]  $
is the sum of the indistinction probabilities.

All this carries over from QM/sets to full QM where it provides \textit{an
interpretation of the entries in a density matrix}. Let $\rho=\sum_{i=1}%
^{m}\lambda_{i}\left\vert \psi_{i}\right\rangle \left\langle \psi
_{i}\right\vert $ be an $n\times n$ density matrix in its orthogonal
decomposition so the non-negative eigenvalues $\lambda_{i}$ sum to one and the
eigenvectors $\psi_{i}$ are orthonormal. Let $\left\{  \left\vert
j\right\rangle :j=1,...,n\right\}  $ be an orthonormal eigenvector basis for
the whole space so that $\psi_{i}=\sum_{j}\alpha_{ij}\left\vert j\right\rangle
$ and $\sum_{j}\alpha_{ij}\alpha_{ij}^{\ast}=1$ where both sums can be taken
as only over the $j$ such that $\left\vert j\right\rangle $ has the eigenvalue
$\lambda_{i}$ (since $\alpha_{ij}=0$ elsewhere). Previously the square
$\rho_{jk}\left(  \pi\right)  ^{2}$ was the two-draw probability for the
ordered pair of indices $\left(  j,k\right)  $ if they are in the same block,
i.e., are indits of $\pi$, otherwise $\rho_{jk}\left(  \pi\right)  =0$.
Similarly, the absolute square $\rho_{jk}\rho_{jk}^{\ast} $ of that $j,k$
entry of $\rho$ is nonzero only if $\left\vert j\right\rangle $ and
$\left\vert k\right\rangle $ are in the same pure state $\psi_{i}$ so those
probabilities can be interpreted as the \textit{coherence probabilities} for
$\left(  \left\vert j\right\rangle ,\left\vert k\right\rangle \right)  $
cohering together in the same pure state $\psi_{i}$. That is,

\begin{center}
$\rho_{jk}\rho_{jk}^{\ast}=\lambda_{i}\alpha_{ij}\alpha_{ik}^{\ast}\lambda
_{i}\alpha_{ij}^{\ast}\alpha_{ik}=\rho_{jj}\rho_{kk}$
\end{center}

\noindent which is the probability of getting the ordered pair of eigenvectors
$\left(  \left\vert j\right\rangle ,\left\vert k\right\rangle \right)  $ in a
pair of independent nondegenerate measurements in the $\left\{  \left\vert
j\right\rangle \right\}  $ basis--if $\left\vert j\right\rangle $ and
$\left\vert k\right\rangle $ cohere together in the same pure state $\psi_{i}%
$. Thus in full QM, $\operatorname*{tr}\left[  \rho^{2}\right]  $ is the
\textit{total coherence probability} while the logical entropy $h\left(
\rho\right)  =1-\operatorname*{tr}\left[  \rho^{2}\right]  $ is the
\textit{total decoherence probability}. For a pure state, there are no
distinctions or decoherence, so the logical entropy is $0$ in both cases. The
following table then summarizes the lifting-delifting relationship between the
density matrix $\rho\left(  \pi\right)  $ of a partition in QM/sets and (the
orthogonal decomposition presentation of) a density matrix $\rho$ in QM.

\begin{center}%
\begin{tabular}
[c]{|c|c|}\hline
Density matrix: $\rho\left(  \pi\right)  $ in QM over sets & $\rho=\sum
_{i}\lambda_{i}\left\vert \psi_{i}\right\rangle \left\langle \psi
_{i}\right\vert $ in QM over $%
\mathbb{C}
$\\\hline\hline
Disjoint blocks: $B\in\pi$ & Orthogonal eigenvectors: $\left\vert \psi
_{i}\right\rangle $\\\hline
Block probabilities: $p_{B}=\frac{|B|}{\left\vert U\right\vert }$ &
Eigenvalues of $\rho$: $\lambda_{i}$\\\hline
Point probabilities: $\frac{1}{\left\vert U\right\vert }$ & $\lambda_{i}%
\alpha_{ij}\alpha_{ij}^{\ast}=\rho_{jj}$\\\hline
Pure state matrix: $\rho\left(  B\right)  =\left\vert B\right\rangle
\left\langle B\right\vert $ & $\rho\left(  \psi_{i}\right)  =\left\vert
\psi_{i}\right\rangle \left\langle \psi_{i}\right\vert $\\\hline
Density matrix: $\rho\left(  \pi\right)  =\sum_{B\in\pi}p_{B}\rho\left(
B\right)  $ & $\rho=\sum_{i}\lambda_{i}\rho\left(  \psi_{i}\right)  $\\\hline
Prob. $\left(  j,k\right)  $ if indit of $\pi$: $\rho_{jk}\left(  \pi\right)
^{2}=1/\left\vert U\right\vert ^{2}$ & Coherence prob.: $\rho_{jk}\rho
_{jk}^{\ast}=\rho_{jj}\rho_{kk}$\\\hline
Logical entropy: $h\left(  \rho\left(  \pi\right)  \right)
=1-\operatorname*{tr}\left[  \rho\left(  \pi\right)  ^{2}\right]  $ &
$h\left(  \rho\right)  =1-\operatorname*{tr}\left[  \rho^{2}\right]  $\\\hline
$h\left(  \rho\left(  \pi\right)  \right)  $ = total distinction probability &
$h\left(  \rho\right)  $ = total decoherence prob.\\\hline
Pure state: $h\left(  \rho\left(  B_{i}\right)  \right)  =0$ (no dits) &
$h\left(  \rho\left(  \psi_{i}\right)  \right)  =0$ (no decoherence)\\\hline
\end{tabular}

Density matrices QM/sets and in QM
\end{center}

Previously we formulated a probability calculus for QM/sets and then noted
that it was just the usual logical finite probability theory (in a
"non-commutative" version) so that reflects back to give a better
understanding of the usual probability calculus in full QM. Now we have
formulated the notion of density matrices in QM/sets, and then we noted that
it was just a reformulation of logical information theory using the density
matrix formalism. Then that reflects back to full QM so that we can now
provide an interpretation of the off-diagonal entries in a density matrix
$\rho$ as coherence probabilities (like the indistinction probabilities in the
set case). And then the quantum logical entropy is the total decoherence probability.

\subsection{Density matrices and expectations}

Given an attribute $f:U=\left\{  u_{1},...,u_{n}\right\}  \rightarrow%
\mathbb{R}
$, the matrix representing this attribute in QM/sets is:

\begin{center}
$f=%
\begin{bmatrix}
f(1) & 0 & \cdots & 0\\
0 & f\left(  2\right)  & \cdots & 0\\
\vdots & \vdots & \ddots & \vdots\\
0 & 0 & \cdots & f\left(  n\right)
\end{bmatrix}
$.
\end{center}

Given a subset $S\subseteq U$, the "density matrix" for that state has, with
some column and row interchanges, a constant $\left\vert S\right\vert
\times\left\vert S\right\vert $ block with the values $1/\left\vert
S\right\vert $ and zeros elsewhere:

\begin{center}
$\rho\left(  S\right)  =%
\begin{bmatrix}
\frac{1}{\left\vert S\right\vert } & \cdots & \frac{1}{\left\vert S\right\vert
} & 0 & \cdots & 0\\
\vdots & \ddots & \vdots & \vdots & \ddots & \vdots\\
\frac{1}{\left\vert S\right\vert } & \cdots & \frac{1}{\left\vert S\right\vert
} & 0 & \cdots & 0\\
0 & \cdots & 0 & 0 & \cdots & 0\\
\vdots & \ddots & \vdots & \vdots & \ddots & \vdots\\
0 & \cdots & 0 & 0 & \cdots & 0
\end{bmatrix}
$.
\end{center}

\noindent Then, as in full QM, we have the result that the average value of an
operator $f$ in a state given by a density matrix $\rho\left(  S\right)  $ is
the trace of the product $f\rho\left(  S\right)  $:

\begin{center}
$\operatorname*{tr}\left[  f\rho\left(  S\right)  \right]  =\frac
{1}{\left\vert S\right\vert }\sum_{u\in S}f\left(  u\right)  =\frac
{1}{\left\vert S\right\vert }\sum_{u\in U}f\left(  u\right)  \left\langle
S|_{U}\left\{  u\right\}  \right\rangle \left\langle \left\{  u\right\}
|_{U}S\right\rangle $

$=\frac{1}{\left\vert S\right\vert }\left\langle S|_{U}f\upharpoonright
()\sum_{u}\left\vert \left\{  u\right\}  \right\rangle \left\langle \left\{
u\right\}  \right\vert _{U}|S\right\rangle =\frac{\left\langle S|_{U}%
f\upharpoonright()|S\right\rangle }{\left\langle S|_{U}S\right\rangle
}=\left\langle f\right\rangle _{S}$
\end{center}

\noindent where $f\upharpoonright\left\vert \left\{  u\right\}  \right\rangle
=f\left(  u\right)  \left\vert \left\{  u\right\}  \right\rangle $ and
$\sum_{u}\left\vert \left\{  u\right\}  \right\rangle \left\langle \left\{
u\right\}  \right\vert _{U}=I$.

\subsection{Measuring measurement in QM/sets}

A real-valued "observable" is a set attribute $f:U\rightarrow%
\mathbb{R}
$ which defines an inverse-image partition $\left\{  f^{-1}\left(  r\right)
\right\}  $. Recall from the logic of partitions that the blocks of the join
$\pi\vee\sigma$ of two partitions $\pi=\left\{  B\right\}  $ and
$\sigma=\left\{  C\right\}  $ are the non-empty intersections $B\cap C$. This
action of the join operation could be considered as a set of projection
operators $\left\{  B\cap()\right\}  _{B\in\pi}$ acting on the blocks
$C\in\sigma$--or on a single subset $S\subseteq U$. The partition
$f^{-1}=\left\{  f^{-1}\left(  r\right)  \right\}  $ acts as a set of
projection operators $f^{-1}\vee()=\left\{  f^{-1}\left(  r\right)
\cap()\right\}  $ on the "pure-state" $S$ to partition it into the parts
$f^{-1}\vee(S)=\left\{  f^{-1}\left(  r\right)  \cap S\right\}  $.

What is the "law of motion" to describe the change in the density matrix
resulting from a measurement? Given the density matrix $\rho\left(  S\right)
$ of the "pure state" $S$, the density matrix $\hat{\rho}\left(  S\right)  $
resulting from the measurement of the observable $f$ is the "mixed state"
density matrix $\rho\left(  \pi\right)  $ for the partition given by the join
operation $\pi=f^{-1}\vee\left(  S\right)  $. Thus the "law of motion" is the
join operation on partitions. That is the canonical way that distinctions are
made to move to a more refined partition.

Let's put the previous measurement of the state $S=U$ using the non-degenerate
attribute $f(a)=1,f(b)=2$, and $f(c)=3$ in this form using density matrices.
The pre-measurement density matrix is the previous $\rho\left(  U\right)  $,
the constant matrix with all entries $1/3$. The three projection operators to
the eigenspaces of the $f$-attribute in the $U$ -basis are now:

\begin{center}
$P_{1}=%
\begin{bmatrix}
1 & 0 & 0\\
0 & 0 & 0\\
0 & 0 & 0
\end{bmatrix}
$, $P_{2}=%
\begin{bmatrix}
0 & 0 & 0\\
0 & 1 & 0\\
0 & 0 & 0
\end{bmatrix}
$, and $P_{3}=%
\begin{bmatrix}
0 & 0 & 0\\
0 & 0 & 0\\
0 & 0 & 1
\end{bmatrix}
$
\end{center}

\noindent instead of $\left\{  f^{-1}\left(  r\right)  \cap\left(  {}\right)
\right\}  _{r=1,2,3}$ in the non-matrix version. Hence the "density matrix"
for the projection to the eigenspace for $\lambda=1$ is obtained by first
projecting the state $P_{1}\left\vert U\right\rangle $ (like $f^{-1}\left(
1\right)  \cap\left(  U\right)  =\{a\}$ in the non-matrix version) and then
forming the "density matrix"

\begin{center}
$\left(  P_{1}\left\vert U\right\rangle \right)  \left(  P_{1}\left\vert
U\right\rangle \right)  ^{t}=P_{1}\rho\left(  U\right)  P_{1}=P_{1}%
\begin{bmatrix}
\frac{1}{\sqrt{3}}\\
\frac{1}{\sqrt{3}}\\
\frac{1}{\sqrt{3}}%
\end{bmatrix}%
\begin{bmatrix}
\frac{1}{\sqrt{3}} & \frac{1}{\sqrt{3}} & \frac{1}{\sqrt{3}}%
\end{bmatrix}
P_{1}$

$=P_{1}%
\begin{bmatrix}
\frac{1}{3} & \frac{1}{3} & \frac{1}{3}\\
\frac{1}{3} & \frac{1}{3} & \frac{1}{3}\\
\frac{1}{3} & \frac{1}{3} & \frac{1}{3}%
\end{bmatrix}
P_{1}=%
\begin{bmatrix}
\frac{1}{3} & \frac{1}{3} & \frac{1}{3}\\
0 & 0 & 0\\
0 & 0 & 0
\end{bmatrix}
P_{1}=%
\begin{bmatrix}
\frac{1}{3} & 0 & 0\\
0 & 0 & 0\\
0 & 0 & 0
\end{bmatrix}
$
\end{center}

\noindent so doing the same for the other eigenvalues and summing gives the
mixed state density matrix $\hat{\rho}\left(  U\right)  $ that results from
the measurement:

\begin{center}
$\hat{\rho}\left(  U\right)  =\sum_{i=1}^{3}P_{i}\rho\left(  U\right)  P_{i}=%
\begin{bmatrix}
\frac{1}{3} & 0 & 0\\
0 & \frac{1}{3} & 0\\
0 & 0 & \frac{1}{3}%
\end{bmatrix}
$.
\end{center}

\noindent The \textit{main result} is that this standard diagonal density
matrix representing the result of a non-degenerate measurement is the density
matrix $\rho\left(  \pi\right)  $ of the partition formed by the join-action:

\begin{center}
$\pi=f^{-1}\vee\left(  U\right)  =\left\{  \left\{  a\right\}  ,\left\{
b\right\}  ,\left\{  c\right\}  \right\}  \vee\left\{  a,b,c\right\}
=\left\{  \left\{  a\right\}  ,\left\{  b\right\}  ,\left\{  c\right\}
\right\}  =\mathbf{1}$.
\end{center}

\noindent Since it was a non-degenerate measurement, all the distinctions were
made so all the off-diagonal terms are $0$. Each of the off-diagonal terms was
"decohered" by the nondegenerate measurement so the post-measurement
"amplitude" of $\left(  i,j\right)  $ still "cohering" is $0$. The density
matrix version of the

\begin{center}
$\rho\left(  U\right)  \overset{measurement}{\longrightarrow}\hat{\rho}\left(
U\right)  =\rho\left(  f^{-1}\vee\left(  U\right)  \right)  $

\textit{Measurement as join-action}
\end{center}

\noindent allows us, as usual, to state the general result of a measurement
without assuming a particular outcome.\footnote{Note that this set-version of
"decoherence" means actual reduction of state, not a "for all practical
purposes" or FAPP \cite{bell:againstm} reduction.}

The general result is that the logical entropy increase resulting from a
measurement is the sum of the new distinction probabilities created by the
join, which is the sum of the squared amplitudes of the off-diagonal
indistinction amplitudes in the density matrix that were zeroed or "decohered"
by the measurement.

In the example, the six off-diagonal amplitudes of $\frac{1}{3}$ were all
zeroed so the change in logical entropy is: $6\times\left(  \frac{1}%
{3}\right)  ^{2}=\frac{6}{9}=\frac{2}{3}$.

\begin{center}
$\rho\left(  U\right)  =%
\begin{bmatrix}
\frac{1}{3} & \frac{1}{3} & \frac{1}{3}\\
\frac{1}{3} & \frac{1}{3} & \frac{1}{3}\\
\frac{1}{3} & \frac{1}{3} & \frac{1}{3}%
\end{bmatrix}
\overset{measurement}{\longrightarrow}\hat{\rho}\left(  U\right)  =%
\begin{bmatrix}
\frac{1}{3} & 0 & 0\\
0 & \frac{1}{3} & 0\\
0 & 0 & \frac{1}{3}%
\end{bmatrix}
$.
\end{center}

\noindent In terms of sets, there are no distinctions in the indiscrete
partition $\mathbf{0=}\left\{  U\right\}  $ so $h\left(  \mathbf{0}\right)
=\frac{\left\vert \operatorname*{dit}\left(  \mathbf{0}\right)  \right\vert
}{\left\vert U\times U\right\vert }=0$. Measurement by a non-degenerate
attribute $f$ gives the discrete partition $\mathbf{1}=f^{-1}\vee\left(
U\right)  $ where the distinctions are the ordered pairs $\left(  a,b\right)
$, $\left(  a,c\right)  $, and $\left(  b,c\right)  $ together with the three
opposite ordered pairs $\left(  b,a\right)  $, $\left(  c,a\right)  $, and
$\left(  c,b\right)  $ so the logical entropy is $h\left(  \mathbf{1}\right)
=\frac{\left\vert \operatorname*{dit}\left(  \mathbf{1}\right)  \right\vert
}{\left\vert U\times U\right\vert }=\frac{6}{9}=\frac{2}{3}$. Those ordered
pairs correspond exactly to off-diagonal terms zeroed in the transition
$\rho\left(  U\right)  \rightarrow\hat{\rho}$ and $h\left(  \mathbf{1}\right)
=1-\operatorname*{tr}\left[  \hat{\rho}^{2}\right]  =1-\left(  \frac{1}%
{9}+\frac{1}{9}+\frac{1}{9}\right)  =\frac{2}{3}$.

In this manner, the density matrices of QM/sets capture the set-based
operations of logical information theory, and that, in turn, shows \textit{how
to interpret the density matrices of full QM} in terms of coherence and
decoherence probabilities. The usual notion of von Neumann entropy in quantum
information theory provides no such information-theoretic term-by-term
interpretation of density matrices, not to mention of the process of
measurement. In this manner, QM/sets shows, from the information-theoretic
viewpoint, the essence at the logical level of what is going on in the full
QM, i.e., QM/sets shows the "logic" of QM. The further development of the
classical or quantum information theory using logical entropy is beyond the
scope of this introductory paper \cite{ell:distinctions}.

\section{Quantum computation theory in QM/sets}

\subsection{Qubits over 2 and non-singular gates}

In QM over $%
\mathbb{C}
$, a \textit{quantum bit} or \textit{qubit} is a non-zero (normalized) vector
in $%
\mathbb{C}
^{2}$. A standard orthonormal basis is denoted $\left\vert 0\right\rangle $
and $\left\vert 1\right\rangle $ so a qubit can be any (normalized)
superposition $\alpha\left\vert 0\right\rangle +\beta\left\vert 1\right\rangle
$ for $\alpha,\beta\in%
\mathbb{C}
$. In QM/sets, i.e., QM over $%
\mathbb{Z}
_{2}$, a \textit{qubit over }$\mathit{2}$ or \textit{qubit/}$\mathit{2}$ is
any non-zero vector in $%
\mathbb{Z}
_{2}^{2}$ which for a given basis $\left\vert 0\right\rangle $ and $\left\vert
1\right\rangle $ would have the form $\alpha\left\vert 0\right\rangle
+\beta\left\vert 1\right\rangle $ for $\alpha,\beta\in%
\mathbb{Z}
_{2}$. As previously noted, Schumacher and Westmoreland (S\&W)
\cite{schum:modal} restrict their treatment of Dirac's brackets to take values
in the base field of $%
\mathbb{Z}
_{2}$ which precludes a probability calculus so they develop a modal
interpretation ($0$ = impossible and $1$ = possible). Hence they call a
non-zero vector in $%
\mathbb{Z}
_{2}^{2}$ a "mobit" and call the resulting theory "modal quantum theory."
Since our different treatment of the brackets yields a full probability
calculus in QM/sets, we will not use the "modal" terminology but,
nevertheless, their "mobit" is the same as our "qubit/$2$."

In $%
\mathbb{C}
^{2}$, there is a continuum of qubits $\alpha\left\vert 0\right\rangle
+\beta\left\vert 1\right\rangle $ for $\alpha,\beta\in%
\mathbb{C}
$ but in $%
\mathbb{Z}
_{2}^{2}$, there are only $3$ qubits/$2$, namely $\left\vert 0\right\rangle $,
$\left\vert 1\right\rangle $, and $\left\vert 0\right\rangle +\left\vert
1\right\rangle $. Hence a qubit/$2$ can be seen as the simplest possible
extension beyond the classical bit with the two possibilities $\left\vert
0\right\rangle $ and $\left\vert 1\right\rangle $ by adding the superposition
$\left\vert 0\right\rangle +\left\vert 1\right\rangle $.\footnote{Here we are
following the mild conceptual sloppiness common in the field of referring to
any binary option as a "classical bit" when the bit as defined in Shannon's
information theory is actually an \textit{equiprobable} binary option. The
comparable notion in logical information theory is a \textit{distinction} or
\textit{dit} of a partition $\pi$ on $U$ which is exactly defined as an
ordered pair $\left(  u,u^{\prime}\right)  $ elements distinguished by $\pi$
in the sense of the elements being in distinct blocks of $\pi$.} As already
noted in our treatment of Bell's Theorem in QM/sets, there are only three
basis sets for $%
\mathbb{Z}
_{2}^{2}$; any two of non-zero vectors are a basis with the third as their superposition.

In QM/sets (as in S\&W's modal quantum theory), the dynamics are given by
non-singular transformations which may be represented as non-singular zero-one
matrices (which have non-zero determinants mod $2$). A qubit over $2 $,
$\alpha\left\vert 0\right\rangle +\beta\left\vert 1\right\rangle $, is
represented in the standard basis $\left\vert 0\right\rangle $ and $\left\vert
1\right\rangle $ by the column vector $[\alpha,\beta]^{t}$.

The non-singular transformations are the \textit{gates} that may be used in an
algorithm for quantum computing over $2$ (QC/$2$). The two one-qubit gates
that carry over from quantum computing over $%
\mathbb{C}
$ are the:

\begin{center}
\textit{identity} $I=%
\begin{bmatrix}
1 & 0\\
0 & 1
\end{bmatrix}
$ and \textit{negation} $X=%
\begin{bmatrix}
0 & 1\\
1 & 0
\end{bmatrix}
$.
\end{center}

\noindent The four other one-qubit/$2$ gates in QC/$2$ are non-singular but
when interpreted as matrices in $%
\mathbb{C}
^{2}$ are not unitary. In particular, there is no requirement that a gate
preserves the norm of a vector. One one-qubit/$2$ gate puts $\left\vert
0\right\rangle $ into the superposition $\left\vert 0\right\rangle +\left\vert
1\right\rangle $ and leaves $\left\vert 1\right\rangle $ the same:

\begin{center}
$H_{0}=%
\begin{bmatrix}
1 & 0\\
1 & 1
\end{bmatrix}
$.
\end{center}

\noindent Similarly another one-qubit/$2$ gate puts $\left\vert 1\right\rangle
$ into the superposition and leaves $\left\vert 0\right\rangle $ the same:

\begin{center}
$H_{1}=%
\begin{bmatrix}
1 & 1\\
0 & 1
\end{bmatrix}
$.
\end{center}

\noindent And finally the other two one-qubit/$2$ gates are their negations:

\begin{center}
$XH_{0}=%
\begin{bmatrix}
0 & 1\\
1 & 0
\end{bmatrix}%
\begin{bmatrix}
1 & 0\\
1 & 1
\end{bmatrix}
=%
\begin{bmatrix}
1 & 1\\
1 & 0
\end{bmatrix}
$ and $XH_{1}=%
\begin{bmatrix}
0 & 1\\
1 & 0
\end{bmatrix}%
\begin{bmatrix}
1 & 1\\
0 & 1
\end{bmatrix}
=%
\begin{bmatrix}
0 & 1\\
1 & 1
\end{bmatrix}
$.
\end{center}

\noindent These six one-qubit/$2$ gates are the only non-singular
transformations $%
\mathbb{Z}
_{2}^{2}\rightarrow%
\mathbb{Z}
_{2}^{2}$.

As we will see below, some problems like the simplest Deutsch problem of
determining if a single-variable Boolean function is balanced or constant can
be solved in QC/$2$ solely with one-qubit/$2$ gates, whereas the usual
solution to that problem in quantum computing over $%
\mathbb{C}
$ uses two-qubit gates (four dimensional matrices). This is not as paradoxical
as it may seem if we recall that quantum computing over $2$ allows
non-singular gates whereas the gates over $%
\mathbb{C}
$ have to be unitary.

In representing these gates in the standard basis, we will use the standard
Alice-Bob convention that the first or top one-qubit/$2$ (on the left) belongs
to Alice and the second or bottom one-qubit/$2$ (on the right) belongs to Bob
so the four basis vectors are: $\left\vert 0_{A}\right\rangle \otimes
\left\vert 0_{B}\right\rangle =\left\vert 0_{A}0_{B}\right\rangle $,
$\left\vert 0_{A}\right\rangle \otimes\left\vert 1_{B}\right\rangle
=\left\vert 0_{A}1_{B}\right\rangle $, $\left\vert 1_{A}\right\rangle
\otimes\left\vert 0_{B}\right\rangle =\left\vert 1_{A}0_{B}\right\rangle $,
and $\left\vert 1_{A}\right\rangle \otimes\left\vert 1_{B}\right\rangle
=\left\vert 1_{A}1_{B}\right\rangle $ (they are arranged in that order in the
column vectors).

One two-qubit gate that carries over from quantum computing over $%
\mathbb{C}
$ is the \textit{controlled negation} gate:

\begin{center}
$Cnot_{A}=%
\begin{bmatrix}
1 & 0 & 0 & 0\\
0 & 1 & 0 & 0\\
0 & 0 & 0 & 1\\
0 & 0 & 1 & 0
\end{bmatrix}
$
\end{center}

\noindent which may be represented as acting on Alice's top line and Bob's
bottom line:

\begin{center}
$%
\begin{array}
[c]{c}%
\rightarrow\bullet\rightarrow\\
\mid\\
\rightarrow\oplus\rightarrow
\end{array}
$.
\end{center}

\noindent The action of a gate is specified by how it acts on the basis
vectors. For either case $\left\vert 0_{A}0_{B}\right\rangle $ and $\left\vert
0_{A}1_{B}\right\rangle $ where Alice's qubit/$2$ is $\left\vert
0_{A}\right\rangle $, the gate acts like the identity. But in the cases
$\left\vert 1_{A}0_{B}\right\rangle $ and $\left\vert 1_{A}1_{B}\right\rangle
$ where Alice's qubit/$2$ is $\left\vert 1_{A}\right\rangle $, then Bob's
qubit/$2$ is negated so that $\left\vert 1_{A}0_{B}\right\rangle
\rightarrow\left\vert 1_{A}1_{B}\right\rangle $ and $\left\vert 1_{A}%
1_{B}\right\rangle \rightarrow\left\vert 1_{A}0_{B}\right\rangle $. In this
case, Alice's qubit/$2$ is said to be the \textit{controlling} qubit/$2$
(indicated by the subscript on $Cnot_{A}$) and Bob's the \textit{target}
qubit/$2$.

The controlling and target roles are reversed in the gate:

\begin{center}
$Cnot_{B}=%
\begin{bmatrix}
1 & 0 & 0 & 0\\
0 & 0 & 0 & 1\\
0 & 0 & 1 & 0\\
0 & 1 & 0 & 0
\end{bmatrix}
$ represented as $%
\begin{array}
[c]{c}%
\rightarrow\oplus\rightarrow\\
\mid\\
\rightarrow\bullet\rightarrow
\end{array}
$
\end{center}

\noindent where if Bob's qubit/$2$ is $\left\vert 0_{B}\right\rangle $, then
it acts like the identity, but if Bob's qubit/$2$ is $\left\vert
1_{B}\right\rangle $, then Alice's qubit/$2$ is negated.

In a two-qubit/$2$ system, if a one-qubit/$2$ gate is to be applied to only
one line, then tensor product of matrices is used. For instance to apply
$H_{0}$ only to Bob's line, the two-qubit/$2$ gate is:

\begin{center}
$I\otimes H_{0}=%
\begin{bmatrix}
1 & 0 & 0 & 0\\
1 & 1 & 0 & 0\\
0 & 0 & 1 & 0\\
0 & 0 & 1 & 1
\end{bmatrix}
$ represented as $%
\begin{array}
[c]{c}%
\longrightarrow\\
\rightarrow\fbox{$H_0$}\rightarrow
\end{array}
$.
\end{center}

\subsection{Teleportation of a qubit/$2$ with $1$ classical bit}

S\&W's treatment \cite{schum:modal} of the no-cloning theorem and superdense
coding would work the same in QC/$2$ so we will not repeat it here. But after
their treatment of superdense coding (of two classical bits), they remark:
"The same set of entangled mobit states and single-mobit transformations can
also be used to accomplish the MQT analogue of quantum teleportation."
\cite[p. 924]{schum:modal} But that MQT (modal quantum theory) analogue of the
usual quantum teleportation in full QM is somewhat odd since there are only
three possible non-zero qubits/$2$ or mobits, and two classical bits suffice
to transmit the identity of four different states--without entanglement having
anything to do with it--if Alice knew which of the three mobits she had. It
would be more in the spirit of quantum teleportation to transmit a qubit/$2$
(or mobit) using only one classical bit so that the entanglement has a real
role. That is what we do.

In contrast with the usual two-bit teleportation scheme (\cite{bennett:telep},
\cite[pp. 26-28]{nc:qcqi}), Alice only has one line instead of two, and she
starts off with the qubit/$2$ $\left\vert \psi\right\rangle =\alpha\left\vert
0_{A}\right\rangle +\beta\left\vert 1_{A}\right\rangle $ to be teleported to
Bob, while Bob starts with the usual $\left\vert 0_{B}\right\rangle $, so the
initial state in the two-qubit/$2$ system is $\left\vert \varphi
_{0}\right\rangle =\left(  \alpha\left\vert 0_{A}\right\rangle +\beta
\left\vert 1_{A}\right\rangle \right)  \otimes\left\vert 0_{B}\right\rangle
=\alpha\left\vert 0_{A}0_{B}\right\rangle +\beta\left\vert 1_{A}%
0_{B}\right\rangle $. The circuit diagram for the one-bit teleportation
protocol is:%

\begin{center}
\includegraphics[
height=1.8323in,
width=3.5052in
]%
{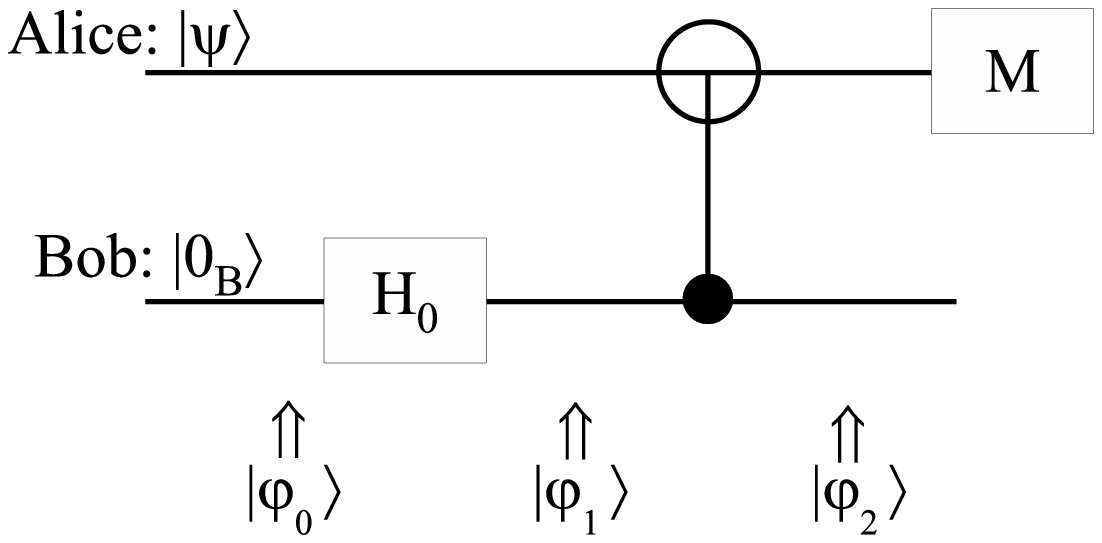}%
\end{center}

\begin{center}
Figure 10: Teleportation scheme for a qubit/$2$ using $1$ classical bit
\end{center}

\noindent where \fbox{$M$} refers to Alice measuring her qubit/$2$. Alice and
Bob start off together. First Bob applies the $H_{0}$ gate to his line (i.e.,
$I\otimes H_{0}$ is applied to both lines) to put Bob's state in the
superposition $\left\vert 0_{B}\right\rangle +\left\vert 1_{B}\right\rangle $:

\begin{center}
$\left\vert \varphi_{1}\right\rangle =\left(  I\otimes H_{0}\right)
\begin{bmatrix}
\alpha\\
0\\
\beta\\
0
\end{bmatrix}
=%
\begin{bmatrix}
1 & 0 & 0 & 0\\
1 & 1 & 0 & 0\\
0 & 0 & 1 & 0\\
0 & 0 & 1 & 1
\end{bmatrix}%
\begin{bmatrix}
\alpha\\
0\\
\beta\\
0
\end{bmatrix}
=\allowbreak%
\begin{bmatrix}
\alpha\\
\alpha\\
\beta\\
\beta
\end{bmatrix}
$

$=\alpha\left(  \left\vert 0_{A}0_{B}\right\rangle +\left\vert 0_{A}%
1_{B}\right\rangle \right)  +\beta\left(  \left\vert 1_{A}0_{B}\right\rangle
+\left\vert 1_{A}1_{B}\right\rangle \right)  $

$=\left(  \alpha\left\vert 0_{A}\right\rangle +\beta\left\vert 1_{A}%
\right\rangle \right)  \otimes\left(  \left\vert 0_{B}\right\rangle
+\left\vert 1_{B}\right\rangle \right)  $.
\end{center}

\noindent That non-entangled mutual state is then entangled by applying the
$Cnot_{B}$ gate:

\begin{center}
$\left\vert \varphi_{2}\right\rangle =%
\begin{bmatrix}
1 & 0 & 0 & 0\\
0 & 0 & 0 & 1\\
0 & 0 & 1 & 0\\
0 & 1 & 0 & 0
\end{bmatrix}%
\begin{bmatrix}
\alpha\\
\alpha\\
\beta\\
\beta
\end{bmatrix}
=\allowbreak%
\begin{bmatrix}
\alpha\\
\beta\\
\beta\\
\alpha
\end{bmatrix}
$

$=\alpha\left(  \left\vert 0_{A}0_{B}\right\rangle +\left\vert 1_{A}%
1_{B}\right\rangle \right)  +\beta\left(  \left\vert 0_{A}1_{B}\right\rangle
+\left\vert 1_{A}0_{B}\right\rangle \right)  $

$=\left\vert 0_{A}\right\rangle \otimes\left(  \alpha\left\vert 0_{B}%
\right\rangle +\beta\left\vert 1_{B}\right\rangle \right)  +\left\vert
1_{A}\right\rangle \otimes\left(  \beta\left\vert 0_{B}\right\rangle
+\alpha\left\vert 1_{B}\right\rangle \right)  $.
\end{center}

\noindent Then Bob and Alice "separate" (like a pair of particles in the EPR
experiment) so their only connection is the entangled state--and a classical
communication channel for one classical bit. Without further operations, Alice
then measures her line and gets either a $\left\vert 0_{A}\right\rangle $ or
$\left\vert 1_{A}\right\rangle $. If she gets $\left\vert 0_{A}\right\rangle
$, then the state on Bob's line is $\alpha\left\vert 0_{B}\right\rangle
+\beta\left\vert 1_{B}\right\rangle $ so that $\left\vert \psi\right\rangle
=\alpha\left\vert 0_{A}\right\rangle +\beta\left\vert 1_{A}\right\rangle $ has
been teleported to Bob. If Alice gets $\left\vert 1_{A}\right\rangle $ then
Bob's state is $\beta\left\vert 0_{B}\right\rangle +\alpha\left\vert
1_{B}\right\rangle $ so he only need apply the negation gate $X$ to get
$\alpha\left\vert 0_{B}\right\rangle +\beta\left\vert 1_{B}\right\rangle $.
Hence Alice only has to send one classical bit with $0$ = "do nothing" and $1$
= "apply $X$" in order to tell Bob how to get the teleported state on his
line. Taking $M$ as the classical bit sent by Alice and $X^{0}=I$, then the
instruction to Bob is to apply $X^{M}$ to his state to get the teleported state.

Replace the non-unitary but non-singular $H_{0}$ by the unitary
\textit{Hadamard matrix}

\begin{center}
$H=\frac{1}{\sqrt{2}}%
\begin{bmatrix}
1 & 1\\
1 & -1
\end{bmatrix}
$
\end{center}

\noindent and the protocol will teleport a full qubit $\left\vert
\psi\right\rangle =\alpha\left\vert 0_{A}\right\rangle +\beta\left\vert
1_{A}\right\rangle \in%
\mathbb{C}
^{2}$ with \textit{one} classical bit. That (little known) protocol is called
$X$\textit{-teleportation}, was developed by Charles Bennett, and analyzed,
along with some other single-bit teleportation schemes, by Zhou, Leung, and
Chuang \cite{zhou:1bit-telep}.

\subsection{Deutsch's simplest problem in QC/2}

Deutsch's simplest problem is that of determining if a given Boolean function
$y=f\left(  x\right)  $ is \textit{balanced} in the sense of being one-one or
is \textit{constant} (two-to-one). An equivalent classification of the four
unary Boolean functions is whether their \textit{parity} in the sense of the
mod $2$ sum of their values $f\left(  0\right)  +f\left(  1\right)  $ is odd
(balanced) or even (constant)--which is called the\textit{\ parity
satisfiability problem} or \textit{Parity SAT }\cite{vv:np-usat}. In the usual
treatment of Deutsch's problem in quantum computation over $%
\mathbb{C}
$, the gates $U_{f}$ that evaluate the function are $4\times4$ gates which are
unitary. But in quantum computing over $2$, the gates need only be
non-singular. A scheme to encode the four functions in non-singular evaluation
$2\times2$ gates is:

\begin{center}
$E_{f}=X^{f\left(  1\right)  }H_{f\left(  0\right)  }$
\end{center}

\noindent so the four function evaluation gates are:

\begin{center}
$f=X$ so $f\left(  0\right)  =1$ and $f(1)=0$: $E_{f}=X^{0}H_{1}=%
\begin{bmatrix}
1 & 1\\
0 & 1
\end{bmatrix}
=%
\begin{bmatrix}
f\left(  0\right)  & f(1)+1\\
f(1) & f\left(  0\right)
\end{bmatrix}
$;

$f=I$ so $f\left(  0\right)  =0$ and $f\left(  1\right)  =1$: $E_{f}%
=X^{1}H_{0}=%
\begin{bmatrix}
1 & 1\\
1 & 0
\end{bmatrix}
=%
\begin{bmatrix}
f(1) & f\left(  0\right)  +1\\
f\left(  0\right)  +1 & f(1)+1
\end{bmatrix}
$;

$f=0$ so $f\left(  0\right)  =0$ and $f\left(  1\right)  =0$: $E_{f}%
=X^{0}H_{0}=%
\begin{bmatrix}
1 & 0\\
1 & 1
\end{bmatrix}
=%
\begin{bmatrix}
f\left(  0\right)  +1 & f(1)\\
f(1)+1 & f\left(  0\right)  +1
\end{bmatrix}
$;

$f=1$ so $f\left(  0\right)  =1$ and $f\left(  1\right)  =1$: $E_{f}%
=X^{1}H_{1}=%
\begin{bmatrix}
0 & 1\\
1 & 1
\end{bmatrix}
=%
\begin{bmatrix}
f(1)+1 & f\left(  0\right) \\
f\left(  0\right)  & f(1)
\end{bmatrix}
$.
\end{center}

Then it is evident that the mod $2$ sum across the rows is the same for all
four cases:

\begin{center}
$E_{f}%
\begin{bmatrix}
1\\
1
\end{bmatrix}
=%
\begin{bmatrix}
f\left(  0\right)  +f\left(  1\right)  +1\\
f\left(  0\right)  +f\left(  1\right)
\end{bmatrix}
$
\end{center}

\noindent so we only need measure that one-qubit/$2$ line to determine the
function's parity. If the result is $\left\vert 0\right\rangle $, then
$f\left(  0\right)  +f\left(  1\right)  +1=1$ (and $f\left(  0\right)
+f\left(  1\right)  =0)$ so the parity is even (or function is constant) and
if the result is $\left\vert 1\right\rangle $, then $f\left(  0\right)
+f\left(  1\right)  =1$ so the parity is odd (or function is balanced). Hence
the circuit diagram for the QC/$2$ algorithm is:

\begin{center}
$%
\begin{array}
[c]{ccccccc}%
\left\vert 0\right\rangle  & \longrightarrow & \fbox{$H_0$} & \longrightarrow
& \fbox{$E_f$} & \longrightarrow & \fbox{$M$}%
\end{array}
$

QC/$2$ algorithm for the Deutsch problem or Parity SAT problem for unary
Boolean functions
\end{center}

\noindent and the matrix operation giving the one-qubit/$2$ to be measured is:

\begin{center}
$X^{f\left(  1\right)  }H_{f\left(  0\right)  }H_{0}%
\begin{bmatrix}
1\\
0
\end{bmatrix}
=X^{f\left(  1\right)  }H_{f\left(  0\right)  }%
\begin{bmatrix}
1 & 0\\
1 & 1
\end{bmatrix}%
\begin{bmatrix}
1\\
0
\end{bmatrix}
=X^{f\left(  1\right)  }H_{f\left(  0\right)  }%
\begin{bmatrix}
1\\
1
\end{bmatrix}
=%
\begin{bmatrix}
f\left(  0\right)  +f\left(  1\right)  +1\\
f\left(  0\right)  +f\left(  1\right)
\end{bmatrix}
$.
\end{center}

This is the same Deutsch problem usually solved by a two-qubit circuit in full
quantum computing over $%
\mathbb{C}
$. In either case, two classical function evaluations are needed to determine
the parity of the sum of the functions values so the "quantum speedup" is seen
in the quantum algorithm in QC/$2$ or full QC only requiring one function evaluation.

\subsection{The general Parity SAT problem solved in QC/2}

The generalization to $n$-ary Boolean functions $f:%
\mathbb{Z}
_{2}^{n}\rightarrow%
\mathbb{Z}
_{2}$ is simple for the problem of determining the parity of the function
where the parity is determined by the mod $2$ sum of the function's $2^{n}$
values. To keep the notation manageable, we will consider the case of $n=2$
which will make the pattern clear.

The function evaluation matrices $E_{f}$ for binary Boolean functions
$y=f\left(  x_{1},x_{2}\right)  $ may be taken as:

\begin{center}
$E_{f}=X^{f\left(  0,1\right)  }H_{f\left(  0,0\right)  }\otimes
X^{f(1,1)}H_{f\left(  1,0\right)  }$.
\end{center}

Consider the binary Boolean function of the truth-functional conditional or
implication $x_{1}\Rightarrow x_{2}$ where (simplifying $f\left(  0,0\right)
$ to $f_{00}$ etc.) $f_{00}=f_{01}=f_{11}=1$ but $f_{10}=0$, the function
evaluation matrix is:

\begin{center}
$X^{f\left(  0,1\right)  }H_{f\left(  0,0\right)  }\otimes X^{f(1,1)}%
H_{f\left(  1,0\right)  }=X^{1}H_{1}\otimes X^{1}H_{0}$

$=%
\begin{bmatrix}
0 & 1\\
1 & 1
\end{bmatrix}
\otimes%
\begin{bmatrix}
1 & 1\\
1 & 0
\end{bmatrix}
=%
\begin{bmatrix}
f_{01}+1 & f_{00}\\
f_{00} & f_{01}%
\end{bmatrix}
\otimes%
\begin{bmatrix}
f_{11} & f_{10}+1\\
f_{10}+1 & f_{11}+1
\end{bmatrix}
$

$=%
\begin{bmatrix}
\left(  f_{01}+1\right)
\begin{bmatrix}
f_{11} & f_{10}+1\\
f_{10}+1 & f_{11}+1
\end{bmatrix}
& f_{00}%
\begin{bmatrix}
f_{11} & f_{10}+1\\
f_{10}+1 & f_{11}+1
\end{bmatrix}
\\
f_{00}%
\begin{bmatrix}
f_{11} & f_{10}+1\\
f_{10}+1 & f_{11}+1
\end{bmatrix}
& f_{01}%
\begin{bmatrix}
f_{11} & f_{10}+1\\
f_{10}+1 & f_{11}+1
\end{bmatrix}
\end{bmatrix}
$

$=%
\begin{bmatrix}
\left(  f_{01}+1\right)  f_{11} & \left(  f_{01}+1\right)  \left(
f_{10}+1\right)  & f_{00}f_{11} & f_{00}\left(  f_{10}+1\right) \\
\left(  f_{01}+1\right)  \left(  f_{10}+1\right)  & \left(  f_{01}+1\right)
\left(  f_{11}+1\right)  & f_{00}\left(  f_{10}+1\right)  & f_{00}\left(
f_{11}+1\right) \\
f_{00}f_{11} & f_{00}\left(  f_{10}+1\right)  & f_{01}f_{11} & f_{01}\left(
f_{10}+1\right) \\
f_{00}\left(  f_{10}+1\right)  & f_{00}\left(  f_{11}+1\right)  &
f_{01}\left(  f_{10}+1\right)  & f_{01}\left(  f_{11}+1\right)
\end{bmatrix}
$.
\end{center}

The key to any quantum algorithm is the clever use of superposition to extract
the needed information. In this case, the superposition just adds up each row,
which after some simplification, yields the two-qubit/$2$ column vector:

\begin{center}
$%
\begin{bmatrix}
\left(  f_{00}+f_{01}+1\right)  \left(  f_{10}+f_{11}+1\right) \\
\left(  f_{00}+f_{01}+1\right)  \left(  f_{10}+f_{11}\right) \\
\left(  f_{00}+f_{01}\right)  \left(  f_{10}+f_{11}+1\right) \\
\left(  f_{00}+f_{01}\right)  \left(  f_{10}+f_{11}\right)
\end{bmatrix}
$.
\end{center}

\noindent And \textit{regardless} of the binary function (calculated above for
the implication), the above column vector of the row sums in terms of the
function values is always the \textit{same}!\footnote{The proof is just an
elaboration on the fact that the row sums of the tensor product of two
matrices is the product of the row sums of the two matrices.} Note further
that regardless of the values of the function, only one row has sum of $1$ and
the others sum to $0$. Hence we only need to measure that two-qubit/$2$ to
determine the parity of the function. Moreover the sum of values of the
function occur in pairs with the first variable fixed, e.g., $f_{00}+f_{01}$
and $f_{10}+f_{11}$, so each row sum also contains the information about the
parity of those unary functions $f\left(  0,x_{2}\right)  $ and $f\left(
1,x_{2}\right)  $. Hence the significance of the row sums being $1$ is:

\begin{center}
$%
\begin{bmatrix}
\left(  f_{00}+f_{01}+1\right)  \left(  f_{10}+f_{11}+1\right) \\
\left(  f_{00}+f_{01}+1\right)  \left(  f_{10}+f_{11}\right) \\
\left(  f_{00}+f_{01}\right)  \left(  f_{10}+f_{11}+1\right) \\
\left(  f_{00}+f_{01}\right)  \left(  f_{10}+f_{11}\right)
\end{bmatrix}
=%
\begin{bmatrix}
1=EE\\
1=EO\\
1=OE\\
1=OO
\end{bmatrix}
$.
\end{center}

For instance, if the measurement gives $\left\vert 10\right\rangle $, then the
entry in the third row $\left(  f_{00}+f_{01}\right)  \left(  f_{10}%
+f_{11}+1\right)  $ is $1$ which can only happen if each factor is $1$ so the
unary function $f\left(  0,x_{2}\right)  $ is odd and the unary function
$f\left(  1,x_{2}\right)  $ is even.\footnote{It could also be arranged for
the pairs to represent the other two unary functions $f\left(  x_{1},0\right)
$ and $f\left(  x_{1},1\right)  $ by changing the functional evaluation matrix
to: $X^{f\left(  1,0\right)  }H_{f\left(  0,0\right)  }\otimes X^{f\left(
1,1\right)  }H_{f\left(  0,1\right)  }$.} Hence the $1$ in the third row
signifies $1=OE$. The parity of the whole binary function is immediately
determined by the parity of those two unary functions since the sum of all the
values is only even in the $EE$ and $OO$ cases (since the rule for adding even
and odd numbers is: $E+E=E=O+O$), and is otherwise odd (since $E+O=O=O+E$).

Starting with the initial state $\left\vert 00\right\rangle $, the gate
$H_{0}\otimes H_{0}$ gives the superposition $\left[  1,1,1,1\right]  ^{t}$
and the evaluation of the function gate $E_{f}$ at that superposition takes
the row sums of the evaluation matrix to yield the column vector to be
measured. Hence the circuit diagram of two-qubit/$2$ gates is:

\begin{center}
$%
\begin{array}
[c]{ccccccc}%
\left\vert 00\right\rangle  & \longrightarrow & \fbox{$H_0\otimes H_0$} &
\longrightarrow & \fbox{$E_f$} & \longrightarrow & \fbox{$M$}%
\end{array}
$

QC/$2$ algorithm for parity problem for binary Boolean functions.
\end{center}

This $n=2$ example indicates the pattern for the general case which uses
$n$-qubit/$2$ gates in $%
\mathbb{Z}
_{2}^{2}\otimes...\otimes%
\mathbb{Z}
_{2}^{2}$ ($n$ times) $=%
\mathbb{Z}
_{2}^{2^{n}}$:

\begin{center}
$%
\begin{array}
[c]{ccccccc}%
\left\vert 0...0\right\rangle  & \longrightarrow & \fbox{$H_0^\otimes n$} &
\longrightarrow & \fbox{$E_f$} & \longrightarrow & \fbox{$M$}%
\end{array}
$

QC/$2$ algorithm for Parity SAT problem for $n$-ary Boolean functions.
\end{center}

The Unambiguous SAT problem is--when one is given or "promised" that a Boolean
function has at most one case where it is satisfied--to find if it is
satisfied or not. The solution to the Parity SAT problem also solves the
Unambiguous SAT problem since "even" means no satisfying cases and "odd" means
one satisfying case.\footnote{Hanson et al. \cite{hansonsabry:dqt} give a
somewhat more complicated algorithm that solves the Unambiguous SAT problem in
QC/$2$.}

\noindent The quantum speedup is particularly clear since classically each of
the $2^{n}$ values of an $n$-ary Boolean function needs to be evaluated to
determine the parity of the sum of the values, but the QC/$2$ algorithm only
makes one functional evaluation for any $n$.

\section{Concluding overview}

QM/sets is the set version of the mathematics of quantum mechanics--without
any specifically physical concepts (e.g., the Hamiltonian or DeBroglie
relations). The connection between the two mathematical theories is the
sets-to-vector-spaces bridge (or ladder) provided by the basis principle and
used particularly by Weyl, but also by von Neumann and many others as it is
essentially part of the mathematical folklore.

In the context of toy models of QM on vector spaces over finite fields (i.e.,
"modal quantum theory" of Schumacher and Westmoreland \cite{schum:modal} or
the better-named "discrete quantum theory" of Hanson et al.
\cite{hansonsabry:dqt}), the special case of the base field $%
\mathbb{Z}
_{2}$ stands out since vectors can then be interpreted as a natural
mathematical objects, i.e., sets. It is \textit{only} this special case of
base field $%
\mathbb{Z}
_{2}$ that engages the sets-to-vector-spaces bridge of the lifting program.
Thus the notion of a partition of a set lifts to a direct sum decomposition of
a vector space, a numerical attribute on a set lifts to a linear operator on
the space, the inverse-image set partition given by the numerical attribute
lifts to the direct sum decomposition given by the eigenspaces of a
(diagonalizable) linear operator, and so forth.

The set version of some QM concept, result, or model represents the simplest
($%
\mathbb{Z}
_{2}$ is base field) essentials or "logic" of the matter, and in that
old-fashioned sense, QM/sets is proposed as the "logic" of QM. Thus QM/sets is
not only of pedagogical importance by showing the essential logic of the
subject; it provides a treatment of many aspects of "quantum weirdness" using
simple set concepts and thus it adds to the conceptual understanding (and
demystification) of QM. For instance, the probability calculus of QM distills
down in QM/sets to the usual Laplace-Boole calculus of logical finite
probability theory (reformulated in a "non-commutative" fashion over the
vector space $%
\mathbb{Z}
_{2}^{\left\vert U\right\vert }$ which allows different bases instead of just
one set $U$ of outcomes). And quantum entanglement distills down in QM/sets to
joint probability distributions on the direct product of two sets being
correlated rather than independent, and Bell's Theorem carries over to sets by
showing that the probabilities involved in QM/sets measurements could not come
from an independent joint distribution.

Quantum information theory based on QM/sets is essentially the logical
information theory defined by the normalized counting measure on partitions
(represented as partition relations or apartness relations) just as logical
probability theory is defined by the normalized counting measure on subsets
(events) of a universe set of outcomes. The normalized counting measure on
partitions is the notion of logical entropy \cite{ell:distinctions} that
provides a new logical foundation for information theory. Shannon's notion of
entropy is a higher-level concept adapted to the theory of communication (as
Shannon always named the theory \cite{shannon:comm}). The notion of logical
entropy of a partition can be formulated in terms of "delifted" density
matrices and it provides an exact interpretation of the entries in a density
matrix in terms of indistinction probabilities (and in terms of "coherence
probabilities" in the relifted version). The Shannon notion of entropy lifted
to quantum information theory as von Neumann entropy provides no such logical
analysis of a state (pure or mixed) represented in a density matrix; it is
suited for analyzing the quantum communications protocols lifted to QM from
Shannon's theory of communication through classical channels.

In quantum computing in QM/sets or QC/$2$, the coefficients in the gates are
only from $%
\mathbb{Z}
_{2}$ but the gates need only be non-singular (unitarity is only defined on
inner product spaces and vector spaces over finite fields have no inner
products). In addition to the no-cloning theorem and superdense coding, there
is a simple protocol for teleporting a qubit/$2$ using only one classical bit
(that foreshadows a little-known single-bit protocol in full QM
\cite{zhou:1bit-telep}). As an example of a quantum computing algorithm over
$2$ (or $%
\mathbb{Z}
_{2}$), the simplest Deutsch problem is reformulated as the Parity SAT problem
for unary Boolean functions, and is solved by a simple algorithm using only
one-qubit/$2$ gates. This then generalizes immediately to a QC/$2$ algorithm
solving the general Parity SAT problem (to determine the parity of sum of
values of an $n$-ary Boolean function). As expected, the algorithm is so
simple that the key role of superposition is obvious, and that superposition
gives the quantum speedup of a single function evaluation in contrast with the
$2^{n}$ evaluations needed classically.\footnote{This shows that the quantum
speedup has nothing to do with the greater power of calculations in the
complex numbers $%
\mathbb{C}
$ as opposed to classical computing using $%
\mathbb{Z}
_{2}$, since QC/$2$ is also restricted to $%
\mathbb{Z}
_{2}$.}

Finally, quantum mechanics over sets or QM/sets is part of a research program
that arose out of the recent development of the logic of partitions
(\cite{ell:partitions} and \cite{ell:intropartlogic}), the logic that is
mathematically dual to the ordinary Boolean logic of subsets (usually
mis-specified as the special case of "propositional" logic). This research
program ultimately aims to interpret quantum mechanics using the notion of
objective indefiniteness \cite{ell:objindef}. Quantum mechanics over sets is a
key part of that program since the fundamental QM notions such as eigenstates,
superpositions of eigenstates, and measurement in vector spaces are distilled
down into the "definite" (i.e., singleton) subsets, the "indefinite" (i.e.,
multiple element) subsets that "superpose" (i.e., collect together) a number
of definite elements, and the join-action of a set partition (inverse-image of
a numerical attribute) on a "pure" indefinite subset to create a "mixed state"
of more definite subsets.

\end{document}